\documentclass[prd,nopacs,preprintnumbers,twocolumn,amsmath,nofootinbib,amssymb]{revtex4}
\usepackage{rotating}
\usepackage{natbib}
\usepackage{longtable,pdflscape}
\usepackage{graphicx,color,dcolumn,booktabs,bm}
\usepackage{epsfig,dsfont,amssymb,amsmath,amsfonts,amsbsy,mathrsfs}
\usepackage{feynmf}   
\usepackage{slashed}  
\usepackage{cases}
\usepackage{xcolor}
\usepackage{float}
\usepackage{makecell}
\usepackage{multirow}
\usepackage{array}
\usepackage{gensymb}
\usepackage{txfonts}
\usepackage{overpic}
\usepackage{epstopdf}
\usepackage{indentfirst}
\usepackage{threeparttable}



\usepackage[margin=1in,letterpaper]{geometry} 

\graphicspath{{Figures/}} %

\usepackage{hyperref}
\hypersetup{colorlinks,citecolor=blue,anchorcolor=red,menucolor=red, linkcolor=red,filecolor=red,runcolor=red,urlcolor=blue,frenchlinks=red}

\makeatletter
\@addtoreset{equation}{section}
\makeatother
\renewcommand{\theequation}{\arabic{section}.\arabic{equation}}
\allowdisplaybreaks

\newcolumntype{I}{!{\vrule width 0.9pt}}

\begin{document}

\title{Nonleptonic decays of the double-charmed baryon $\Xi_{cc}^{+}$ with dynamically determined baryon wave functions}
\author{Yu-Shuai Li$^{1}$}\email{liysh@upc.edu.cn}
\author{Hong-Tao An$^{2}$}
\affiliation{$^1$ College of Science, China University of Petroleum, Qingdao, Shandong 266580, China\\
$^2$ Department of Physics and Center for High Energy Physics, Tsinghua University, Beijing 100084, China}

\begin{abstract}
Motivated by the recent observation of the double-charmed baryon $\Xi_{cc}^{+}$ by the LHCb Collaboration, we systematically investigate the two-body nonleptonic decays of the $\Xi_{cc}^{+}$ into a single-charmed baryon and a light pseudoscalar meson. In our calculation, both factorizable and nonfactorizable decay amplitudes are evaluated within the framework of nonrelativistic quark model combined with the baryon wave functions dynamically determined by numerically solving the three-body Schr\"{o}dinger equation with a nonrelativistic potential model. In particular, the nonfactorizable $W$-exchange amplitudes are estimated under pole model by considering $J^{P}=1/2^{\pm}$ low-lying intermediate charmed-baryon poles, and our results show that these contributions are comparable to or even dominate the factorizable amplitudes in several decay channels. Moreover, assuming a lifetime of $\tau_{\Xi_{cc}^{+}}=47~\text{fs}$, we predict relatively large branching fractions for several CF decays, including $\mathcal{B}(\Xi_{cc}^{+}\to\Xi_{c}^{\prime0}\pi^{+})=8.90\%$, $\mathcal{B}(\Xi_{cc}^{+}\to\Omega_{c}^{0}K^{+})=7.09\%$, $\mathcal{B}(\Xi_{cc}^{+}\to\Sigma_{c}^{+}\bar{K}^{0})=6.92\%$, $\mathcal{B}(\Xi_{cc}^{+}\to\Xi_{c}^{+}\pi^{0})=5.28\%$, and $\mathcal{B}(\Xi_{cc}^{+}\to\Sigma_{c}^{++}K^{-})=5.17\%$. These channels are promising candidates for future experimental searches at the LHCb.
\end{abstract}

\maketitle

\section{Introduction}
\label{sec1}

Very recently, the LHCb Collaboration reported the first observation of the singly charged double-charmed baryon $\Xi_{cc}^{+}$ in the $\Xi_{cc}^{+}\to\Lambda_{c}^{+}K^{-}\pi^{+}$ decay mode by using Run 3 data~\cite{LHCb:2026pxn}. The $\Xi_{cc}^{+}$ mass was measured to be
\begin{equation*}
m_{\Xi_{cc}^{+}}=(3619.97\pm0.83\pm0.26^{+1.90}_{-1.30})~\text{MeV},
\end{equation*}
and the masses difference with respect to the doubly charged double-charmed baryon $\Xi_{cc}^{++}$ is determined as~\cite{LHCb:2026pxn}
\begin{equation*}
\delta_{m}=m_{\Xi_{cc}^{+}}-m_{\Xi_{cc}^{++}}=(-1.77\pm0.84\pm0.15^{+1.90}_{-1.30})~\text{MeV}.
\end{equation*}
Such small masses difference is consistent with the theoretical expectation that the ground states $\Xi_{cc}^{++}(ccu)$ and $\Xi_{cc}^{+}(ccd)$ constitute an isospin doublet, and thereby providing a clear identification of the newly observed $\Xi_{cc}^{+}$ as the isospin partner of the $\Xi_{cc}^{++}$ observed in 2017 by the LHCb Collaboration~\cite{LHCb:2017iph}. More important, this observation finally resolves the long-standing experimental ambiguity surrounding the ground-state $\Xi_{cc}^{+}$.

More than two decades ago, the SELEX Collaboration reported a $\Xi_{cc}^{+}$ candidate, denoted as $\Xi_{cc}^{+}(3520)$, with a mass around 3520 MeV, in the $\Lambda_{c}^{+}K^{-}\pi^{+}$~\cite{SELEX:2002wqn} and $pD^{+}K^{-}$ final states~\cite{SELEX:2004lln}. However, neither signal was subsequently confirmed by the FOCUS~\cite{Ratti:2003ez}, BaBar~\cite{BaBar:2006bab}, Belle~\cite{Belle:2013htj} and LHCb~\cite{LHCb:2013hvt,LHCb:2019gqy} collaborations. In 2017, the ground state doubly charged double-charmed $\Xi_{cc}^{++}$ was first observed by the LHCb Collaboration in $\Lambda_{c}^{+}K^{-}\pi^{+}\pi^{+}$ final state~\cite{LHCb:2017iph}, and then confirmed in $\Xi_{cc}^{++}\to\Xi_{c}^{+}\pi^{+}$~\cite{LHCb:2018pcs}, $D^{+}pK^{-}\pi^{+}$~\cite{LHCb:2019ybf}, $\Xi_{c}^{\prime+}\pi^{+}$~\cite{LHCb:2022rpd}, and $\Xi_{c}^{0}\pi^{+}\pi^{+}$~\cite{LHCb:2025shu} decay modes. Its mass was established as~\cite{LHCb:2019epo}
\begin{equation*}
m_{\Xi_{cc}^{++}}=(3621.55\pm0.23\pm0.30)~\text{MeV}.
\end{equation*}
Puzzlingly, the mass splitting between the reported $\Xi_{cc}^{+}(3520)$ and the well-established $\Xi_{cc}^{++}$ is about 100 MeV, which significantly deviates from the expected size of isospin symmetry-breaking in any other baryon systems.

\begin{figure}[htbp]
\centering
  \includegraphics[width=80mm]{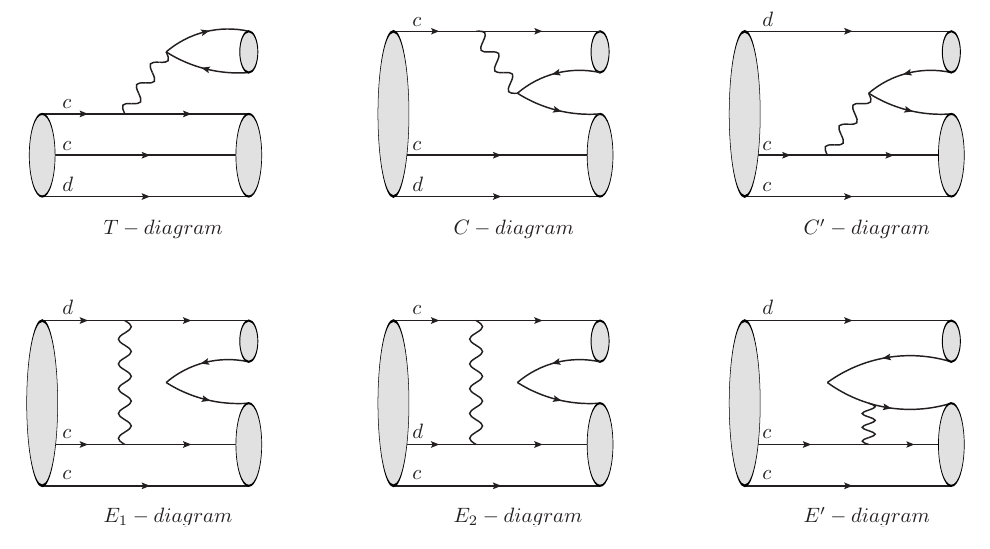}\\
  \caption{The Feynman diagrams of two-body nonleptonic decays of the double-charmed baryon $\Xi_{cc}^{+}$.}
  \label{fig:FeynmanDiagrams}
\end{figure}

The double-charmed baryon $\Xi_{cc}^{+}$ provides a unique platform for studying the interplay between heavy and light-quark dynamics in baryon systems. Moreover, as a ground state with $J^{P}=1/2^{+}$, the $\Xi_{cc}^{+}$ can decay only through the weak interaction. Therefore, studies of its nonleptonic decays are also helpful for probing weak decay mechanisms and QCD nonperturbative effects. Theoretically, the two-body nonleptonic decay amplitudes receive contributions from the following tree-level Feynman diagrams: the external $W$-emission ($T$) and internal $W$-emission ($C$) diagrams, which are factorizable, as well as the inner $W$-emission ($C^{\prime}$) and $W$-exchange ($E_{1,2},E^{\prime}$) diagrams, which are nonfactorizable~\cite{Cheng:2021qpd}. The corresponding Feynman diagrams are displayed in Fig.~\ref{fig:FeynmanDiagrams}, while the contributing Feynman diagrams for our concerned nonleptonic decays are summarized in Table~\ref{tab:FeynmanDiagrams}. We note that although the $C^{\prime}$ and $E^{\prime}$ diagrams correspond to the same topology, they represent distinct contributions to the decay amplitudes in our calculation. In addition, penguin diagram contributions also play important roles in double-charmed baryon nonleptonic decays~\cite{Liu:2025hbf}. However, in the present analysis, we retain only the contributions from current-current operators, while those from penguin operators are neglected.

\begin{table*}[htbp]
\centering
\caption{The contributing Feynman diagrams for concerned two-body nonleptonic decays of $\Xi_{cc}^{+}$.}
\label{tab:FeynmanDiagrams}
\renewcommand\arraystretch{1.15}
\begin{tabular*}{140mm}{l@{\extracolsep{\fill}}clclc}
\toprule[1pt]
\toprule[0.5pt]
CF decay  &diagrams  &SCS decay  &diagrams  &DCS decay  &diagrams\\
\midrule[0.5pt]
$\Xi_{cc}^{+}\to\Xi_{c}^{(\prime)0}\pi^{+}$  &$T,E_{1}$  &$\Xi_{cc}^{+}\to\Sigma_{c}^{0}\pi^{+}$  &$T,E_{1}$  &$\Xi_{cc}^{+}\to\Sigma_{c}^{0}K^{+}$  &$T$\\
$\Xi_{cc}^{+}\to\Lambda_{c}^{+}\bar{K}^{0}$  &$C,E_{2}$  &$\Xi_{cc}^{+}\to\Xi_{c}^{(\prime)0}K^{+}$  &$T,E_{1}$  &$\Xi_{cc}^{+}\to\Lambda_{c}^{+}K^{0}$  &$C,C^{\prime},E^{\prime}$\\
$\Xi_{cc}^{+}\to\Sigma_{c}^{+}\bar{K}^{0}$  &$C,E_{2}$  &$\Xi_{cc}^{+}\to\Lambda_{c}^{+}\pi^{0}$  &$C,C^{\prime},E_{1},E_{2},E^{\prime}$  &$\Xi_{cc}^{+}\to\Sigma_{c}^{+}K^{0}$  &$C,C^{\prime},E^{\prime}$\\
$\Xi_{cc}^{+}\to\Xi_{c}^{(\prime)+}\pi^{0}$  &$C^{\prime},E_{1},E^{\prime}$  &$\Xi_{cc}^{+}\to\Sigma_{c}^{+}\pi^{0}$  &$C,C^{\prime},E_{1},E_{2},E^{\prime}$  &$$  &$$\\
$\Xi_{cc}^{+}\to\Xi_{c}^{(\prime)+}\eta^{(\prime)}$  &$C^{\prime},E_{1},E_{2},E^{\prime}$  &$\Xi_{cc}^{+}\to\Lambda_{c}^{+}\eta^{(\prime)}$  &$C,C^{\prime},E_{1},E_{2},E^{\prime}$  &$$  &$$\\
$\Xi_{cc}^{+}\to\Sigma_{c}^{++}K^{-}$  &$E_{2}$  &$\Xi_{cc}^{+}\to\Sigma_{c}^{+}\eta^{(\prime)}$  &$C,C^{\prime},E_{1},E_{2},E^{\prime}$  &$$  &$$\\
$\Xi_{cc}^{+}\to\Omega_{c}^{0}K^{+}$  &$E_{1}$  &$\Xi_{cc}^{+}\to\Xi_{c}^{(\prime)+}K^{0}$  &$C^{\prime},E_{2},E^{\prime}$  &$$  &$$\\
& &$\Xi_{cc}^{+}\to\Sigma_{c}^{++}\pi^{-}$  &$E_{2}$  &$$  &$$\\
\bottomrule[0.5pt]
\bottomrule[1pt]
\end{tabular*}
\end{table*}

Under the naive factorization assumption, the factorizable $T$ and $C$ amplitudes can often be easily evaluated as the product of baryon transition matrix element and meson decay constant, i.e.,
\begin{equation}
\mathcal{M}_{T,C}=\langle\mathcal{B}_{f}P\vert\mathcal{H}_{\text{eff}}\vert\mathcal{B}_{i}\rangle
\sim a_{1,2}f_{P}\times[f_{1},g_{1}],
\end{equation}
where $a_{1,2}$ are the effective Wilson coefficients, $f_{P}$ is the pseudoscalar-meson decay constant, and $f_{1}$ and $g_{1}$ represent the form factors of baryon transition matrix element. In contrast, the nonfactorizable amplitudes, particularly those arising from $W$-exchange diagrams are more challenging to quantify. The experimental measurement of the purely $W$-exchange process $\Lambda_{c}^{+}\to\Xi^{0}K^{+}$~\cite{BESIII:2023wrw} provides direct evidence that the nonfactorizable $W$-exchange contributions can be sizable and therefore cannot be neglected in charmed baryon nonleptonic decays. Moreover, the nonleptonic decay amplitudes involve intrinsically nonperturbative baryon transition matrix elements, thereby posing a significant theoretical challenge to their reliable calculations. To tackle the theoretical challenges posed by nonfactorizable amplitudes and nonperturbative effects, various theoretical approaches have been developed to study the nonleptonic decays of double-charmed baryons, including pole model~\cite{Sharma:2017txj,Cheng:2020wmk,Liu:2022igi,Liu:2023dvg,Li:2025mxz,Li:2025noz}, the covariant confined quark model~\cite{Gutsche:2018msz,Gutsche:2019iac,Gutsche:2019wgu,Ivanov:2020iaq}, light-cone sum rules~\cite{Shi:2022kfa}, and current algebra approach~\cite{Groote:2021pxt}. The final-state-interaction (FSI) effects are also taken into account in Refs.~\cite{Yu:2017zst,Jiang:2018oak,Wang:2022wrb,Hu:2024uia,Liu:2025hbf}. Besides, the irreducible-representation approach (IRA)~\cite{Wang:2017azm,Shi:2017dto,Geng:2017mxn,Wang:2019dls} and the topological diagram approach (TDA)~\cite{Liu:2025hbf} based on SU(3)$_{F}$ symmetry analyses has been widely employed to explore relations among different decay amplitudes. In addition, the isospin sum rules~\cite{Luo:2023vbx} and SU(3) sum rules~\cite{Wang:2022kwe} were also employed to establish relations among two- or three-body nonleptonic decay amplitudes.

As an effective phenomenological framework for evaluating nonfactorizable $W$-exchange amplitudes, pole model has been widely used in studies of charmed-baryon nonleptonic decays~\cite{Cheng:1991sn,Cheng:1992ff,Cheng:1993gf,Uppal:1994pt,Sharma:2017txj,Cheng:2018hwl,Zou:2019kzq,Cheng:2020wmk,Hu:2020nkg,Meng:2020euv,Niu:2020gjw,Niu:2021qcc,Zeng:2022egh,Cheng:2022kea,Cheng:2022jbr,Liu:2022igi,Ivanov:2023wir,Liu:2023dvg,Li:2025alu,Li:2025mxz,Li:2025noz,Niu:2025lgt}. In pole model, the $W$-exchange decay amplitudes are described through intermediate baryon-pole states. More specifically, the parity-violating (PV) decay amplitudes arise predominantly from low-lying $J^{P}=1/2^{-}$ baryon poles, whereas the parity-conserving (PC) ones are dominated by $J^{P}=1/2^{+}$ ground-state baryon poles~\cite{Cheng:1991sn}. Meanwhile, the nonperturbative parameters of baryon transition matrix elements were estimated within the bag model~\cite{Cheng:2018hwl,Zou:2019kzq,Hu:2020nkg,Meng:2020euv,Cheng:2022kea,Cheng:2022jbr,Liu:2022igi}, the nonrelativistic quark model (NRQM)~\cite{Niu:2020gjw,Niu:2021qcc,Zeng:2022egh,Niu:2025lgt,Li:2025alu,Li:2025mxz,Li:2025noz}, the covariant confined quark model~\cite{Ivanov:2023wir}, or constrained using flavor symmetry~\cite{Liu:2023dvg}. Despite these theoretical efforts, substantial model dependence remains in the calculations of nonfactorizable contributions and nonperturbative parameters.

In this work, we investigate the two-body nonleptonic decays $\Xi_{cc}^{+}\to\mathcal{B}_{c}P$, where $\mathcal{B}_{c}$ and $P$ represent a charmed baryon and a light pseudoscalar meson, respectively. Particular attention is paid to the role of $W$-exchange contributions, and to identifying the most promising $\Xi_{cc}^{+}$ baryon decay channels for future experimental measurements. In our numerical calculation, the relevant decay amplitudes are evaluated within the framework of NRQM, and particularly the nonfactorizable contributions from $W$-exchange diagrams are treated using pole model. As well known, a major theoretical uncertainty in calculations of nonleptonic decay amplitudes arises from the nonperturbative parameters in baryon transition matrix elements, which are sensitive to underlying baryon wave functions. Addressing this challenge, we employ the baryon wave function dynamically determined by solving the three-quark Schr\"{o}dinger equation with a nonrelativistic quark potential, rather than adopt a simply Gaussian ansatz. Thus, our calculation reduces the sensitivity of the decay amplitudes on ad hoc wave function choices, thereby mitigating the associated model uncertainty. The parameters of the nonrelativistic quark potential can be constrained by charmed baryon spectroscopy, allowing the wave functions of all baryons entering the decay amplitudes, including the initial, final, and intermediate baryon states, to be determined consistently within the same dynamical framework. Furthermore, the obtained baryon spectrum also provides a natural criterion for truncating the sum over intermediate baryon states in the $W$-exchange amplitudes, in which the dominant contributions are expected to arise from the ground and low-lying excited states.

This paper is organized as follows. In Sec.~\ref{sec2}, we derive the weak decay amplitudes within the framework of NRQM. Then in Sec.~\ref{sec3}, we introduce the nonrelativistic quark potential, and determine the relevant charmed baryon wave functions using Gaussian expansion method. In Sec.~\ref{sec4}, we present the numerical results for the decay amplitudes, and future discuss the branching fractions and asymmetry parameters $\alpha$. Finally, a brief summary is given in Sec.~\ref{sec5}.

\section{Weak decay amplitudes in the nonrelativistic quark model}
\label{sec2}

The effective Hamiltonian relevant to the weak decays considered in this work is given by
\begin{equation}
\mathcal{H}_{\text{eff}}=\frac{G_{F}}{\sqrt{2}}V_{\text{CKM}}V_{\text{CKM}}^{\prime*}
\big{(}c_{1}\hat{\mathcal{O}}_{1}+c_{2}\hat{\mathcal{O}}_{2}\big{)},
\end{equation}
where $G_{F}=1.16637\times10^{-5}~\text{GeV}^{-2}$ is the Fermi coupling constant, $V_{\text{CKM}}V_{\text{CKM}}^{\prime}$ denotes the relevant product of Cabibbo-Kobayashi-Maskawa (CKM) matrix elements, i.e.,
\begin{equation}
V_{\text{CKM}}V_{\text{CKM}}^{\prime*}=\Bigg{\{}
\begin{array}{ll}
V_{cs}V_{ud}^{*}   &\text{for}\ \text{CF},\\
V_{cs}V_{us}^{*}\ \text{or}\ V_{cd}V_{ud}^{*}  &\text{for}\ \text{SCS},\\
V_{cd}V_{us}^{*}   &\text{for}\ \text{DCS}.\\
\end{array}\\
\end{equation}
In the Wolfenstein parameterization up to $O(\lambda^{2})$, the CKM matrix elements are written as
\begin{equation*}
\begin{split}
V_{ud}=&1-\lambda^{2}/2,~~V_{us}=\lambda,\\
V_{cd}=&-\lambda,~~V_{cs}=1-\lambda^{2}/2
\end{split}
\end{equation*}
with $\lambda=0.22501$~\cite{ParticleDataGroup:2026mpi}. $c_{1,2}$ are Wilson coefficients, and in the numerical calculations, we adopt $c_{1}=1.346$ and $c_{2}=-0.636$ at the renormalization scale $\mu=1.25~\text{GeV}$~\cite{Cheng:2020wmk}. For the four-fermion operators, taking the CF decay as an example, we define
\begin{equation}
\hat{\mathcal{O}}_{1}=(\bar{u}d)(\bar{s}c),~~~\hat{\mathcal{O}}_{2}=(\bar{s}d)(\bar{u}c),
\end{equation}
respectively, where $(\bar{q}_{1}q_{2})\equiv\bar{q}_{1}\gamma_{\mu}(1-\gamma_{5})q_{2}$.

Based on the effective Hamiltonian introduced above, we next evaluate the $W$-emission and $W$-exchange amplitudes within the NRQM.

\subsection{The decay amplitudes of $W$-emission diagrams}
\label{sec2.1}

\begin{figure}[htbp]\centering
  \includegraphics[width=55mm]{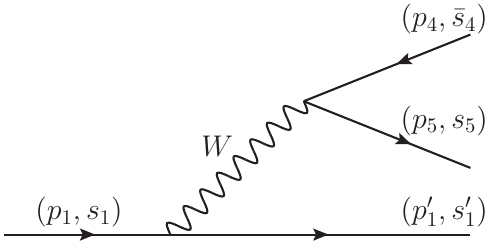}\\
  \caption{The kinematic conventions of $W$-emission process.}
  \label{fig:13}
\end{figure}

Following the conventions shown in Fig.~\ref{fig:13}, the effective Hamiltonian for the $W$-emission process can be written as
\begin{equation}
\begin{split}
\mathcal{H}_{1\to3}=&\frac{G_{F}}{\sqrt{2}}V_{\text{CKM}}V_{\text{CKM}}^{\prime*}\frac{\beta}{(2\pi)^{3}}\delta^{3}(\pmb{p}_{1}-\pmb{p}_{1}^{\prime}-\pmb{p}_{4}-\pmb{p}_{5})\\
&\times\hat{\alpha}_{1}^{(-)}\hat{I}_{P}\Big{[}c_{1}(\mu)\hat{\mathcal{O}}_{1}+c_{2}(\mu)\hat{\mathcal{O}}_{2}\Big{]},
\label{eq:Heff123}
\end{split}
\end{equation}
where $\pmb{p}_{i}$ denotes the three-momentum of $i$th quark, and $\hat{\alpha}_{1}^{(-)}\vert{c}\rangle=\vert{s}\rangle~(\text{or}~\vert{d}\rangle)$ is the flavor-changing operator acting on quark 1. The factor $\beta$ is taken to be 2.

In the nonrelativistic limit, the effective Hamiltonian~\eqref{eq:Heff123} associated with the operator $\hat{\mathcal{O}}_{1}$ can be decomposed into parity-conserving (PC) and parity-violating (PV) parts as~\cite{Niu:2020gjw,Niu:2021qcc,Niu:2025lgt}
\begin{widetext}
\begin{equation}
\begin{split}
\mathcal{H}_{1,1\to3}^{\text{PC}}=&\frac{G_{F}}{\sqrt{2}}V_{\text{CKM}}V_{\text{CKM}}^{\prime*}\frac{\beta}{(2\pi)^{3}}\delta^{3}(\pmb{p}_{1}-\pmb{p}_{1}^{\prime}-\pmb{p}_{4}-\pmb{p}_{5})
\Bigg{\{}\langle{s_{z,1}^{\prime}}\vert{I}\vert{s_{z,1}}\rangle\langle{s_{z,5}\bar{s}_{z,4}}\vert{\pmb{\sigma}}\vert{0}\rangle\Bigg{(}\frac{\pmb{p}_{5}}{2m_{5}}+\frac{\pmb{p}_{4}}{2m_{4}}\Bigg{)}\\
&-\Bigg{[}\Bigg{(}\frac{\pmb{p}_{1}^{\prime}}{2m_{1}^{\prime}}+\frac{\pmb{p}_{1}}{2m_{1}}\Bigg{)}\langle{s_{z,1}^{\prime}}\vert{I}\vert{s_{z,1}}\rangle-i\langle{s_{z,1}^{\prime}}\vert{\pmb{\sigma}}\vert{s_{z,1}}\rangle\times\Bigg{(}\frac{\pmb{p}_{1}}{2m_{1}}-\frac{\pmb{p}_{1}^{\prime}}{2m_{1}^{\prime}}\Bigg{)}\Bigg{]}\langle{s_{z,5}\bar{s}_{z,4}}\vert{\pmb{\sigma}}\vert{0}\rangle\\
&-\langle{s_{z,1}^{\prime}}\vert{\pmb{\sigma}}\vert{s_{z,1}}\rangle\Bigg{[}\Bigg{(}\frac{\pmb{p}_{5}}{2m_{5}}+\frac{\pmb{p}_{4}}{2m_{4}}\Bigg{)}\langle{s_{z,5}\bar{s}_{z,4}}\vert{I}\vert{0}\rangle-i\langle{s_{z,5}\bar{s}_{z,4}}\vert{\pmb{\sigma}}\vert{0}\rangle\times\Bigg{(}\frac{\pmb{p}_{4}}{2m_{4}}-\frac{\pmb{p}_{5}}{2m_{5}}\Bigg{)}\Bigg{]}\\
&+\langle{s_{z,1}^{\prime}}\vert{\pmb{\sigma}}\vert{s_{z,1}}\rangle\Bigg{(}\frac{\pmb{p}_{1}^{\prime}}{2m_{1}^{\prime}}+\frac{\pmb{p}_{1}}{2m_{1}}\Bigg{)}\langle{s_{z,5}\bar{s}_{z,4}}\vert{I}\vert{0}\rangle\Bigg{\}}\hat{\alpha}_{1}^{(-)}\hat{I}_{P},\\
\mathcal{H}_{1,1\to3}^{\text{PV}}=&\frac{G_{F}}{\sqrt{2}}V_{\text{CKM}}V_{\text{CKM}}^{\prime*}\frac{\beta}{(2\pi)^{3}}\delta^{3}(\pmb{p}_{1}-\pmb{p}_{1}^{\prime}-\pmb{p}_{4}-\pmb{p}_{5})
\Bigg{(}-\langle{s_{z,1}^{\prime}}\vert{I}\vert{s_{z,1}}\rangle\langle{s_{z,5}\bar{s}_{z,4}}\vert{I}\vert{0}\rangle\\
&+\langle{s_{z,1}^{\prime}}\vert{\pmb{\sigma}}\vert{s_{z,1}}\rangle\langle{s_{z,5}\bar{s}_{z,4}}\vert{\pmb{\sigma}}\vert{0}\rangle\Bigg{)}\hat{\alpha}_{1}^{(-)}\hat{I}_{P},
\label{eq:H13}
\end{split}
\end{equation}
where $I$ is the $2\times2$ identity matrix, while $\pmb{\sigma}=(\sigma^{1},\sigma^{2},\sigma^{3})$ represent the Pauli matrices. Besides, the $\vert{s_{z,i}}\rangle$ is used as an abbreviation for $\vert{s_{i},s_{z,i}}\rangle$, representing the spin of the $i$th quark. The effective Hamiltonian $\mathcal{H}_{2,1\to3}^{\text{PC},\text{PV}}$ associated with $\hat{\mathcal{O}}_{2}$ can be obtained from Eq.~\eqref{eq:H13} through the replacements $(p_{1}^{\prime},m_{1}^{\prime})\leftrightarrows(p_{5},m_{5})$ and $s_{z,1}^{\prime}{\leftrightarrows}s_{z,5}$.

For the $W$-emission diagrams ($T$, $C$, and $C^{\prime}$), the corresponding decay amplitudes for the $\mathcal{B}_{i}(\pmb{P}_{i};J^{i},J_{z}^{i})\to\mathcal{B}_{f}(\pmb{P}_{f};J^{f},J_{z}^{f})+P(\pmb{k})$ process are given by
\begin{eqnarray}
\label{eq:Tamplitude}
\mathcal{M}_{T,\text{PC}(\text{PV})}^{J_{z}^{f},J_{z}^{i}}&=&\langle{\mathcal{B}_{f}(\pmb{P}_{f};J^{f},J_{z}^{f})P(\pmb{k})}{\vert}
\Big{[}c_{1}(\mu)\mathcal{H}_{1,1\to3}^{\text{PC}(\text{PV})}+\frac{c_{2}(\mu)}{N_{c}}\mathcal{H}_{2,1\to3}^{\text{PC}(\text{PV})}\Big{]}
{\vert}{\mathcal{B}_{i}(\pmb{P}_{i};J^{i},J_{z}^{i})}\rangle,\\
\label{eq:Camplitude}
\mathcal{M}_{C,\text{PC}(\text{PV})}^{J_{z}^{f},J_{z}^{i}}&=&\langle{\mathcal{B}_{f}(\pmb{P}_{f};J^{f},J_{z}^{f})P(\pmb{k})}{\vert}
\Big{[}\frac{c_{1}(\mu)}{N_{c}}\mathcal{H}_{1,1\to3}^{\text{PC}(\text{PV})}+c_{2}(\mu)\mathcal{H}_{2,1\to3}^{\text{PC}(\text{PV})}\Big{]}
{\vert}{\mathcal{B}_{i}(\pmb{P}_{i};J^{i},J_{z}^{i})}\rangle,\\
\label{eq:Cprimamplitude}
\mathcal{M}_{C^{\prime},\text{PC}(\text{PV})}^{J_{z}^{f},J_{z}^{i}}&=&\langle{\mathcal{B}_{f}(\pmb{P}_{f};J^{f},J_{z}^{f})P(\pmb{k})}{\vert}
\Big{[}\frac{c_{1}(\mu)}{N_{c}}\mathcal{H}_{1,1\to3}^{\text{PC}(\text{PV})}-\frac{c_{2}(\mu)}{N_{c}}\mathcal{H}_{2,1\to3}^{\text{PC}(\text{PV})}\Big{]}
{\vert}{\mathcal{B}_{i}(\pmb{P}_{i};J^{i},J_{z}^{i})}\rangle,
\end{eqnarray}
where $J^{i}=J^{f}=1/2$, and $N_{c}=3$ denotes the number of colors with the factor $1/N_{c}$ accounting for color suppression. The three-momentum of the initial double-charmed baryon is $\pmb{P}_{i}=\pmb{p}_{1}+\pmb{p}_{2}+\pmb{p}_{3}$, while the ones of the final charmed baryon and pseudoscalar meson are determined as $\pmb{P}_{f}=\pmb{p}_{1}^{\prime}+\pmb{p}_{2}+\pmb{p}_{3}~(\pmb{P}_{f}=\pmb{p}_{2}+\pmb{p}_{3}+\pmb{p}_{5},~\pmb{P}_{f}=\pmb{p}_{1}^{\prime}+\pmb{p}_{2}+\pmb{p}_{5})$ and $\pmb{k}=\pmb{p}_{4}+\pmb{p}_{5}~(\pmb{k}=\pmb{p}_{1}^{\prime}+\pmb{p}_{4},~\pmb{k}=\pmb{p}_{3}+\pmb{p}_{4})$ for $T(C,C^{\prime})$ diagrams, respectively, by the conventions in Fig.~\ref{fig:13}. In the rest frame of initial state, $\pmb{P}_{i}=0$, and the momentum conservation implies $\pmb{P}_{f}=-\pmb{k}$.

Taking the $T$ diagram as an example, the decay amplitude~\eqref{eq:Tamplitude} can be decomposed into spin-flavor and spatial matrix elements as
\begin{equation}
\begin{split}
\mathcal{M}_{T,\text{PC}(\text{PV})}^{J_{z}^{f},J_{z}^{i}}=&\big{\langle}{\mathcal{B}_{f}(\pmb{P}_{f};J^{f},J_{z}^{f})P(\pmb{k})}{\vert}
\Big{[}c_{1}(\mu)\mathcal{H}_{1,1\to3}^{\text{PC}(\text{PV})}+\frac{c_{2}(\mu)}{N_{c}}\mathcal{H}_{2,1\to3}^{\text{PC}(\text{PV})}\Big{]}
{\vert}{\mathcal{B}_{i}(\pmb{P}_{i};J^{i},J_{z}^{i})}\big{\rangle}\\
=&\sum_{S_{z}^{f},L_{z}^{f},S_{z}^{i},L_{z}^{i}}\langle{S^{f},S_{z}^{f};L^{f},L_{z}^{f}}\vert{J^{f},J_{z}^{f}}\rangle\langle{S^{i},S_{z}^{i};L^{i},L_{z}^{i}}\vert{J^{i},J_{z}^{i}}\rangle\\
&\times\Big{[}
c_{1}(\mu)\big{\langle}\phi_{f}\chi_{f}^{S^{f},S_{z}^{f}};\phi_{P}\chi_{P}\vert\hat{\mathcal{O}}_{1,1\to3}^{\text{spin}}\hat{\mathcal{O}}_{1,1\to3}^{\text{flavor}}\vert\phi_{i}\chi_{i}^{S^{i},S_{z}^{i}}\big{\rangle}
\big{\langle}\psi_{P}(\pmb{k})\psi_{L^{f},L_{z}^{f}}(\pmb{P}_{f})\vert{\hat{\mathcal{O}}_{1,1\to3}^{\text{spatial}}}\vert\psi_{L^{i},L_{z}^{i}}(\pmb{P}_{i})\big{\rangle}\\
&+\frac{c_{2}(\mu)}{N_{c}}\big{\langle}\phi_{f}\chi_{f}^{S^{f},S_{z}^{f}};\phi_{P}\chi_{P}\vert\hat{\mathcal{O}}_{2,1\to3}^{\text{spin}}\hat{\mathcal{O}}_{2,1\to3}^{\text{flavor}}\vert\phi_{i}\chi_{i}^{S^{i},S_{z}^{i}}\big{\rangle}
\big{\langle}\psi_{P}(\pmb{k})\psi_{L^{f},L_{z}^{f}}(\pmb{P}_{f})\vert{\hat{\mathcal{O}}_{2,1\to3}^{\text{spatial}}}\vert\psi_{L^{i},L_{z}^{i}}(\pmb{P}_{i})\big{\rangle}
\Big{]},
\label{eq:amplitudes1}
\end{split}
\end{equation}
\end{widetext}
where $\langle{S,S_{z};L,L_{z}}\vert{J,J_{z}}\rangle$ is the Clebsch-Gordan coefficient, while $\phi$, $\chi$ and $\psi$ denote the flavor, spin and spatial wave functions of the corresponding hadrons, respectively. The color-overlap factor has been incorporated into the factor $1/N_{c}$. The concrete expressions of decay amplitudes for the $C$ and $C^{\prime}$ diagrams can be decomposed in an analogous manner. For clarity and conciseness, their explicit expressions are not repeated here.

The spin-flavor matrix elements appearing in Eq.~\eqref{eq:amplitudes1} can be evaluated using the spin and flavor wave functions of the initial and final baryons and pseudoscalar meson. The details of the calculation, together with the resulting spin-flavor matrix element, are collected in Appendix~\ref{app01}. After separating out the spin-flavor contributions, the remaining spatial overlap integrals can be written as
\begin{equation}
\begin{split}
\mathcal{I}&_{i,T}^{L^{f},L_{z}^{f};L^{i},L_{z}^{i}}=\langle{\psi_{P}(\pmb{k})\psi_{L^{f},L_{z}^{f}}(\pmb{P}_{f})}\vert\hat{\mathcal{O}}_{i,1\to3}^{\text{spatial}}(\pmb{p}_{i})\vert{\psi_{L^{i},L_{z}^{i}}(\pmb{P}_{i})}\rangle\\
=&\int{d\pmb{p}_{1}d\pmb{p}_{2}d\pmb{p}_{3}d\pmb{p}_{1}^{\prime}d\pmb{p}_{4}d\pmb{p}_{5}}\delta^{3}(\pmb{p}_{1}-\pmb{p}_{4}-\pmb{p}_{5}-\pmb{p}_{1}^{\prime})\\
&\times\delta^{3}(\pmb{P}_{i}{\!}-{\!}\pmb{p}_{1}{\!}-{\!}\pmb{p}_{2}{\!}-{\!}\pmb{p}_{3})
\delta^{3}(\pmb{P}_{f}{\!}-{\!}\pmb{p}_{1}^{\prime}{\!}-{\!}\pmb{p}_{2}{\!}-{\!}\pmb{p}_{3})\\
&\times\delta^{3}(\pmb{k}{\!}-{\!}\pmb{p}_{5}{\!}-{\!}\pmb{p}_{4})\psi_{L^{f},L_{z}^{f}}^{*}(\pmb{p}_{1}^{\prime},\pmb{p}_{3},\pmb{p}_{2})\psi_{L^{i},L_{z}^{i}}(\pmb{p}_{1},\pmb{p}_{2},\pmb{p}_{3})\\
&\times\mathcal{O}_{i,1\to3}^{\text{spatial}}(\pmb{p}_{i})\psi^{*}_{P}(\pmb{p}_{5},\pmb{p}_{4}),
\end{split}
\end{equation}
\begin{equation}
\begin{split}
\mathcal{I}&_{i,C}^{L^{f},L_{z}^{f};L^{i},L_{z}^{i}}=\langle{\psi_{P}(\pmb{k})\psi_{L^{f},L_{z}^{f}}(\pmb{P}_{f})}\vert\hat{\mathcal{O}}_{i,1\to3}^{\text{spatial}}(\pmb{p}_{i})\vert{\psi_{L^{i},L_{z}^{i}}(\pmb{P}_{i})}\rangle\\
=&\int{d\pmb{p}_{1}d\pmb{p}_{2}d\pmb{p}_{3}d\pmb{p}_{1}^{\prime}d\pmb{p}_{4}d\pmb{p}_{5}}\delta^{3}(\pmb{p}_{1}-\pmb{p}_{4}-\pmb{p}_{5}-\pmb{p}_{1}^{\prime})\\
&\times\delta^{3}(\pmb{P}_{i}{\!}-{\!}\pmb{p}_{1}{\!}-{\!}\pmb{p}_{2}{\!}-{\!}\pmb{p}_{3})
\delta^{3}(\pmb{P}_{f}{\!}-{\!}\pmb{p}_{5}{\!}-{\!}\pmb{p}_{2}{\!}-{\!}\pmb{p}_{3})\\
&\times\delta^{3}(\pmb{k}{\!}-{\!}\pmb{p}_{1}^{\prime}{\!}-{\!}\pmb{p}_{4})\psi_{L^{f},L_{z}^{f}}^{*}(\pmb{p}_{3},\pmb{p}_{5},\pmb{p}_{2})\psi_{L^{i},L_{z}^{i}}(\pmb{p}_{1},\pmb{p}_{2},\pmb{p}_{3})\\
&\times\mathcal{O}_{i,1\to3}^{\text{spatial}}(\pmb{p}_{i})\psi^{*}_{P}(\pmb{p}_{1}^{\prime},\pmb{p}_{4}),
\end{split}
\end{equation}
\begin{equation}
\begin{split}
\mathcal{I}&_{i,C^{\prime}}^{L^{f},L_{z}^{f};L^{i},L_{z}^{i}}=\langle{\psi_{P}(\pmb{k})\psi_{L^{f},L_{z}^{f}}(\pmb{P}_{f})}\vert\hat{\mathcal{O}}_{i,1\to3}^{\text{spatial}}(\pmb{p}_{i})\vert{\psi_{L^{i},L_{z}^{i}}(\pmb{P}_{i})}\rangle\\
=&\int{d\pmb{p}_{1}d\pmb{p}_{2}d\pmb{p}_{3}d\pmb{p}_{1}^{\prime}d\pmb{p}_{4}d\pmb{p}_{5}}\delta^{3}(\pmb{p}_{1}-\pmb{p}_{4}-\pmb{p}_{5}-\pmb{p}_{1}^{\prime})\\
&\times\delta^{3}(\pmb{P}_{i}{\!}-{\!}\pmb{p}_{1}{\!}-{\!}\pmb{p}_{2}{\!}-{\!}\pmb{p}_{3})
\delta^{3}(\pmb{P}_{f}{\!}-{\!}\pmb{p}_{1}^{\prime}{\!}-{\!}\pmb{p}_{5}{\!}-{\!}\pmb{p}_{2})\\
&\times\delta^{3}(\pmb{k}{\!}-{\!}\pmb{p}_{3}{\!}-{\!}\pmb{p}_{4})\psi_{L^{f},L_{z}^{f}}^{*}(\pmb{p}_{1}^{\prime},\pmb{p}_{5},\pmb{p}_{2})\psi_{L^{i},L_{z}^{i}}(\pmb{p}_{1},\pmb{p}_{2},\pmb{p}_{3})\\
&\times\mathcal{O}_{i,1\to3}^{\text{spatial}}(\pmb{p}_{i})\psi^{*}_{P}(\pmb{p}_{3},\pmb{p}_{4}),
\end{split}
\end{equation}
where $i=1,2$ labels the two four-fermion operator structures. The operator $\mathcal{O}_{i,1\to3}^{\text{spatial}}(\pmb{p}_{i})$ represents the spatial 
part of the corresponding weak interaction operator, and it is the function of $\pmb{p}_{i}/(2m_{i})$ for $\text{PC}$ amplitudes, whereas $\mathcal{O}_{i,1\to3}^{\text{spatial}}(\pmb{p}_{i})\equiv1$ for $\text{PV}$ amplitudes. The evaluation of these spatial overlap integrals requires the spatial wave functions of both charmed baryons and pseudoscalar mesons involved in decay processes. For the pseudoscalar meson, we employ the simple-harmonic-oscillator wave function~\cite{Niu:2020gjw,Niu:2021qcc}
\begin{equation}
\psi_{P}(\pmb{k}_{1},\pmb{k}_{2})=\frac{1}{\pi^{3/4}R_{P}^{3/2}}\exp\Big{[}-\frac{(\pmb{k}_{1}-\pmb{k}_{2})^{2}}{8R_{P}^{2}}\Big{]},
\end{equation}
where $R_{P}$ is a phenomenological parameter. In the numerical calculations, we take $R_{\pi}=0.28~\text{GeV}$, $R_{K}=0.50~\text{GeV}$, $R_{\eta}=0.40~\text{GeV}$, and $R_{\eta^{\prime}}=0.90~\text{GeV}$~\cite{Niu:2025lgt,Li:2025noz}. 

For the spatial wave function of charmed baryon, we adopt the expression as
\begin{equation}
\begin{split}
\psi_{L,L_{z}}(\pmb{k}_{1},\pmb{k}_{2},\pmb{k}_{3})=&\sum_{n_{\rho},n_{\lambda}}\sum_{m_{\rho},m_{\lambda}}\langle{l_{\rho},m_{\rho};l_{\lambda},m_{\lambda}}\vert{L,L_{z}}\rangle\\
\times&{C^{(n_{\rho},n_{\lambda})}}\psi_{n_{\rho},l_{\rho},m_{\rho}}(\pmb{\rho})\psi_{n_{\lambda},l_{\lambda},m_{\lambda}}(\pmb{\lambda}),
\label{eq:wavefunctions}
\end{split}
\end{equation}
where the Jacobi momenta of $\rho$ and $\lambda$-modes are defined as
\begin{equation}
\begin{split}
\pmb{\rho}=&\frac{m_{1}\pmb{k}_{2}-m_{2}\pmb{k}_{1}}{m_{1}+m_{2}},\\
\pmb{\lambda}=&\frac{(m_{1}+m_{2})\pmb{k}_{3}-m_{3}(\pmb{k}_{1}+\pmb{k}_{2})}{m_{1}+m_{2}+m_{3}},
\end{split}
\end{equation}
respectively. In constructing the charmed baryon wave functions, we treat a single-charmed baryon as a bound state composed of a light-quark pair and a charm quark, whereas a double-heavy baryon is treated as a bound state composed of a $cc$ subsystem and a light quark. Besides, the $\psi_{n,l,m}(\pmb{k})$ is adopted as Gaussian basis function~\cite{Hiyama:2003cu}
\begin{equation}
\begin{split}
\psi_{n,l,m}(\pmb{k})=&N_{nl}~k^{l}~e^{-\nu_{n}k^{2}}Y_{lm}(\hat{\pmb{k}}),\\
N_{nl}=&\Bigg{(}\frac{2^{l+2}(2\nu_{n})^{l+3/2}}{\sqrt{\pi}(2l+1)!!}\Bigg{)}^{1/2},
\end{split}
\end{equation}
where $k\equiv\vert\pmb{k}\vert$, $\hat{\pmb{k}}=\pmb{k}/k$ is the corresponding unit vector, and $\nu_{n}$ denotes the Gaussian size parameter. In Eq.~\eqref{eq:wavefunctions}, the $C^{(n_{\rho},n_{\lambda})}$ denotes the Gaussian expansion coefficients, which will be determined dynamically for the relevant charmed baryons by solving the three-quark Schr\"{o}dinger equation, as described in Sec.~\ref{sec3}.

After synthesizing the above discussion, especially combining the Eq.~\eqref{eq:relation11}, the nonvanishing decay amplitudes satisfy
\begin{equation}
\begin{split}
\mathcal{M}_{\text{PC}}^{-1/2,-1/2}=&-\mathcal{M}_{\text{PC}}^{1/2,1/2},\\
\mathcal{M}_{\text{PV}}^{-1/2,-1/2}=&\mathcal{M}_{\text{PV}}^{1/2,1/2},
\label{eq:relation1}
\end{split}
\end{equation}
for all three $W$-emission diagrams $T$, $C$, and $C^{\prime}$. In addition, we also have the relation $\mathcal{M}_{\text{PV},\text{PC}}^{-1/2,1/2}=\mathcal{M}_{\text{PV},\text{PC}}^{1/2,-1/2}=0$. Therefore, only one independent PC and one independent PV amplitude need to be evaluated explicitly for each $W$-emission diagram.\\

\subsection{The decay amplitudes of $W$-exchange diagrams in pole model}
\label{sec2.2}

\begin{figure}[htbp]\centering
  \includegraphics[width=55mm]{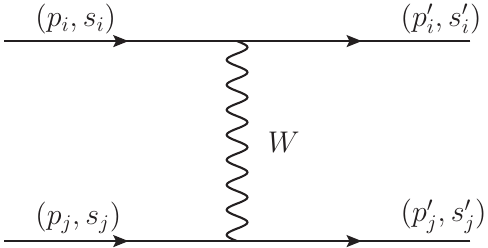}\\
  \caption{The kinematic conventions of $W$-exchange process.}
  \label{fig:22}
\end{figure}

We next turn to the evaluation of the decay amplitudes of $W$-exchange diagrams. Following the conventions shown in Fig.~\ref{fig:22}, the effective Hamiltonian for $W$-exchange process can be written as
\begin{equation}
\begin{split}
\mathcal{H}_{2\to2}=&\frac{G_{F}}{\sqrt{2}}V_{\text{CKM}}V_{\text{CKM}}^{\prime*}
\frac{1}{(2\pi)^{3}}\delta^{3}(\pmb{p}_{i}^{\prime}+\pmb{p}_{j}^{\prime}-\pmb{p}_{i}-\pmb{p}_{j})\\
&\times\Big{[}c_{1}(\mu)\hat{\mathcal{O}}_{1}+c_{2}(\mu)\hat{\mathcal{O}}_{2}\Big{]}.
\end{split}
\end{equation}

In the nonrelativistic limit, the contribution associated with $\hat{\mathcal{O}}_{1}$ can be decomposed into PC and PV parts as~\cite{Niu:2020gjw,Niu:2021qcc,Niu:2025lgt}
\begin{widetext}
\begin{equation}
\begin{split}
\mathcal{H}_{1,2\to2}^{\text{PC}}=&\frac{G_{F}}{\sqrt{2}}V_{\text{CKM}}V_{\text{CKM}}^{\prime*}\frac{1}{(2\pi)^{3}}\sum_{i{\neq}j}\hat{\alpha}_{i}^{(-)}\hat{\beta}_{j}^{(+)}\delta^{3}(\pmb{p}_{i}^{\prime}+\pmb{p}_{j}^{\prime}-\pmb{p}_{i}-\pmb{p}_{j})(1-\langle{s_{z,i}^{\prime}}\vert{\pmb{\sigma}_{i}}\vert{s_{z,i}}\rangle\langle{s_{z,j}^{\prime}}\vert{\pmb{\sigma}_{j}}\vert{s_{z,j}}\rangle),\\
\mathcal{H}_{1,2\to2}^{\text{PV}}=&\frac{G_{F}}{\sqrt{2}}V_{\text{CKM}}V_{\text{CKM}}^{\prime*}\frac{1}{(2\pi)^{3}}\sum_{i{\neq}j}\hat{\alpha}_{i}^{(-)}\hat{\beta}_{j}^{(+)}\delta^{3}(\pmb{p}_{i}^{\prime}+\pmb{p}_{j}^{\prime}-\pmb{p}_{i}-\pmb{p}_{j})\\
&\times\Bigg{\{}-\big{(}\langle{s_{z,i}^{\prime}}\vert{\pmb{\sigma}_{i}}\vert{s_{z,i}}\rangle-\langle{s_{z,j}^{\prime}}\vert{\pmb{\sigma}_{j}}\vert{s_{z,j}}\rangle\big{)}\Bigg{[}\Bigg{(}\frac{\pmb{p}_{i}}{2m_{i}}-\frac{\pmb{p}_{j}}{2m_{j}}\Bigg{)}+\Bigg{(}\frac{\pmb{p}_{i}^{\prime}}{2m_{i}^{\prime}}-\frac{\pmb{p}_{j}^{\prime}}{2m_{j}^{\prime}}\Bigg{)}\Bigg{]}\\
&+i\big{(}\langle{s_{z,i}^{\prime}}\vert{\pmb{\sigma}_{i}}\vert{s_{z,i}}\rangle\times\langle{s_{z,j}^{\prime}}\vert{\pmb{\sigma}_{j}}\vert{s_{z,j}}\rangle\big{)}\Bigg{[}\Bigg{(}\frac{\pmb{p}_{i}}{2m_{i}}-\frac{\pmb{p}_{j}}{2m_{j}}\Bigg{)}-\Bigg{(}\frac{\pmb{p}_{i}^{\prime}}{2m_{i}^{\prime}}-\frac{\pmb{p}_{j}^{\prime}}{2m_{j}^{\prime}}\Bigg{)}\Bigg{]}\Bigg{\}},
\label{eq:Hamilton22}
\end{split}
\end{equation}
\end{widetext}
where $\hat{\alpha}_{i}^{(-)}$ and $\hat{\beta}_{j}^{(+)}$ are the flavor-changing operators acting on $i$th and $j$th quarks, respectively. Specifically, their actions for the relevant $W$-exchange transitions are summarized as follows:
\begin{equation}
\begin{array}{c|cc}
\hline
\text{Transition} & \hat{\alpha}_{i}^{(-)} & \hat{\beta}_{j}^{(+)} \\
\hline
cd\to su & |c\rangle\to|s\rangle & |d\rangle\to|u\rangle \\
cd\to du & |c\rangle\to|d\rangle & |d\rangle\to|u\rangle \\
cs\to su & |c\rangle\to|s\rangle & |s\rangle\to|u\rangle \\
cs\to du & |c\rangle\to|d\rangle & |s\rangle\to|u\rangle \\
\hline
\end{array}
\end{equation}
In addition, the effective Hamiltonian $\mathcal{H}_{2,2\to2}^{\text{PC},\text{PV}}$ associated with $\hat{\mathcal{O}}_{2}$ can be obtained by taking the replacements $(p_{i}^{\prime},m_{i}^{\prime})\leftrightarrows(p_{j}^{\prime},m_{j}^{\prime})$ and $s_{z,i}^{\prime}{\leftrightarrows}s_{z,j}^{\prime}$ in $\mathcal{H}_{1,2\to2}^{\text{PC},\text{PV}}$.

\begin{figure}[htbp]\centering
  \includegraphics[width=80mm]{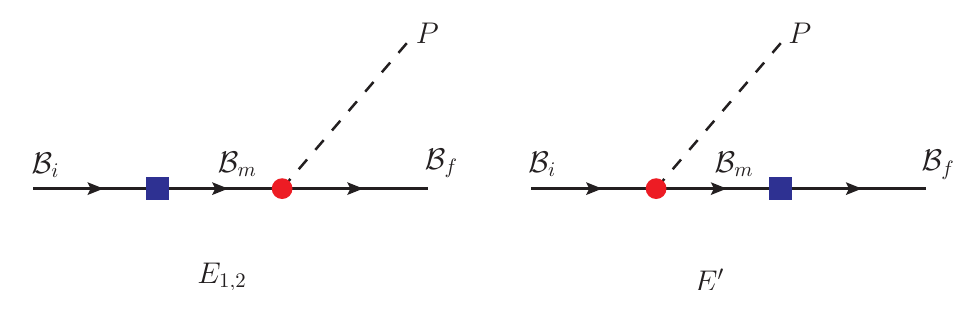}\\
  \caption{The corresponding pole diagrams of $W$-exchange processes, where the blue squares and red circles represents weak and strong interaction vertices, respectively.}
  \label{fig:W}
\end{figure}

In this work, we employ the pole model to estimate the nonfactorizable $W$-exchange amplitudes. Within the assumption of pole model, the $W$-exchange amplitude can be expressed as a sum over intermediate baryon-pole contributions. The contribution of an intermediate baryon state $\mathcal{B}_{m}$ to the $W$-exchange amplitude can be written as~\cite{Niu:2020gjw,Niu:2021qcc,Niu:2025lgt}
\begin{widetext}
\begin{equation}
\begin{split}
i\mathcal{M}_{\mathcal{B}_{m}-pole}^{J_{z}^{f},J_{z}^{i}}=\langle{\mathcal{B}_{f}(\pmb{P}_{f};J^{f},J_{z}^{f})}\vert{\mathcal{H}_{P}}\vert{\mathcal{B}_{m}(\pmb{P}_{i};J^{i},J_{z}^{i})}\rangle\frac{i}{\slashed{p}_{\mathcal{B}_{m}}{\!}-{\!}m_{\mathcal{B}_{m}}{\!}+{\!}\frac{i}{2}\Gamma_{\mathcal{B}_{m}}}
\langle{\mathcal{B}_{m}(\pmb{P}_{i};J^{i},J_{z}^{i})}\vert{\mathcal{H}_{2\to2}}\vert{\mathcal{B}_{i}(\pmb{P}_{i};J^{i},J_{z}^{i})}\rangle,\\
\label{eq:amplitudes21}
\end{split}
\end{equation}
\begin{equation}
\begin{split}
i\mathcal{M}_{\mathcal{B}_{m}-pole}^{J_{z}^{f},J_{z}^{i}}=\langle{\mathcal{B}_{f}(\pmb{P}_{f};J^{f},J_{z}^{f})}\vert{\mathcal{H}_{2\to2}}\vert{\mathcal{B}_{m}(\pmb{P}_{f};J^{f},J_{z}^{f})}\rangle\frac{i}{\slashed{p}_{\mathcal{B}_{m}}{\!}-{\!}m_{\mathcal{B}_{m}}{\!}+{\!}\frac{i}{2}\Gamma_{\mathcal{B}_{m}}}
\langle{\mathcal{B}_{m}(\pmb{P}_{f};J^{f},J_{z}^{f})}\vert{\mathcal{H}_{P}}\vert{\mathcal{B}_{i}(\pmb{P}_{i};J^{i},J_{z}^{i})}\rangle,
\label{eq:amplitudes22}
\end{split}
\end{equation}
\end{widetext}
for $E_{1,2}$ and $E^{\prime}$ diagrams, respectively, as illustrated in Fig.~\ref{fig:W}. Here, we work in the rest frame of the initial baryon, with $\pmb{P}_{i}=0$, $\pmb{P}_{f}=-\pmb{k}$, and $\pmb{k}=(0,0,k)$. The symbol $\mathcal{B}_{m}$ denotes an intermediate baryon state with quantum numbers of $J^{P}=1/2^{+}$ or $1/2^{-}$. Particularly in the present calculation, the relevant intermediate baryon states are $\mathcal{B}_{m}=(\Xi_{c}^{(\prime)+},\Lambda_{c}^{+},\Sigma_{c}^{+})$ for $E_{1,2}$ diagrams, and $\mathcal{B}_{m}=(\Xi_{cc}^{+},\Omega_{cc}^{+})$ for $E^{\prime}$ diagrams, including the corresponding positive and negative-parity states when allowed.

In the numerical calculation, the approximation of propagator
\begin{equation}
\frac{1}{\slashed{p}-m+i\Gamma/2}\simeq\frac{2m}{p^{2}-m^{2}+i\Gamma m},
\end{equation}
is employed. Moreover, considering that the concerned intermediate charmed baryons are narrow, and $m\Gamma\ll\vert{p^{2}-m^{2}}\vert$, we further neglect the $i\Gamma{m}$ term in the numerical calculation.

The interaction Hamiltonian $\mathcal{H}_{P}$ describing the coupling between quarks and pseudoscalar meson is taken as
\begin{equation}
\mathcal{H}_{P}=\sum_{j}{\int}d{\pmb{x}}\frac{1}{f_{P}}\bar{q}_{j}(\pmb{x})\gamma_{\mu}\gamma_{5}q_{j}(\pmb{x})\partial^{\mu}\phi_{P}(\pmb{x}),
\end{equation}
where $f_{P}$ is the pseudoscalar meson decay constant, while $q(\pmb{x})$ and $\phi_{P}$ denote the quark and pseudoscalar meson fields, respectively. In the nonrelativistic limit, this effective Hamiltonian can be reduced as~\cite{Niu:2020gjw,Niu:2021qcc,Niu:2025lgt}
\begin{equation}
\begin{split}
\mathcal{H}_{P}=&\frac{1}{\sqrt{(2\pi)^{3}2\omega_{P}}}\sum_{j}\frac{1}{f_{P}}\Bigg{[}\omega_{P}\bigg{(}\frac{\pmb{\sigma}\cdot\pmb{p}_{f}^{j}}{2m_{f}}+\frac{\pmb{\sigma}\cdot\pmb{p}_{i}^{j}}{2m_{i}}\bigg{)}-\pmb{\sigma}\cdot\pmb{k}\Bigg{]}\hat{I}_{P}^{j}\\
&\times\delta^{3}(\pmb{p}_{f}^{j}+\pmb{k}-\pmb{p}_{i}^{j}),
\end{split}
\end{equation}
where $\omega_{P}$ and $\pmb{k}$ are the energy and three-momentum of the pseudoscalar meson in the rest frame of the initial state, respectively. The $\pmb{p}_{i}^{j}$ and $\pmb{p}_{f}^{j}$ are the initial and final three-momentum of the $j$th quark, respectively. The operator $\hat{I}_{P}^{j}$, acting on the $j$th quark, is given explicitly by~\cite{Niu:2025lgt}
\begin{equation}
\begin{split}
\hat{I}_{P}
=\left\{
\begin{array}{ll}
b_{d}^{\dagger}b_{u}, &\text{for}~\pi^{+},\\
b_{u}^{\dagger}b_{d}, &\text{for}~\pi^{-},\\
{[}b_{u}^{\dagger}b_{u}-b_{d}^{\dagger}b_{d}{]}/\sqrt{2}, &\text{for}~\pi^{0},\\
b_{s}^{\dagger}b_{u}, &\text{for}~K^{+},\\
b_{u}^{\dagger}b_{s}, &\text{for}~K^{-},\\
b_{s}^{\dagger}b_{d}, &\text{for}~K^{0},\\
b_{d}^{\dagger}b_{s}, &\text{for}~\bar{K}^{0},\\
\frac{\cos\zeta}{\sqrt{2}}[b_{u}^{\dagger}b_{u}+b_{d}^{\dagger}b_{d}]-\sin\zeta[b_{s}^{\dagger}b_{s}], &\text{for}~\eta,\\
\frac{\sin\zeta}{\sqrt{2}}[b_{u}^{\dagger}b_{u}+b_{d}^{\dagger}b_{d}]+\cos\zeta[b_{s}^{\dagger}b_{s}], &\text{for}~\eta^{\prime}.\\
\end{array}
\right.
\end{split}
\end{equation}
Here $\zeta=39.3^{\circ}$ is the $\eta-\eta^{\prime}$ mixing angle, defined by
\begin{equation}
\left(\begin{array}{c}
\eta\\
\eta^{\prime}
\end{array}\right)
=\left(\begin{array}{cc}
\cos{\zeta} & -\sin{\zeta}\\
\sin{\zeta} & \cos{\zeta}
\end{array}\right)
\left(\begin{array}{c}
\eta_{q}\\
\eta_{s}
\end{array}\right),
\end{equation}
where $\eta_{q}=(u\bar{u}+d\bar{d})/\sqrt{2}$ and $\eta_{s}=s\bar{s}$.

Taking the pole contribution in Eq.~\eqref{eq:amplitudes21} as an example, the baryon weak transition matrix elements of PC and PV amplitudes can be evaluated by producing the integrations as
\begin{widetext}
\begin{equation}
\begin{split}
\langle{\mathcal{B}_{m}(\pmb{P}_{i};J^{i},J_{z}^{i})}\vert{\mathcal{H}_{1(2),2\to2}^{\text{PC}}}\vert&{\mathcal{B}_{i}(\pmb{P}_{i};J^{i},J_{z}^{i})}\rangle=
(\pm1)\times\frac{G_{F}}{\sqrt{2}}V_{\text{CKM}}V_{\text{CKM}}^{\prime*}\frac{6}{(2\pi)^{3}}\int{d\pmb{p}_{1}}{d\pmb{p}_{2}}{d\pmb{p}_{3}}{d\pmb{p}_{1}^{\prime}}{d\pmb{p}_{2}^{\prime}}{d\pmb{p}_{3}^{\prime}}\delta^{3}(\pmb{p}_{3}^{\prime}-\pmb{p}_{3})\\
&\times\sum_{S_{z}^{i},S_{z}^{m}}\sum_{L_{z}^{i},L_{z}^{m}}\langle{S^{m},S_{z}^{m};L^{m},L_{z}^{m}}\vert{J^{i},J_{z}^{i}}\rangle\langle{S^{i},S_{z}^{i};L^{i},L_{z}^{i}}\vert{J^{i},J_{z}^{i}}\rangle\\
&\times\langle{\mathcal{B}_{m}(S^{m},S_{z}^{m})}{\vert}\hat{\alpha}_{1}^{(-)}\hat{\beta}_{2}^{(+)}\hat{O}_{1(2),2\to2}^{\text{spin}}{\vert}{\mathcal{B}_{i}(S^{i},S_{z}^{i})}\rangle\delta^{3}(\pmb{p}_{1}^{\prime}+\pmb{p}_{2}^{\prime}-\pmb{p}_{1}-\pmb{p}_{2})\\
&\times\psi_{L^{m},L_{z}^{m}}^{*}(\pmb{p}_{1}^{\prime},\pmb{p}_{2}^{\prime},\pmb{p}_{3}^{\prime})\psi_{L^{i},L_{z}^{i}}(\pmb{p}_{1},\pmb{p}_{2},\pmb{p}_{3})
\delta^{3}(\pmb{P}_{i}-\pmb{p}_{1}-\pmb{p}_{2}-\pmb{p}_{3})\delta^{3}(\pmb{P}_{i}-\pmb{p}_{1}^{\prime}-\pmb{p}_{2}^{\prime}-\pmb{p}_{3}^{\prime}),
\label{eq:amplitudes2}
\end{split}
\end{equation}
\begin{equation}
\begin{split}
\langle{\mathcal{B}_{m}(\pmb{P}_{i};J^{i},J_{z}^{i})}\vert{\mathcal{H}_{1(2),2\to2}^{\text{PV}}}\vert&{\mathcal{B}_{i}(\pmb{P}_{i};J^{i},J_{z}^{i})}\rangle=
(\pm1)\times\frac{G_{F}}{\sqrt{2}}V_{\text{CKM}}V_{\text{CKM}}^{\prime*}\frac{6}{(2\pi)^{3}}\int{d\pmb{p}_{1}}{d\pmb{p}_{2}}{d\pmb{p}_{3}}{d\pmb{p}_{1}^{\prime}}{d\pmb{p}_{2}^{\prime}}{d\pmb{p}_{3}^{\prime}}\delta^{3}(\pmb{p}_{3}^{\prime}-\pmb{p}_{3})\\
&\times\sum_{S_{z}^{i},S_{z}^{m}}\sum_{L_{z}^{i},L_{z}^{m}}\langle{S^{m},S_{z}^{m};L^{m},L_{z}^{m}}\vert{J^{i},J_{z}^{i}}\rangle\langle{S^{i},S_{z}^{i};L^{i},L_{z}^{i}}\vert{J^{i},J_{z}^{i}}\rangle\\
&\times\langle{\mathcal{B}_{m}(S^{m},S_{z}^{m})}\vert\hat{\alpha}_{1}^{(-)}\hat{\beta}_{2}^{(+)}\hat{\mathcal{O}}_{1(2),2\to2}^{\text{spin}}\vert{\mathcal{B}_{i}(S^{i},S_{z}^{i})}\rangle\delta^{3}(\pmb{p}_{1}^{\prime}+\pmb{p}_{2}^{\prime}-\pmb{p}_{1}-\pmb{p}_{2})\\
&\times\psi_{L^{m},L_{z}^{m}}^{*}(\pmb{p}_{1}^{\prime},\pmb{p}_{2}^{\prime},\pmb{p}_{3}^{\prime})\mathcal{O}_{1(2),2\to2}^{\text{spatial}}(\pmb{p}_{i})\psi_{L^{i},L_{z}^{i}}(\pmb{p}_{1},\pmb{p}_{2},\pmb{p}_{3})
\delta^{3}(\pmb{P}_{i}-\pmb{p}_{1}-\pmb{p}_{2}-\pmb{p}_{3})\delta^{3}(\pmb{P}_{i}-\pmb{p}_{1}^{\prime}-\pmb{p}_{2}^{\prime}-\pmb{p}_{3}^{\prime}),
\label{eq:amplitudes3}
\end{split}
\end{equation}
respectively. Here, the factor $\pm1$ arises from the color wave function overlap, while the factor $6$ accounts for the flavor symmetry factor.

For the strong transition matrix element $\langle{\mathcal{B}_{f}(\pmb{P}_{f})}\vert{\mathcal{H}_{P}}\vert{\mathcal{B}_{m}(\pmb{P}_{i})}\rangle$, one obtains
\begin{equation}
\begin{split}
\langle{\mathcal{B}_{f}(\pmb{P}_{f};J^{f},J_{z}^{f})}\vert{\mathcal{H}_{P}}\vert&{\mathcal{B}_{m}(\pmb{P}_{i};J^{i},J_{z}^{i})}\rangle=\frac{1}{\sqrt{(2\pi)^{3}2\omega_{P}}}\frac{1}{f_{P}}
\int{d\pmb{p}_{1}}{d\pmb{p}_{2}}{d\pmb{p}_{3}}{d\pmb{p}_{1}^{\prime}}{d\pmb{p}_{2}^{\prime}}{d\pmb{p}_{3}^{\prime}}\delta^{3}(\pmb{p}_{1}^{\prime}-\pmb{p}_{1})\delta^{3}(\pmb{p}_{2}^{\prime}-\pmb{p}_{2})\\
&\times\sum_{S_{z}^{f},S_{z}^{m}}\sum_{L_{z}^{f},L_{z}^{m}}\langle{S^{f},S_{z}^{f};L^{f},L_{z}^{f}}\vert{J^{f},J_{z}^{f}}\rangle\langle{S^{m},S_{z}^{m};L^{m},L_{z}^{m}}\vert{J^{i},J_{z}^{i}}\rangle\\
&\times\langle{\mathcal{B}_{f}(S^{f},S_{z}^{f})}\vert\hat{I}_{P}(-\pmb{\sigma}\cdot\pmb{k})\vert{\mathcal{B}_{m}(S^{m},S_{z}^{m})}\rangle{\frac{\omega_{P}+2(m_{1}+m_{2}+m_{3})}{2(m_{1}+m_{2}+m_{3})}}\delta^{3}(\pmb{p}_{3}^{\prime}-\pmb{p}_{3}+\pmb{k})\\
&\times\psi_{L^{f},L_{z}^{f}}^{*}(\pmb{p}_{1}^{\prime},\pmb{p}_{2}^{\prime},\pmb{p}_{3}^{\prime})\psi_{L^{m},L_{z}^{m}}(\pmb{p}_{1},\pmb{p}_{2},\pmb{p}_{3})
\delta^{3}(\pmb{P}_{i}-\pmb{p}_{1}-\pmb{p}_{2}-\pmb{p}_{3})\delta^{3}(\pmb{P}_{f}-\pmb{p}_{1}^{\prime}-\pmb{p}_{2}^{\prime}-\pmb{p}_{3}^{\prime}).
\label{eq:amplitudes4}
\end{split}
\end{equation}
\end{widetext}
With the $z$ axis chosen along the momentum of the pseudoscalar meson, $\pmb{k}=(0,0,k)$, the spin-flavor matrix element can be reduced to
\begin{equation}
\begin{split}
\langle{\mathcal{B}^{\prime}(S^{\prime},S_{z}^{\prime})}&\vert\hat{I}_{P}(-\pmb{\sigma}\cdot\pmb{k})\vert{\mathcal{B}(S,S_{z})}\rangle\\
&=\langle{\mathcal{B}^{\prime}(S^{\prime},S_{z}^{\prime})}\vert\hat{I}_{P}\sigma_{z}\vert{\mathcal{B}(S,S_{z})}\rangle\times{-k}.
\label{eq:gA}
\end{split}
\end{equation}

The spin-flavor matrix elements involved in Eqs.~(\ref{eq:amplitudes2}-\ref{eq:gA}) are evaluated using the baryon spin-flavor wave functions, with the calculation details and numerical results are presented in Appendix~\ref{app02}.

Analogously, the concrete expression for the pole contribution in Eq.~\eqref{eq:amplitudes22} can be obtained. Combining the above results, the $W$-exchange amplitudes satisfy
\begin{equation}
\begin{split}
\mathcal{M}_{\text{PC}}^{-1/2,-1/2}=&-\mathcal{M}_{\text{PC}}^{1/2,1/2},~\mathcal{M}_{\text{PV}}^{-1/2,-1/2}=\mathcal{M}_{\text{PV}}^{1/2,1/2},\\
\mathcal{M}_{\text{PV},\text{PC}}^{-1/2,1/2}=&\mathcal{M}_{\text{PV},\text{PC}}^{1/2,-1/2}=0.
\label{eq:relation2}
\end{split}
\end{equation}

\section{Charmed baryon spectroscopy and wave functions}
\label{sec3}

In this section, we determine the spatial wave functions of the charmed baryons involved in the decay amplitudes.
To this end, we employ a nonrelativistic Hamiltonian~\cite{Copley:1979wj,Pervin:2007wa,Roberts:2007ni}
\begin{equation}
\mathcal{H}=\sum_{i=1,2,3}\Big{(}m_{i}+\frac{p_{i}^{2}}{2m_{i}}\Big{)}+\sum_{i<j}V_{ij},
\end{equation}
to describe the three-quark baryon system, where $m_{i}$ and $p_{i}$ are mass and momentum of the $i$th quark, respectively. The nonrelativistic quark-quark interaction is decomposed as
\begin{equation}
V_{ij}=V_{ij}^{\text{conf}}+V_{ij}^{\text{hyp}}+V_{ij}^{\text{so(cm)}}+V_{ij}^{\text{so(tp)}},
\end{equation}
where the confinement potential $V_{ij}^{\text{conf}}$, the hyperfine potential $V_{ij}^{\text{hyp}}$, the color-magnetic spin-orbit potential $V_{ij}^{\text{so(cm)}}$, and the Thomas-precession spin-orbit potential $V_{ij}^{\text{so(tp)}}$ take the following explicit forms~\cite{Luo:2023sra,Luo:2023sne,Peng:2024pyl,Zhou:2025fpp,Li:2025noz}
\begin{align}
\label{eq:potentialmodel1}
&V_{ij}^{\text{conf}}=-\frac{2}{3}\frac{\alpha_{s}}{r_{ij}}+\frac{b}{2}r_{ij}+\frac{C}{2},\\
\label{eq:potentialmodel2}
&V_{ij}^{\text{hyp}}=\frac{2\alpha_{s}}{3m_{i}m_{j}}\bigg{[}\frac{8\pi}{3}\tilde{\delta}(r_{ij})\pmb{s}_{i}\cdot\pmb{s}_{j}{\!}+{\!}\frac{1}{r_{ij}^{3}}\Big{(}\frac{3\pmb{s}_{i}{\!}\cdot{\!}\pmb{r}_{ij}\pmb{s}_{j}{\!}\cdot{\!}\pmb{r}_{ij}}{r_{ij}^{2}}{\!}-{\!}\pmb{s}_{i}\cdot\pmb{s}_{j}\Big{)}\bigg{]},\\
\label{eq:potentialmodel3}
&V_{ij}^{\text{so(cm)}}=\frac{2\alpha_{s}}{3r_{ij}^{3}}\Big{(}\frac{\pmb{r}_{ij}\times\pmb{p}_{i}\cdot\pmb{s}_{i}}{m_{i}^{2}}-\frac{\pmb{r}_{ij}\times\pmb{p}_{j}\cdot\pmb{s}_{j}}{m_{j}^{2}}\nonumber\\
&\qquad\quad-\frac{\pmb{r}_{ij}\times\pmb{p}_{j}\cdot\pmb{s}_{i}-\pmb{r}_{ij}\times\pmb{p}_{i}\cdot\pmb{s}_{j}}{m_{i}m_{j}}\Big{)},\\
\label{eq:potentialmodel4}
&V_{ij}^{\text{so(tp)}}=-\frac{1}{2r_{ij}}\frac{\partial{V_{ij}^{\text{conf}}}}{\partial{r_{ij}}}\Big{(}\frac{\pmb{r_{ij}}\times\pmb{p}_{i}\cdot\pmb{s}_{i}}{m_{i}^{2}}-\frac{\pmb{r_{ij}}\times\pmb{p}_{j}\cdot\pmb{s}_{j}}{m_{j}^{2}}\Big{)}.
\end{align}

The nonrelativistic quark potential introduced in Eqs.~(\ref{eq:potentialmodel1}-\ref{eq:potentialmodel4}) is characterized by four phenomenological parameters: the effective one-gluon-exchange coupling constant $\alpha_{s}$, the linear confinement strength $b$, the constant term $C$, and the smearing parameter $\sigma$ which is introduced by
\begin{equation}
\tilde{\delta}(r_{ij})=\frac{\sigma^{3}}{\pi^{3/2}}\exp(-\sigma^{2}r_{ij}^{2}).
\end{equation}
This parameters can be constrained by the well-established charmed-baryon spectrum, thereby providing a spectroscopy-based dynamical input for the baryon wave functions used in the calculations of decay amplitudes. The values of these parameters adopted in the present calculation are summarized in Table~\ref{tab:PotentialParameters}. In addition, constituent-quark masses are taken as $m_{u,d}=370~\text{MeV}$, $m_{s}=600~\text{MeV}$, and $m_{c}=1880~\text{MeV}$~\cite{Luo:2023sra,Luo:2023sne,Peng:2024pyl}.

\begin{table}[htbp]\centering
\caption{Parameters used in the nonrelativistic potential model.}
\label{tab:PotentialParameters}
\renewcommand\arraystretch{1.15}
\begin{tabular*}{76mm}{c@{\extracolsep{\fill}}ccc}
\toprule[1pt]
\toprule[0.5pt]
Parameter   &Value   &Parameter   &Value\\
\midrule[0.5pt]
$\alpha_{s}$            &$0.560$   &$b~(\text{GeV}^{2})$   &$0.125$\\
$\sigma~(\text{GeV})$   &$1.600$   &$C~(\text{GeV})$       &$-0.644$\\
\bottomrule[0.5pt]
\bottomrule[1pt]
\end{tabular*}
\end{table}

With the quark potential specified above, the masses and spatial wave functions of baryons can be obtained by solving the three-quark Schr\"{o}dinger equation with the nonrelativisitc Hamiltonian $\mathcal{H}$
\begin{equation}
\mathcal{H}\vert\Psi_{J,M_{J}}\rangle=E\vert\Psi_{J,M_{J}}\rangle,
\label{eq:Schrodinger}
\end{equation}
in which the baryon wave function $\Psi_{J,M_{J}}$ is constructed as a combination of color, spin, spatial, and flavor terms:
\begin{equation}
\begin{split}
\Psi_{J,M_{J}}=&\sum_{\alpha}C^{(\alpha)}\Psi_{J,M_{J}}^{(\alpha)},\\
\Psi_{J,M_{J}}^{(\alpha)}=&\chi^{\text{color}}
\Big{[}\big{[}[s_{q_{1}}s_{q_{2}}]_{S_{\rho}}s_{q_{3}}\big{]}_{S}\psi_{l_{\rho},l_{\lambda},L}^{\text{spatial}}(\pmb{r}_{\rho},\pmb{r}_{\lambda})\Big{]}_{J,M_{J}}
\psi^{\text{flavor}}.
\end{split}
\end{equation}
Here, $\alpha$ collectively represents all possible quantum numbers, and $C^{(\alpha)}$ are the corresponding expansion coefficients.

\begin{figure}[htbp]
\centering
  \includegraphics[width=50mm]{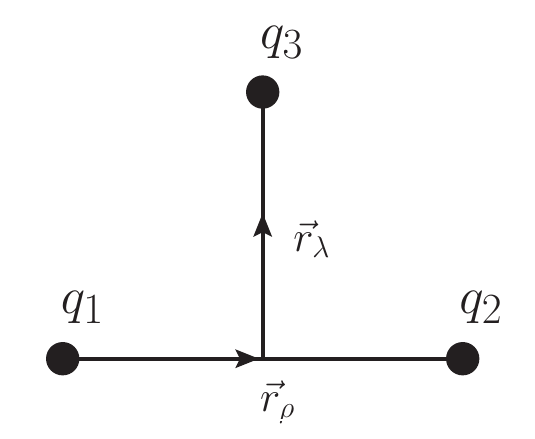}\\
  \caption{Definition of the Jacobi coordinates $\pmb{r}_{\rho}$ and $\pmb{r}_{\lambda}$ for charmed baryon. We set $q_{1}q_{2}q_{3}=(qqc)$ for single charmed baryons and $q_{1}q_{2}q_{3}=(ccq)$ for double-charmed baryons.}
  \label{fig:Jacobi}
\end{figure}

The internal motion of the three-quark system is conveniently described by the Jacobi coordinates $\pmb{r}_{\rho}$ and $\pmb{r}_{\lambda}$ (as shown in Fig.~\ref{fig:Jacobi}), defined as
\begin{equation}
\begin{split}
\pmb{r}_{\rho}=&\pmb{r}_{2}-\pmb{r}_{1},\\
\pmb{r}_{\lambda}=&\pmb{r}_{3}-\frac{m_{1}\pmb{r}_{1}+m_{2}\pmb{r}_{2}}{m_{1}+m_{2}}.
\end{split}
\end{equation}
For the single charmed baryon, we choose $q_{1}q_{2}q_{3}=(qqc)$, so that the $\rho$-mode describes the relative motion of the two light quarks, whereas for the double-charmed baryon, we choose $q_{1}q_{2}q_{3}=(ccq)$, and the corresponding $\rho$-mode describes the relative motion of the two charmed quarks.

According to Fig.~\ref{fig:Jacobi}, the spatial part of charmed baryon wave function can be constructed by coupling the $\rho$ and $\lambda$-modes, i.e.,
\begin{equation}
\begin{split}
\psi_{l_{\rho},l_{\lambda},L}^{\text{spatial}}(\pmb{r}_{\rho},\pmb{r}_{\lambda})=&\sum_{m_{\rho},m_{\lambda}}\langle{l_{\rho},m_{\rho};l_{\lambda},m_{\lambda}}\vert{L,m_{\rho}+m_{\lambda}}\rangle\\
&\times\psi_{l_{\rho},m_{\rho}}(\pmb{r}_{\rho})\psi_{l_{\lambda},m_{\lambda}}(\pmb{r}_{\lambda}),
\label{eq:wavefunction2}
\end{split}
\end{equation}
in which, the spatial wave functions $\psi_{l_{\rho},m_{\rho}}$ and $\psi_{l_{\lambda},m_{\lambda}}$ are expanded in the infinitesimally-shift Gaussian basis~\cite{Hiyama:2003cu,Hiyama:2018ivm}
\begin{equation}
\begin{split}
\phi_{nlm}^{G}(\pmb{r})=&\phi^{G}_{nl}(r)~Y_{lm}(\pmb{\hat{r}})\\
=&\sqrt{\frac{2^{l+2}(2\nu_{n})^{l+3/2}}{\sqrt{\pi}(2l+1)!!}}\lim_{\varepsilon\rightarrow0}
\frac{1}{(\nu_{n}\varepsilon)^l}\sum_{k=1}^{k_{\text{max}}}C_{lm,k}e^{-\nu_{n}\big{(}\pmb{r}-\varepsilon\pmb{D}_{lm,k}\big{)}^2},
\label{eq:Gaussianbasis}
\end{split}
\end{equation}
using Gaussian expansion method. The Gaussian size parameter $\nu_{n}$ is chosen according to a geometric progression~\cite{Li:2021qod,Luo:2022cun,Li:2025alu}
\begin{equation}
\nu_{n}=1/r^2_{n},~~~r_{n}=r_{min}~a^{n-1}
\end{equation}
with
\begin{equation}
a=\Big{(}r_{max}/r_{min}\Big{)}^{\frac{1}{n_{max}-1}}.
\end{equation}
In the numerical calculations, the Gaussian size parameters for the $\rho$-mode are chosen as $(r_{\rho_{min}},r_{\rho_{max}},n_{\rho_{max}})=(0.2~\text{fm},2.0~\text{fm},6)$, and the same set of parameters is adopted for the $\lambda$-mode. This choice enables the Gaussian basis to effectively describe both the short- and long-distance components of the baryon wave functions. The corresponding momentum-space Gaussian basis can be obtained from Eq.~\eqref{eq:Gaussianbasis} through the replacements $r\to{p}$ and $\nu_{n}\to1/(4\nu_{n})$.

Assisted by Gaussian expansion method~\cite{Hiyama:2003cu,Hiyama:2018ivm}, the three-quark Schr\"{o}dinger equation \eqref{eq:Schrodinger} can be transformed into a generalized eigenvalue problem
\begin{equation}
\Big{(}T^{\alpha^{\prime},\alpha}+V^{\alpha^{\prime},\alpha}\Big{)}C^{(\alpha)}=EN^{\alpha^{\prime},\alpha}C^{(\alpha)},
\label{eq:Schrodinger2}
\end{equation}
by applying the variational principle. Here, $T^{\alpha^{\prime},\alpha}$, $V^{\alpha^{\prime},\alpha}$, and $N^{\alpha^{\prime},\alpha}$ represent the free-Hamiltonian, interaction-potential, and normalization matrix elements, respectively, which are defined as
\begin{equation}
\begin{split}
T^{\alpha^{\prime},\alpha}=&\Big{\langle}\Psi_{{J},{M_J}}^{(\alpha^{\prime})}\big{\vert}
\Bigg{[}\sum_{i=1,2,3}\Big{(}m_{i}+\frac{p_{i}^{2}}{2m_{i}}\Big{)}\Bigg{]}
\big{\vert}\Psi_{{J},{M_J}}^{(\alpha)}\Big{\rangle},\\
V^{\alpha^{\prime},\alpha}=&\Big{\langle}\Psi_{{J},{M_J}}^{(\alpha^{\prime})}\big{\vert}
\sum_{i<j}V_{ij}
\big{\vert}\Psi_{{J},{M_J}}^{(\alpha)}\Big{\rangle},\\
N^{\alpha^{\prime},\alpha}=&\Big{\langle}\Psi_{{J},{M_J}}^{(\alpha^{\prime})}
\big{\vert}\Psi_{{J},{M_J}}^{(\alpha)}\Big{\rangle}.
\label{eq:matrixelement}
\end{split}
\end{equation}
In Eq.~\eqref{eq:Schrodinger2}, the eigenvalues $E$ are the charmed baryon masses, while the corresponding eigenvectors $C^{(\alpha)}$ specify the Gaussian-basis decomposition of the spatial wave functions.

To illustrate the performance of the nonrelativistic potential model with the adopted parameters, we present the masses of the relevant single and double-charmed baryons in Table~\ref{tab:mass1}, together with the available experiment values for comparison.

\begin{table}[htbp]
\centering
\caption{The comparison of theoretical and experimental values of masses of charmed baryons.}
\label{tab:mass1}
\renewcommand\arraystretch{1.05}
\begin{tabular*}{76mm}{l@{\extracolsep{\fill}}cccc}
\toprule[1pt]
\toprule[0.5pt]
\multirow{2}*{State}   &Quantum numbers   &This work   &Expt.~\cite{ParticleDataGroup:2026mpi}\\
&$(l_{\rho},l_{\lambda},L,S_{\rho},S,J^{P})$   &$[\text{MeV}]$   &$[\text{MeV}]$\\
\midrule[0.5pt]
$\Lambda_{c}(1S)$             &$(0,0,0,0,\frac{1}{2},\frac{1}{2}^{+})$    &$2287$                  &$2286$\\
$\Lambda_{c}(2S)$             &$(0,0,0,0,\frac{1}{2},\frac{1}{2}^{+})$    &$2796$                  &$2765$\\
$\Lambda_{c}(1^{2}P_{\lambda})$  &$(0,1,1,0,\frac{1}{2},\frac{1}{2}^{-})$    &$2598$               &$2595$\\
\specialrule{0em}{1pt}{1pt}
$\Sigma_{c}(1S)$              &$(0,0,0,1,\frac{1}{2},\frac{1}{2}^{+})$    &$2470$                  &$2455$\\
$\Sigma_{c}(2S)$              &$(0,0,0,1,\frac{1}{2},\frac{1}{2}^{+})$    &$2959$                  &\\
$\Sigma_{c}(1^{2}P_{\lambda})$  &$(0,1,1,1,\frac{1}{2},\frac{1}{2}^{-})$    &$2798$                &\\
$\Sigma_{c}(1^{4}P_{\lambda})$  &$(0,1,1,1,\frac{3}{2},\frac{1}{2}^{-})$    &$2788$                &\\
\specialrule{0em}{1pt}{1pt}
$\Xi_{c}(1S)$                 &$(0,0,0,0,\frac{1}{2},\frac{1}{2}^{+})$    &$2484$                  &$2470$\\
$\Xi_{c}(2S)$                 &$(0,0,0,0,\frac{1}{2},\frac{1}{2}^{+})$    &$2966$                  &$2970$\\
$\Xi_{c}(1^{2}P_{\lambda})$   &$(0,1,1,0,\frac{1}{2},\frac{1}{2}^{-})$    &$2783$                  &$2790$\\
\specialrule{0em}{1pt}{1pt}
$\Xi_{c}^{\prime}(1S)$        &$(0,0,0,1,\frac{1}{2},\frac{1}{2}^{+})$    &$2589$                  &$2580$\\
$\Xi_{c}^{\prime}(2S)$        &$(0,0,0,1,\frac{1}{2},\frac{1}{2}^{+})$    &$3061$                  &\\
$\Xi_{c}^{\prime}(1^{2}P_{\lambda})$   &$(0,1,1,1,\frac{1}{2},\frac{1}{2}^{-})$    &$2909$         &\\
$\Xi_{c}^{\prime}(1^{4}P_{\lambda})$   &$(0,1,1,1,\frac{3}{2},\frac{1}{2}^{-})$    &$2895$         &\\
\specialrule{0em}{1pt}{1pt}
$\Omega_{c}(1S)$              &$(0,0,0,1,\frac{1}{2},\frac{1}{2}^{+})$    &$2701$                  &$2695$\\
$\Omega_{c}(2S)$              &$(0,0,0,1,\frac{1}{2},\frac{1}{2}^{+})$    &$3165$                  &\\
$\Omega_{c}(1^{2}P_{\lambda})$  &$(0,1,1,1,\frac{1}{2},\frac{1}{2}^{-})$    &$3019$                &\\
$\Omega_{c}(1^{4}P_{\lambda})$  &$(0,1,1,1,\frac{3}{2},\frac{1}{2}^{-})$    &$3001$                &\\
\specialrule{0em}{1pt}{1pt}
$\Xi_{cc}(1S)$                &$(0,0,0,1,\frac{1}{2},\frac{1}{2}^{+})$    &$3650$                  &$3621$~\cite{LHCb:2019epo}\\
$\Xi_{cc}(2S)$                &$(0,0,0,1,\frac{1}{2},\frac{1}{2}^{+})$    &$4065$                  &\\
$\Xi_{cc}(1^{2}P_{\rho})$     &$(1,0,1,0,\frac{1}{2},\frac{1}{2}^{-})$    &$3966$                  &\\
$\Xi_{cc}(1^{2}P_{\lambda})$  &$(0,1,1,1,\frac{1}{2},\frac{1}{2}^{-})$    &$4065$                  &\\
$\Xi_{cc}(1^{4}P_{\lambda})$  &$(0,1,1,1,\frac{3}{2},\frac{1}{2}^{-})$    &$3975$                  &\\
\specialrule{0em}{1pt}{1pt}
$\Omega_{cc}(1S)$                &$(0,0,0,1,\frac{1}{2},\frac{1}{2}^{+})$    &$3751$               &$3726$~\cite{LHCb:2026lnj}\\
$\Omega_{cc}(2S)$                &$(0,0,0,1,\frac{1}{2},\frac{1}{2}^{+})$    &$4185$               &\\
$\Omega_{cc}(1^{2}P_{\rho})$     &$(1,0,1,0,\frac{1}{2},\frac{1}{2}^{-})$    &$4078$               &\\
$\Omega_{cc}(1^{2}P_{\lambda})$  &$(0,1,1,1,\frac{1}{2},\frac{1}{2}^{-})$    &$4149$               &\\
$\Omega_{cc}(1^{4}P_{\lambda})$  &$(0,1,1,1,\frac{3}{2},\frac{1}{2}^{-})$    &$4093$               &\\
\bottomrule[0.5pt]
\bottomrule[1pt]
\end{tabular*}
\end{table}

Along with the mass eigenvalues, the corresponding eigenvectors $C^{(\alpha)}$ are also obtained for all baryons entering the decay amplitudes, including the initial, final, and intermediate baryon states. For brevity, the explicit values of $C^{(\alpha)}$ are not presented here. These masses if corresponding experimental values are unavailable, and these Gaussian coefficients are subsequently used in evaluation of the decay amplitudes. Therefore, the baryon masses and spatial wave functions required in this work are determined consistently within the same dynamical framework.

\section{Numerical results}
\label{sec4}

With the numerical results of charmed baryon spatial wave functions evaluated in preceding section, we subsequently investigate the two-body nonleptonic decays $\Xi_{cc}^{+}\to\mathcal{B}_{c}P$. The decay amplitudes, branching fractions, and asymmetry parameters $\alpha$ for relevant decay channels will be discussed separately for the CF, SCS, and DCS decays.

With the obtained PC and PV amplitudes, the corresponding decay width can be evaluated by
\begin{equation}
\Gamma(\mathcal{B}_{i}{\to}\mathcal{B}_{f}P)=8\pi^{2}\frac{kE_{f}E_{P}}{m_{i}}\frac{1}{2J_{i}+1}\sum_{s_{z}^{f},s_{z}^{i}}\Big{(}\vert\mathcal{M}_{\text{PC}}^{s_{z}^{f},s_{z}^{i}}\vert^{2}+\vert\mathcal{M}_{\text{PV}}^{s_{z}^{f},s_{z}^{i}}\vert^{2}\Big{)},
\label{eq:Br}
\end{equation}
where, $m_{i}$ is the mass of initial particle $\mathcal{B}_{i}$, and $E_{f}$ and $E_{P}$ denote the energies of $\mathcal{B}_{f}$ and $P$, respectively. $J$ is the total spin of initial state $\mathcal{B}_{i}$, and the factor $1/(2J+1)$ comes from averaging over polarizations of the initial state.

To convert the partial decay widths into branching fractions, the lifetime of $\Xi_{cc}^{+}$ is required. Since the lifetime has not yet been measured experimentally, we refer to theoretical estimates available in the literature. These theoretical predictions, together with the assumed range and baseline value adopted in the recent LHCb analysis, are summarized in Table~\ref{tab:lifetime}. In the present calculation, we adopt $\tau_{\Xi_{cc}^{+}}=47~\text{fs}$ as the reference value.

\begin{table}[htbp]\centering
\caption{Theoretical predictions and benchmark values for the lifetime of the double-charmed baryon $\Xi_{cc}^{+}$.}
\label{tab:lifetime}
\renewcommand\arraystretch{1.05}
\begin{tabular*}{78mm}{l@{\extracolsep{\fill}}cc}
\toprule[1pt]
\toprule[0.5pt]
                                  &method    &$\tau_{\Xi_{cc}^{+}}(\times10^{-13}s)$\\
\midrule[0.5pt]
Kiselev~(1999)~\cite{Kiselev:1998sy}   &HQE+NRQCD    &$1.1\pm0.3$\\
Guberina~(1999)~\cite{Guberina:1999mx} &HQE+diquark model    &$2.2$\\
Likhoded~(1999)~\cite{Likhoded:1999yv} &HQE+NRQM                     &$2.0$\\
Kiselev~(2002)~\cite{Kiselev:2001fw}   &HQET+NRQCD    &$1.6\pm0.5$\\
Chang~(2008)~\cite{Chang:2007xa}       &HQE+NRSHOM    &$2.5$\\
Karliner~(2014)~\cite{Karliner:2014gca} &    &$0.53$\\
Cheng~(2018)~\cite{Cheng:2018mwu}      &HQE+NRQM    &$0.44$\\
Berezhnoy~(2018)~\cite{Berezhnoy:2018bde}  &HQE+NRCQM    &$1.4\pm0.1$\\
Dulibic~(2023)~\cite{Dulibic:2023jeu}  &HQE+NRCQM    &$0.6\pm0.1^{+0.2}_{-0.1}$\\
Cheng~(2026)~\cite{Cheng:2026mlv}       &HQE+bag model    &$0.47^{+0.09}_{-0.08}$\\
\midrule[0.5pt]
\multirow{2}*{LHCb~(2026)~\cite{LHCb:2026pxn}}   &assumed range    &$0.15\sim1.6$\\
                                                 &baseline         &$0.45$\\
This work                               &    &$0.47$\\
\bottomrule[0.5pt]
\bottomrule[1pt]
\end{tabular*}
\end{table}

In addition, we investigate the asymmetry parameter $\alpha$, which is defined as
\begin{equation}
\alpha=\frac{2\text{Re}[(\mathcal{M}_{\text{PV}}^{-1/2,-1/2})^{*}\mathcal{M}_{\text{PC}}^{-1/2,-1/2}]}{\vert\mathcal{M}_{\text{PC}}^{-1/2,-1/2}\vert^{2}+\vert\mathcal{M}_{\text{PV}}^{-1/2,-1/2}\vert^{2}}.
\label{eq:alpha}
\end{equation}
The asymmetry parameter $\alpha$ is affected by both the relative magnitudes and the relative phase of the PC and PV amplitudes, thereby providing information complementary to the branching fraction.

\subsection{The CF decays}
\label{sec4.1}

We first discuss the CF decay modes. By substituting the obtained charmed-baryon spatial wave functions into Eqs.~(\ref{eq:Tamplitude}-\ref{eq:Cprimamplitude}), the corresponding $W$-emission amplitudes contributed by the $T$, $C$ and $C^{\prime}$ diagrams can be evaluated numerically. Similarly, the corresponding $W$-exchange amplitudes associated with the $E_{1,2}$ and $E^{\prime}$ diagrams can be obtained using Eqs.~(\ref{eq:amplitudes21}-\ref{eq:amplitudes22}). Within the pole model treatment, the nonfactorizable $W$-exchange amplitudes are evaluated by retaining the relevant low-lying intermediate baryon states, specifically in this work the single-charmed baryons $\Lambda_{c}^{+}$, $\Sigma_{c}^{+}$, and $\Xi_{c}^{(\prime)+}$ for the $E_{1,2}$ diagrams, and the double-charmed baryons $\Xi_{cc}^{+}$ and $\Omega_{cc}^{+}$ for the $E^{\prime}$ diagram. 

Our numerical results of the individual contributions to the PC and PV decay amplitudes for the CF decays are presented in Table~\ref{tab:CFamplitudes}. As can be seen, the decay amplitudes of $T$ and $C$ diagrams are comparable in magnitude, whereas those from the $C^{\prime}$ diagram are noticeably smaller. In addition, the nonfactorizable $W$-exchange amplitudes provide sizeable contributions in many decays and, in some cases, are even larger than the factorizable $W$-emission amplitudes. This indicates that the contributions from $W$-exchange diagrams cannot be neglected in the charmed baryon nonleptonic decays. Another notable feature is that several pole contributions vanish identically owing to the vanishing spin-flavor matrix elements at the corresponding weak or strong vertices.

\begin{table*}[htbp]\centering
\caption{The PC and PV decay amplitudes $\mathcal{M}_{\text{PC},\text{PV}}^{-1/2,-1/2}$ (in the units of $10^{-4}G_{F}~\text{GeV}^{3/2}$) of CF decays. The symbols $\times~(\text{w})$ and $\times~(\text{s})$ denote vanishing contributions caused by zero spin-flavor matrix elements at the weak and strong interaction vertices, respectively; see Tables~\ref{tab:polefactorPC} and \ref{tab:gA}.}
\label{tab:CFamplitudes}
\renewcommand\arraystretch{1.05}
\begin{tabular*}{170mm}{l@{\extracolsep{\fill}}lclc|llclc}
\toprule[1pt]
\toprule[0.5pt]
\specialrule{0em}{1pt}{1pt}
&\multicolumn{2}{c}{{\normalsize $\mathcal{M}_{\text{PC}}^{-1/2,-1/2}$}}  &\multicolumn{2}{c}{{\normalsize $\mathcal{M}_{\text{PV}}^{-1/2,-1/2}$}} &  &\multicolumn{2}{c}{{\normalsize $\mathcal{M}_{\text{PC}}^{-1/2,-1/2}$}}  &\multicolumn{2}{c}{{\normalsize $\mathcal{M}_{\text{PV}}^{-1/2,-1/2}$}}\\
\specialrule{0em}{1pt}{1pt}
\midrule[0.5pt]
\multirow{1}*{\shortstack{$\Xi_{cc}^{+}\to\Xi_{c}^{0}\pi^{+}$}}   &$T-diagram$   &$-14.19$   &$T-diagram$   &$26.52$
&\multirow{1}*{\shortstack{$\Xi_{cc}^{+}\to\Xi_{c}^{\prime0}\pi^{+}$}}   &$T-diagram$   &$38.39$   &$T-diagram$   &$-15.86$\\
&$\Xi_{c}^{+}(1S)-pole$   &$\times~(\text{s})$   &$\Xi_{c}^{+}(1^{2}P_{\lambda})-pole$   &$\times~(\text{s})$  &  &$\Xi_{c}^{+}(1S)-pole$   &$68.73$   &$\Xi_{c}^{+}(1^{2}P_{\lambda})-pole$   &$-0.90$\\
&$\Xi_{c}^{+}(2S)-pole$   &$\times~(\text{s})$   &$\Xi_{c}^{\prime+}(1^{2}P_{\lambda})-pole$   &$0$  &  &$\Xi_{c}^{+}(2S)-pole$   &$31.64$   &$\Xi_{c}^{\prime+}(1^{2}P_{\lambda})-pole$   &$0$\\
&$\Xi_{c}^{\prime+}(1S)-pole$   &$\times~(\text{w})$   &$\Xi_{c}^{\prime+}(1^{4}P_{\lambda})-pole$   &$0$  &  &$\Xi_{c}^{\prime+}(1S)-pole$   &$\times~(\text{w})$   &$\Xi_{c}^{\prime+}(1^{4}P_{\lambda})-pole$   &$0$\\
&$\Xi_{c}^{\prime+}(2S)-pole$   &$\times~(\text{w})$   &   &  &  &$\Xi_{c}^{\prime+}(2S)-pole$   &$\times~(\text{w})$   &   &\\
&$\text{total}$   &$-14.19$   &$\text{total}$   &$26.52$  &  &$\text{total}$   &$138.76$   &$\text{total}$   &$-16.76$\\
\specialrule{0em}{3pt}{3pt}
\multirow{1}*{\shortstack{$\Xi_{cc}^{+}\to\Lambda_{c}^{+}\bar{K}^{0}$}}   &$C-diagram$   &$-17.62$   &$C-diagram$   &$33.53$
&\multirow{1}*{\shortstack{$\Xi_{cc}^{+}\to\Sigma_{c}^{+}\bar{K}^{0}$}}   &$C-diagram$   &$-44.16$   &$C-diagram$   &$20.32$\\
&$\Xi_{c}^{+}(1S)-pole$   &$\times~(\text{s})$   &$\Xi_{c}^{+}(1^{2}P_{\lambda})-pole$   &$\times~(\text{s})$  &  &$\Xi_{c}^{+}(1S)-pole$   &$-51.81$   &$\Xi_{c}^{+}(1^{2}P_{\lambda})-pole$   &$0.74$\\
&$\Xi_{c}^{+}(2S)-pole$   &$\times~(\text{s})$   &$\Xi_{c}^{\prime+}(1^{2}P_{\lambda})-pole$   &$0$  &  &$\Xi_{c}^{+}(2S)-pole$   &$-20.60$   &$\Xi_{c}^{\prime+}(1^{2}P_{\lambda})-pole$   &$0$\\
&$\Xi_{c}^{\prime+}(1S)-pole$   &$\times~(\text{w})$   &$\Xi_{c}^{\prime+}(1^{4}P_{\lambda})-pole$   &$0$  &  &$\Xi_{c}^{\prime+}(1S)-pole$   &$\times~(\text{w})$   &$\Xi_{c}^{\prime+}(1^{4}P_{\lambda})-pole$   &$0$\\
&$\Xi_{c}^{\prime+}(2S)-pole$   &$\times~(\text{w})$   &   &  &  &$\Xi_{c}^{\prime+}(2S)-pole$   &$\times~(\text{w})$   &   &\\
&$\text{total}$   &$-17.62$   &$\text{total}$   &$33.53$  &  &$\text{total}$   &$-116.57$   &$\text{total}$   &$21.06$\\
\specialrule{0em}{3pt}{3pt}
\multirow{1}*{\shortstack{$\Xi_{cc}^{+}\to\Xi_{c}^{+}\pi^{0}$}}   &$C^{\prime}-diagram$   &$1.65$   &$C^{\prime}-diagram$   &$-6.06$
&\multirow{1}*{\shortstack{$\Xi_{cc}^{+}\to\Xi_{c}^{\prime+}\pi^{0}$}}   &$C^{\prime}-diagram$   &$0$   &$C^{\prime}-diagram$   &$0$\\
&$\Xi_{cc}^{+}(1S)-pole$   &$-38.49$   &$\Xi_{cc}^{+}(1^{2}P_{\rho})-pole$   &$\times~(\text{s})$  &  &$\Xi_{cc}^{+}(1S)-pole$   &$\times~(\text{w})$   &$\Xi_{cc}^{+}(1^{2}P_{\rho})-pole$   &$\times~(\text{s})$\\
&$\Xi_{cc}^{+}(2S)-pole$   &$-1.66$   &$\Xi_{cc}^{+}(1^{2}P_{\lambda})-pole$   &$-1.60$  &  &$\Xi_{cc}^{+}(2S)-pole$   &$\times~(\text{w})$   &$\Xi_{cc}^{+}(1^{2}P_{\lambda})-pole$   &$0$\\
&$\Xi_{c}^{+}(1S)-pole$   &$\times~(\text{s})$   &$\Xi_{cc}^{+}(1^{4}P_{\lambda})-pole$   &$100.51$  &  &$\Xi_{c}^{+}(1S)-pole$   &$48.60$   &$\Xi_{cc}^{+}(1^{4}P_{\lambda})-pole$   &$\times~(\text{w})$\\
&$\Xi_{c}^{+}(2S)-pole$   &$\times~(\text{s})$   &$\Xi_{c}^{+}(1^{2}P_{\lambda})-pole$   &$\times~(\text{s})$  &  &$\Xi_{c}^{+}(2S)-pole$   &$22.37$   &$\Xi_{c}^{+}(1^{2}P_{\lambda})-pole$   &$-0.63$\\
&$\Xi_{c}^{\prime+}(1S)-pole$   &$\times~(\text{w})$   &$\Xi_{c}^{\prime+}(1^{2}P_{\lambda})-pole$   &$0$  &  &$\Xi_{c}^{\prime+}(1S)-pole$   &$\times~(\text{w})$   &$\Xi_{c}^{\prime+}(1^{2}P_{\lambda})-pole$   &$0$\\
&$\Xi_{c}^{\prime+}(2S)-pole$   &$\times~(\text{w})$   &$\Xi_{c}^{\prime+}(1^{4}P_{\lambda})-pole$   &$0$  &  &$\Xi_{c}^{\prime+}(2S)-pole$   &$\times~(\text{w})$   &$\Xi_{c}^{\prime+}(1^{4}P_{\lambda})-pole$   &$0$\\
&$\text{total}$   &$-38.50$   &$\text{total}$   &$92.85$  &  &$\text{total}$   &$70.97$   &$\text{total}$   &$-0.63$\\
\specialrule{0em}{3pt}{3pt}
\multirow{1}*{\shortstack{$\Xi_{cc}^{+}\to\Xi_{c}^{+}\eta$}}   &$C^{\prime}-diagram$   &$-1.67$   &$C^{\prime}-diagram$   &$6.40$
&\multirow{1}*{\shortstack{$\Xi_{cc}^{+}\to\Xi_{c}^{\prime+}\eta$}}   &$C^{\prime}-diagram$   &$0$   &$C^{\prime}-diagram$   &$0$\\
&$\Xi_{cc}^{+}(1S)-pole$   &$30.44$   &$\Xi_{cc}^{+}(1^{2}P_{\rho})-pole$   &$\times~(\text{s})$  &  &$\Xi_{cc}^{+}(1S)-pole$   &$\times~(\text{w})$   &$\Xi_{cc}^{+}(1^{2}P_{\rho})-pole$   &$\times~(\text{s})$\\
&$\Xi_{cc}^{+}(2S)-pole$   &$1.04$   &$\Xi_{cc}^{+}(1^{2}P_{\lambda})-pole$   &$1.23$  &  &$\Xi_{cc}^{+}(2S)-pole$   &$\times~(\text{w})$   &$\Xi_{cc}^{+}(1^{2}P_{\lambda})-pole$   &$0$\\
&$\Xi_{c}^{+}(1S)-pole$   &$\times~(\text{s})$   &$\Xi_{cc}^{+}(1^{4}P_{\lambda})-pole$   &$70.20$  &  &$\Xi_{c}^{+}(1S)-pole$   &$74.87$   &$\Xi_{cc}^{+}(1^{4}P_{\lambda})-pole$   &$\times~(\text{w})$\\
&$\Xi_{c}^{+}(2S)-pole$   &$\times~(\text{s})$   &$\Xi_{c}^{+}(1^{2}P_{\lambda})-pole$   &$\times~(\text{s})$  &  &$\Xi_{c}^{+}(2S)-pole$   &$26.00$   &$\Xi_{c}^{+}(1^{2}P_{\lambda})-pole$   &$-0.87$\\
&$\Xi_{c}^{\prime+}(1S)-pole$   &$\times~(\text{w})$   &$\Xi_{c}^{\prime+}(1^{2}P_{\lambda})-pole$   &$0$  &  &$\Xi_{c}^{\prime+}(1S)-pole$   &$\times~(\text{w})$   &$\Xi_{c}^{\prime+}(1^{2}P_{\lambda})-pole$   &$0$\\
&$\Xi_{c}^{\prime+}(2S)-pole$   &$\times~(\text{w})$   &$\Xi_{c}^{\prime+}(1^{4}P_{\lambda})-pole$   &$0$  &  &$\Xi_{c}^{\prime+}(2S)-pole$   &$\times~(\text{w})$   &$\Xi_{c}^{\prime+}(1^{4}P_{\lambda})-pole$   &$0$\\
&$\text{total}$   &$29.81$   &$\text{total}$   &$77.83$  &  &$\text{total}$   &$100.87$   &$\text{total}$   &$-0.87$\\
\specialrule{0em}{3pt}{3pt}
\multirow{1}*{\shortstack{$\Xi_{cc}^{+}\to\Xi_{c}^{+}\eta^{\prime}$}}   &$C^{\prime}-diagram$   &$-0.89$   &$C^{\prime}-diagram$   &$4.84$
&\multirow{1}*{\shortstack{$\Xi_{cc}^{+}\to\Xi_{c}^{\prime+}\eta^{\prime}$}}   &$C^{\prime}-diagram$   &$0$   &$C^{\prime}-diagram$   &$0$\\
&$\Xi_{cc}^{+}(1S)-pole$   &$25.76$   &$\Xi_{cc}^{+}(1^{2}P_{\rho})-pole$   &$\times~(\text{s})$  &  &$\Xi_{cc}^{+}(1S)-pole$   &$\times~(\text{w})$   &$\Xi_{cc}^{+}(1^{2}P_{\rho})-pole$   &$\times~(\text{s})$\\
&$\Xi_{cc}^{+}(2S)-pole$   &$0.36$   &$\Xi_{cc}^{+}(1^{2}P_{\lambda})-pole$   &$0.80$  &  &$\Xi_{cc}^{+}(2S)-pole$   &$\times~(\text{w})$   &$\Xi_{cc}^{+}(1^{2}P_{\lambda})-pole$   &$0$\\
&$\Xi_{c}^{+}(1S)-pole$   &$\times~(\text{s})$   &$\Xi_{cc}^{+}(1^{4}P_{\lambda})-pole$   &$37.44$  &  &$\Xi_{c}^{+}(1S)-pole$   &$-10.36$   &$\Xi_{cc}^{+}(1^{4}P_{\lambda})-pole$   &$\times~(\text{w})$\\
&$\Xi_{c}^{+}(2S)-pole$   &$\times~(\text{s})$   &$\Xi_{c}^{+}(1^{2}P_{\lambda})-pole$   &$\times~(\text{s})$  &  &$\Xi_{c}^{+}(2S)-pole$   &$-0.52$   &$\Xi_{c}^{+}(1^{2}P_{\lambda})-pole$   &$0.061$\\
&$\Xi_{c}^{\prime+}(1S)-pole$   &$\times~(\text{w})$   &$\Xi_{c}^{\prime+}(1^{2}P_{\lambda})-pole$   &$0$  &  &$\Xi_{c}^{\prime+}(1S)-pole$   &$\times~(\text{w})$   &$\Xi_{c}^{\prime+}(1^{2}P_{\lambda})-pole$   &$0$\\
&$\Xi_{c}^{\prime+}(2S)-pole$   &$\times~(\text{w})$   &$\Xi_{c}^{\prime+}(1^{4}P_{\lambda})-pole$   &$0$  &  &$\Xi_{c}^{\prime+}(2S)-pole$   &$\times~(\text{w})$   &$\Xi_{c}^{\prime+}(1^{4}P_{\lambda})-pole$   &$0$\\
&$\text{total}$   &$25.23$   &$\text{total}$   &$43.08$  &  &$\text{total}$   &$-10.88$   &$\text{total}$   &$0.061$\\
\specialrule{0em}{3pt}{3pt}
\multirow{1}*{\shortstack{$\Xi_{cc}^{+}\to\Sigma_{c}^{++}K^{-}$}}    &$\Xi_{c}^{+}(1S)-pole$   &$-73.27$   &$\Xi_{c}^{+}(1^{2}P_{\lambda})-pole$   &$1.05$  &\multirow{1}*{\shortstack{$\Xi_{cc}^{+}\to\Omega_{c}^{0}K^{+}$}}    &$\Xi_{c}^{+}(1S)-pole$   &$-115.05$   &$\Xi_{c}^{+}(1^{2}P_{\lambda})-pole$   &$0.97$\\
&$\Xi_{c}^{+}(2S)-pole$   &$-29.13$   &$\Xi_{c}^{\prime+}(1^{2}P_{\lambda})-pole$   &$0$  &  &$\Xi_{c}^{+}(2S)-pole$   &$-30.17$   &$\Xi_{c}^{\prime+}(1^{2}P_{\lambda})-pole$   &$0$\\
&$\Xi_{c}^{\prime+}(1S)-pole$   &$\times~(\text{w})$   &$\Xi_{c}^{\prime+}(1^{4}P_{\lambda})-pole$   &$0$  &  &$\Xi_{c}^{\prime+}(1S)-pole$   &$\times~(\text{w})$   &$\Xi_{c}^{\prime+}(1^{4}P_{\lambda})-pole$   &$0$\\
&$\Xi_{c}^{\prime+}(2S)-pole$   &$\times~(\text{w})$   &   &  &  &$\Xi_{c}^{\prime+}(2S)-pole$   &$\times~(\text{w})$   &   &\\
&$\text{total}$   &$-102.40$   &$\text{total}$   &$1.05$  &  &$\text{total}$   &$-145.22$   &$\text{total}$   &$0.97$\\
\bottomrule[0.5pt]
\bottomrule[1pt]
\end{tabular*}
\end{table*}

Numerically in the $W$-exchange amplitudes, the PC parts exhibit a clear hierarchy between $1S$ and $2S$ intermediate pole contributions, that is the $1S$ poles generally provide the dominate contributions, while the contributions from $2S$ poles are reduced. Interestingly, the $2S$ suppression is significantly stronger for double-charmed baryon poles than for single-charmed ones. For instance, in the $\Xi_{cc}^{+}\to\Xi_{c}^{+}\pi^{0}$ decay, the amplitudes arising from the $\Xi_{cc}^{+}(1S)$ and $\Xi_{cc}^{+}(2S)$ poles are $-38.49\times10^{-4}G_{F}~\text{GeV}^{3/2}$ and $-1.66\times10^{-4}G_{F}~\text{GeV}^{3/2}$, respectively, with the $2S$ contribution reduced to only a few percent of the $1S$ one. By contrast, in the $\Xi_{cc}^{+}\to\Xi_{c}^{\prime+}\pi^{0}$ decay, the amplitudes arising from $\Xi_{c}^{+}(1S)$ and $\Xi_{c}^{+}(2S)$ poles are $48.60\times10^{-4}G_{F}~\text{GeV}^{3/2}$ and $22.37\times10^{-4}G_{F}~\text{GeV}^{3/2}$, respectively, with the $2S$ contribution remaining at about $50\%$ of the $1S$ one.

For the PV amplitudes from the $W$-exchange diagrams, which are mediated by $P$-wave intermediate baryons with $J^{P}=1/2^{-}$, a different pattern emerges. In particular, whenever the $\Xi_{cc}^{+}(1^{4}P_{\lambda})$ pole contributes, it tends to dominate the PV amplitudes, as in the $\Xi_{cc}^{+}\to\Xi_{c}^{+}\pi^{0}$ and $\Xi_{c}^{+}\eta^{(\prime)}$ processes. These results indicate that a reliable dynamical description of the internal structure and spatial wave functions of the $P$-wave charmed baryons, particularly the double-charmed baryons, is essential for accurate predictions of the PV amplitudes.

To facilitate a direct comparison with previous theoretical studies, we summarize in Table~\ref{tab:PhysicalObservablesComparison} our numerical results for the factorizable and nonfactorizable contributions to the PV and PC amplitudes of the CF decays, together with the branching fractions and asymmetry parameters $\alpha$. The available predictions from various theoretical approaches are also included for comparison, in which the Refs.~\cite{Cheng:2020wmk,Liu:2023dvg} employed pole model, while the Ref.~\cite{Hu:2024uia} considered FSI effects. In this table, the PV amplitude $A$ and the PC amplitude $B$ are defined through
\begin{equation}
\mathcal{M}[\mathcal{B}_{i}\to\mathcal{B}_{f}+P]=\bar{u}_{f}(A-B\gamma_{5})u_{i},
\end{equation}
and are related to $\mathcal{M}_{\text{PV}}^{-1/2,-1/2}$ and $\mathcal{M}_{\text{PC}}^{-1/2,-1/2}$ presented in Table~\ref{tab:CFamplitudes} by
\begin{equation}
\begin{split}
\mathcal{M}_{\text{PV}}=&\frac{1}{4\pi}\sqrt{\frac{m_{f}}{{\pi}E_{f}E_{P}}}\sqrt{\frac{E_{f}+m_{f}}{2m_{f}}}\chi_{f}^{\dagger}\chi_{i}A,\\
\mathcal{M}_{\text{PC}}=&\frac{1}{4\pi}\sqrt{\frac{m_{f}}{{\pi}E_{f}E_{P}}}\sqrt{\frac{E_{f}+m_{f}}{2m_{f}}}\chi_{f}^{\dagger}\frac{\pmb{\sigma}\cdot\pmb{P}_{f}}{E_{f}+m_{f}}\chi_{i}B,
\end{split}
\end{equation}
respectively.

\begin{table*}[htbp]\centering
\caption{The comparison of factorizable and nonfactorizable amplitudes (in the units of $10^{-2}G_{F}~\text{GeV}^{-2}$), as well as the branching fractions and asymmetry parameters $\alpha$ of CF decays by different theoretical works.}
\label{tab:PhysicalObservablesComparison}
\renewcommand\arraystretch{1.05}
\begin{threeparttable}
\begin{tabular*}{162mm}{l@{\extracolsep{\fill}}cccccccc}
\toprule[1pt]
\toprule[0.5pt]
Modes   &$A^{\text{fac}}$   &$A^{\text{nf}}$   &$A^{\text{total}}$   &$B^{\text{fac}}$   &$B^{\text{nf}}$   &$B^{\text{total}}$   &$\mathcal{B}$   &$\alpha$\\
\midrule[0.5pt]
\multirow{1}*{$\Xi_{cc}^{+}\to\Xi_{c}^{0}\pi^{+}$}&&&&&&&&\\
~This work   &$5.92$  &$0$  &$5.92$  &$-16.89$  &$0$  &$-16.89$  &$0.47\%$  &$-0.83$\\
~Cheng, pole model~\cite{Cheng:2020wmk}  &$8.52$  &$10.79$  &$19.31$  &$-16.46$  &$-0.08$  &$-16.54$  &$3.84\%$  &$-0.31$  \\
~Liu, pole model~\cite{Liu:2023dvg}  &$-5.34$  &$-5.74$  &$-11.08$  &$7.60$  &$0$  &$7.60$  &$1.00\%$  &$-0.25$  \\
~Hu, FSI~\cite{Hu:2024uia}$^{[a]}$  &$9.89$  &$2.07-0.81i$  &$12.00-0.81i$  &$-14.80$  &$-5.55+5.18i$  &$-20.30+5.18i$  &$(1.56^{+0.26}_{-0.22})\%$  &$-0.58^{+0.02}_{-0.02}$  \\
\specialrule{0em}{2pt}{2pt}
\multirow{1}*{$\Xi_{cc}^{+}\to\Xi_{c}^{\prime0}\pi^{+}$}&&&&&&&&\\
~This work   &$-3.38$  &$-0.19$  &$-3.58$  &$49.28$  &$128.83$  &$178.11$  &$8.90\%$  &$-0.24$\\
~Cheng, pole model~\cite{Cheng:2020wmk}  &$5.05$  &$0.04$  &$5.09$  &$-52.31$  &$-17.63$  &$-69.94$  &$1.55\%$  &$-0.73$  \\
~Liu, pole model~\cite{Liu:2023dvg}  &$0$  &$-1.12$  &$-1.12$  &$0$  &$0$  &$0$  &$0.01\%$  &$0$  \\
~Hu, FSI~\cite{Hu:2024uia}  &$-5.27$  &$0.54-0.62i$  &$-4.73-0.62i$  &$42.30$  &$-3.08+2.25i$  &$39.30+2.25i$  &$(0.62^{+0.04}_{-0.03})\%$  &$-0.95^{+0.01}_{-0.00}$  \\
\specialrule{0em}{2pt}{2pt}
\multirow{1}*{$\Xi_{cc}^{+}\to\Lambda_{c}^{+}\bar{K}^{0}$}&&&&&&&&\\
~This work   &$8.08$  &$0$  &$8.08$  &$-20.18$  &$0$  &$-20.18$  &$0.86\%$  &$-0.82$\\
~Cheng, pole model~\cite{Cheng:2020wmk}  &$-3.37$  &$8.90$  &$5.53$  &$5.62$  &$-0.07$  &$5.55$  &$0.31\%$  &$0.40$  \\
~Hu, FSI~\cite{Hu:2024uia}  &$-0.15$  &$2.23-1.41i$  &$2.07-1.41i$  &$0.23$  &$-1.93+14.10i$  &$-1.70+14.10i$  &$(0.15^{+0.16}_{-0.09})\%$  &$-0.65^{+0.06}_{-0.04}$  \\
\specialrule{0em}{2pt}{2pt}
\multirow{1}*{$\Xi_{cc}^{+}\to\Sigma_{c}^{+}\bar{K}^{0}$}&&&&&&&&\\
~This work   &$4.62$  &$0.17$  &$4.79$  &$-57.65$  &$-94.52$  &$-152.17$  &$6.92\%$  &$-0.35$\\
~Cheng, pole model~\cite{Cheng:2020wmk}  &$-2.17$  &$0.04$  &$-2.14$  &$19.37$  &$15.64$  &$35.02$  &$0.38\%$  &$-0.62$  \\
~Hu, FSI~\cite{Hu:2024uia}  &$-0.11$  &$-0.50-0.54i$  &$-0.66-0.54i$  &$0.16$  &$0.58+13.00i$  &$0.81+13.00i$  &$(0.05^{+0.06}_{-0.03})\%$  &$-0.45^{+0.30}_{-0.22}$  \\
\specialrule{0em}{2pt}{2pt}
\multirow{1}*{$\Xi_{cc}^{+}\to\Xi_{c}^{+}\pi^{0}$}&&&&&&&&\\
~This work   &$0$  &$20.72$  &$20.72$  &$0$  &$-45.83$  &$-45.83$  &$5.28\%$  &$-0.71$\\
~Cheng, pole model~\cite{Cheng:2020wmk}  &$0$  &$15.26$  &$15.26$  &$0$  &$-10.49$  &$-10.49$  &$2.38\%$  &$-0.25$  \\
~Liu, pole model~\cite{Liu:2023dvg}  &$0$  &$4.92$  &$4.92$  &$0$  &$-3.84$  &$-3.84$  &$0.20\%$  &$-0.29$  \\
~Hu, FSI~\cite{Hu:2024uia}  &$0$  &$0.74+1.04i$  &$0.74+1.04i$  &$0$  &$-2.39-3.35i$  &$-2.39-3.35i$  &$(0.02^{+0.02}_{-0.01})\%$  &$-0.89^{+0.02}_{-0.02}$  \\
\specialrule{0em}{2pt}{2pt}
\multirow{1}*{$\Xi_{cc}^{+}\to\Xi_{c}^{\prime+}\pi^{0}$}&&&&&&&&\\
~This work   &$0$  &$-0.13$  &$-0.13$  &$0$  &$91.10$  &$91.10$  &$2.29\%$  &$-0.02$\\
~Cheng, pole model~\cite{Cheng:2020wmk}  &$0$  &$0.06$  &$0.06$  &$0$  &$-24.97$  &$-24.97$  &$0.17\%$  &$-0.03$  \\
~Liu, pole model~\cite{Liu:2023dvg}  &$-2.80$  &$-1.59$  &$-4.39$  &$22.31$  &$0$  &$22.31$  &$0.25\%$  &$-0.99$  \\
~Hu, FSI~\cite{Hu:2024uia}  &$0$  &$-0.67+0.13i$  &$-0.67+0.13i$  &$0$  &$-3.42+2.00i$  &$-3.42+2.00i$  &$(0.01^{+0.01}_{-0.01})\%$  &$0.94^{+0.01}_{-0.14}$  \\
\specialrule{0em}{2pt}{2pt}
\multirow{1}*{$\Xi_{cc}^{+}\to\Xi_{c}^{+}\eta$}&&&&&&&&\\
~This work   &$0$  &$17.65$  &$17.65$  &$0$  &$40.56$  &$40.56$  &$3.28\%$  &$0.67$\\
~Cheng, pole model~\cite{Cheng:2020wmk}  &$0$  &$21.75$  &$21.75$  &$0$  &$4.86$  &$4.86$  &$4.18\%$  &$0.07$  \\
~Liu, pole model~\cite{Liu:2023dvg}  &$0$  &$-2.37$  &$-2.37$  &$0$  &$1.85$  &$1.85$  &$0.04\%$  &$-0.26$  \\
\specialrule{0em}{2pt}{2pt}
\multirow{1}*{$\Xi_{cc}^{+}\to\Xi_{c}^{\prime+}\eta$}&&&&&&&&\\
~This work   &$0$  &$-0.19$  &$-0.19$  &$0$  &$153.09$  &$153.09$  &$4.07\%$  &$-0.02$\\
~Cheng, pole model~\cite{Cheng:2020wmk}  &$0$  &$0.09$  &$0.09$  &$0$  &$-17.87$  &$-17.87$  &$0.05\%$  &$-0.07$  \\
~Liu, pole model~\cite{Liu:2023dvg}  &$0$  &$-1.62$  &$-1.62$  &$0$  &$0$  &$0$  &$0.02\%$  &$0$  \\
\specialrule{0em}{2pt}{2pt}
\multirow{1}*{$\Xi_{cc}^{+}\to\Xi_{c}^{+}\eta^{\prime}$}&&&&&&&&\\
~This work   &$0$  &$10.09$  &$10.09$  &$0$  &$55.88$  &$55.88$  &$0.77\%$  &$0.87$\\
\specialrule{0em}{2pt}{2pt}
\multirow{1}*{$\Xi_{cc}^{+}\to\Xi_{c}^{\prime+}\eta^{\prime}$}&&&&&&&&\\
~This work   &$0$  &$0.014$  &$0.014$  &$0$  &$-37.07$  &$-37.07$  &$0.023\%$  &$-0.01$\\
\specialrule{0em}{2pt}{2pt}
\multirow{1}*{$\Xi_{cc}^{+}\to\Sigma_{c}^{++}K^{-}$}&&&&&&&&\\
~This work   &$0$  &$0.24$  &$0.24$  &$0$  &$-133.67$  &$-133.67$  &$5.17\%$  &$-0.02$\\
~Cheng, pole model~\cite{Cheng:2020wmk}  &$0$  &$0.07$  &$0.07$  &$0$  &$22.14$  &$22.14$  &$0.13\%$  &$0.04$  \\
~Liu, pole model~\cite{Liu:2023dvg}  &$0$  &$1.87$  &$1.87$  &$0$  &$0$  &$0$  &$0.03\%$  &$0$  \\
~Hu, FSI~\cite{Hu:2024uia}  &$0$  &$-0.49+0.48i$  &$-0.49+0.48i$  &$0$  &$3.35-9.70i$  &$3.35-9.70i$  &$(0.03^{+0.04}_{-0.02})\%$  &$0.60^{+0.13}_{-0.18}$  \\
\specialrule{0em}{2pt}{2pt}
\multirow{1}*{$\Xi_{cc}^{+}\to\Omega_{c}^{0}K^{+}$}&&&&&&&&\\
~This work   &$0$  &$0.20$  &$0.20$  &$0$  &$-240.77$  &$-240.77$  &$7.09\%$  &$-0.01$\\
~Cheng, pole model~\cite{Cheng:2020wmk}  &$0$  &$0.05$  &$0.05$  &$0$  &$-22.98$  &$-22.98$  &$0.06\%$  &$-0.03$  \\
~Liu, pole model~\cite{Liu:2023dvg}  &$0$  &$-1.87$  &$-1.87$  &$0$  &$0$  &$0$  &$0.02\%$  &$0$  \\
~Hu, FSI~\cite{Hu:2024uia}  &$0$  &$2.20-2.18i$  &$2.20-2.18i$  &$0$  &$-7.29+1.80i$  &$-7.29+1.80i$  &$(0.08^{+0.09}_{-0.05})\%$  &$0.47^{+0.04}_{-0.04}$  \\
\bottomrule[0.5pt]
\bottomrule[1pt]
\end{tabular*}
\begin{tablenotes}
\footnotesize
\item[a] For the results of Ref.~\cite{Hu:2024uia}, only the center values of the decay amplitudes are listed in this table, with the corresponding uncertainties omitted for brevity.
\end{tablenotes}
\end{threeparttable}
\end{table*}

As shown in Table~\ref{tab:PhysicalObservablesComparison}, several CF decay modes are predicted to have large branching fractions, liking
\begin{equation}
\begin{split}
\mathcal{B}(\Xi_{cc}^{+}\to\Xi_{c}^{\prime0}\pi^{+})=&8.90\%,\\
\mathcal{B}(\Xi_{cc}^{+}\to\Sigma_{c}^{+}\bar{K}^{0})=&6.92\%,\\
\mathcal{B}(\Xi_{cc}^{+}\to\Xi_{c}^{+}\pi^{0})=&5.28\%,\\
\mathcal{B}(\Xi_{cc}^{+}\to\Sigma_{c}^{++}K^{-})=&5.17\%,\\
\mathcal{B}(\Xi_{cc}^{+}\to\Omega_{c}^{0}K^{+})=&7.09\%.
\end{split}
\end{equation}
These decay channels therefore constitute promising channels for future experimental searches. The sizable branching fractions of these decay modes can be understood from the structure of their decay amplitudes. For $\Xi_{cc}^{+}\to\Xi_{c}^{\prime0}\pi^{+}$ and $\Xi_{cc}^{+}\to\Sigma_{c}^{+}\bar{K}^{0}$ modes, the factorizable and nonfactorizable contributions to the PC amplitudes interfere constructively, increasing the total PC amplitudes substantially. By contrast, the $\Xi_{cc}^{+}\to\Sigma_{c}^{++}K^{-}$ and $\Xi_{cc}^{+}\to\Omega_{c}^{0}K^{+}$ modes are governed almost entirely by the nonfactorizable $W$-exchange diagrams, with large PC amplitudes generated by the $\Xi_{c}^{+}(1S)$ and $\Xi_{c}^{+}(2S)$ poles.

However, as shown in Table~\ref{tab:PhysicalObservablesComparison}, our predicted branching fractions for the $\Xi_{cc}^{+}\to\Xi_{c}^{\prime0}\pi^{+}$ and $\Xi_{cc}^{+}\to\Omega_{c}^{0}K^{+}$ channels differ significantly from Ref.~\cite{Cheng:2020wmk}. Specifically, we obtain branching fractions of $8.90\%$ and $7.09\%$ for these two channels, respectively, compared with $1.55\%$ and $0.06\%$ obtained in Ref.~\cite{Cheng:2020wmk}. A comparison of the individual decay amplitudes indicates that these discrepancies mainly originate from the nonfactorizable PC contributions, for which we obtain $\vert{B^{\text{nf}}_{\Xi_{cc}^{+}\to\Xi_{c}^{\prime0}\pi^{+}}}\vert=128.83\times10^{-2}G_{F}~\text{GeV}^{-2}$ and $\vert{B^{\text{nf}}_{\Xi_{cc}^{+}\to\Omega_{c}^{0}K^{+}}}\vert=240.77\times10^{-2}G_{F}~\text{GeV}^{-2}$, compared with $17.63\times10^{-2}G_{F}~\text{GeV}^{-2}$ and $22.98\times10^{-2}G_{F}~\text{GeV}^{-2}$ in Ref.~\cite{Cheng:2020wmk}. In Ref.~\cite{Cheng:2020wmk}, the nonfactorizable PC amplitudes were also evaluated using pole model by considering $J^{P}=1/2^{+}$ $\Xi_{c}^{(\prime)+}$ poles. Retaining the dominant contributions from $\Xi_{c}^{+}$ pole, the decay amplitudes can be approximated as
\begin{equation}
\begin{split}
B^{\text{nf}}_{\Xi_{cc}^{+}\to\Xi_{c}^{\prime0}\pi^{+}}\simeq&\frac{1}{f_{\pi}}\Big{(}g_{\Xi_{c}^{\prime0}\Xi_{c}^{+}}^{A(\pi^{+})}\frac{m_{\Xi_{c}^{\prime0}}+m_{\Xi_{c}^{+}}}{m_{\Xi_{cc}^{+}}-m_{\Xi_{c}^{+}}}a_{\Xi_{c}^{+}\Xi_{cc}^{+}}\Big{)},\\
B^{\text{nf}}_{\Xi_{cc}^{+}\to\Omega_{c}^{0}K^{+}}\simeq&\frac{1}{f_{K}}\Big{(}g_{\Omega_{c}^{0}\Xi_{c}^{+}}^{A(K^{+})}\frac{m_{\Omega_{c}^{0}}+m_{\Xi_{c}^{+}}}{m_{\Xi_{cc}^{+}}-m_{\Xi_{c}^{+}}}a_{\Xi_{c}^{+}\Xi_{cc}^{+}}\Big{)}.
\end{split}
\end{equation}
Here, we have neglected the contributions from $\Xi_{c}^{\prime+}$ pole since $a_{\Xi_{c}^{\prime+}\Xi_{cc}^{+}}{\ll}a_{\Xi_{c}^{+}\Xi_{cc}^{+}}$~\cite{Cheng:2020wmk}. The relevant nonperturbative transition matrix elements were evaluated within the MIT bag model. Notably, the present work and Ref.~\cite{Cheng:2020wmk} share essentially the same physical picture for nonfactorizable PC amplitudes, but differ in their numerical evaluations of transition matrix elements, which accounts for the large discrepancies in the predicted amplitudes. The sizable discrepancies highlight the model dependence of nonfactorizable pole amplitudes and motivate further theoretical investigations and future experimental measurements.

The $\Xi_{cc}^{+}\to\Xi_{c}^{+}\pi^{0}$ and $\Xi_{cc}^{+}\to\Xi_{c}^{+}\eta^{(\prime)}$ decay modes share similar dynamical pattern, in which sizable nonfactorizable contributions appear in both the PC and PV amplitudes, dominated by the $\Xi_{cc}^{+}(1S)$ and $\Xi_{cc}^{+}(1^{4}P_{\lambda})$ baryon poles, respectively. Nevertheless, the much larger phase space available for the $\Xi_{c}^{+}\pi^{0}$ final state favors a larger branching fraction as $\mathcal{B}(\Xi_{cc}^{+}\to\Xi_{c}^{+}\pi^{0})=5.28\%$, whereas the heavier $\eta$ and $\eta^{\prime}$ mesons lead to increasingly restricted phase space and hence suppress the corresponding branching fractions as $\mathcal{B}(\Xi_{cc}^{+}\to\Xi_{c}^{+}\eta)=3.28\%$ and $\mathcal{B}(\Xi_{cc}^{+}\to\Xi_{c}^{+}\eta^{\prime})=0.77\%$.

It is also instructive to consider the $\Xi_{cc}^{+}\to\Xi_{c}^{0}\pi^{+}$ and $\Xi_{cc}^{+}\to\Lambda_{c}^{+}\bar{K}^{0}$ decay modes. In our calculation, they receive contributions only from factorizable $T$ and $C$ diagrams, respectively, whereas the nonfactorizable pole contributions vanish owing to the spin-flavor structure of the corresponding weak or strong vertices. This differs markedly from the result of Ref.~\cite{Cheng:2020wmk}, where sizable nonfactorizable PV amplitudes were obtained. In Ref.~\cite{Cheng:2020wmk}, the nonfactorizable PV amplitudes were evaluated using current algebra in the soft-pseudoscalar meson limit. Under this approximation, the contributions from the $J^{P}=1/2^{-}$ intermediate poles are reduced to commutator terms, that is~\cite{Cheng:2020wmk},
\begin{equation}
\begin{split}
A^{\text{nf}}(\Xi_{cc}^{+}\to\Xi_{c}^{0}\pi^{+})=&\frac{1}{f_{\pi}}\big{(}a_{\Xi_{c}^{+}\Xi_{cc}^{+}}\big{)},\\
A^{\text{nf}}(\Xi_{cc}^{+}\to\Lambda_{c}^{+}\bar{K}^{0})=&\frac{1}{f_{K}}\big{(}a_{\Xi_{c}^{+}\Xi_{cc}^{+}}\big{)},
\end{split}
\end{equation}
where the baryon weak transition matrix element is defined as
\begin{equation}
a_{\Xi_{c}^{+}\Xi_{cc}^{+}}=\frac{G_{F}}{2\sqrt{2}}V_{cs}V_{ud}^{*}c_{-}\langle{\Xi_{c}^{+}}\vert\mathcal{O}_{-}\vert{\Xi_{cc}^{+}}\rangle
\end{equation}
with $c_{-}=c_{1}-c_{2}$ and $\mathcal{O}_{-}=(\bar{s}c)(\bar{u}d)-(\bar{s}d)(\bar{u}c)$. As a consequence, the Ref.~\cite{Cheng:2020wmk} obtained sizable nonfactorizable PV amplitudes, $A^{\text{nf}}=10.79\times10^{-2}G_{F}~\text{GeV}^{-2}$ and $8.90\times10^{-2}G_{F}~\text{GeV}^{-2}$ for the $\Xi_{cc}^{+}\to\Xi_{c}^{0}\pi^{+}$ and $\Xi_{cc}^{+}\to\Lambda_{c}^{+}\bar{K}^{0}$ decay modes, respectively, whereas the corresponding contributions vanish in our calculation. This discrepancy highlights the sensitivity of the nonfactorizable PV amplitudes to their theoretical treatment, and motivates further theoretical studies and future experimental measurements to clarify the underlying decay mechanism of the $W$-exchange processes.

\subsection{The SCS and DCS decays}
\label{sec4.2}

We next turn to the SCS and DCS decays, whose decay amplitudes are presented in Tables~\ref{tab:SCSamplitudes} and \ref{tab:DCSamplitudes}, respectively. Compared to the CF decay modes, the SCS and DCS decay amplitudes are suppressed by the CKM matrix elements of approximately $0.2$ and $0.2^{2}$, respectively, leading to an overall suppression of the corresponding branching fractions. The corresponding branching fractions evaluated in this work, together with those reported by Ref.~\cite{Hu:2024uia}, are presented in Table~\ref{tab:SCSDCS}.

\begin{table*}[htbp]\centering
\caption{Same as Table~\ref{tab:CFamplitudes} but for SCS decays.}
\label{tab:SCSamplitudes}
\renewcommand\arraystretch{1.05}
\begin{tabular*}{170mm}{l@{\extracolsep{\fill}}lclc|llclc}
\toprule[1pt]
\toprule[0.5pt]
\specialrule{0em}{1pt}{1pt}
&\multicolumn{2}{c}{{\normalsize $\mathcal{M}_{\text{PC}}^{-1/2,-1/2}$}}  &\multicolumn{2}{c}{{\normalsize $\mathcal{M}_{\text{PV}}^{-1/2,-1/2}$}} &  &\multicolumn{2}{c}{{\normalsize $\mathcal{M}_{\text{PC}}^{-1/2,-1/2}$}}  &\multicolumn{2}{c}{{\normalsize $\mathcal{M}_{\text{PV}}^{-1/2,-1/2}$}}\\
\specialrule{0em}{1pt}{1pt}
\midrule[0.5pt]
\multirow{1}*{\shortstack{$\Xi_{cc}^{+}\to\Sigma_{c}^{0}\pi^{+}$}}    &$T-diagram$   &$-9.82$   &$T-diagram$   &$3.50$
&\multirow{1}*{\shortstack{$\Xi_{cc}^{+}\to\Sigma_{c}^{++}\pi^{-}$}}    &$\Lambda_{c}^{+}(1S)-pole$   &$18.09$   &$\Lambda_{c}^{+}(1^{2}P_{\lambda})-pole$   &$-0.27$\\
&$\Lambda_{c}^{+}(1S)-pole$   &$-18.09$   &$\Lambda_{c}^{+}(1^{2}P_{\lambda})-pole$   &$0.27$   &   &$\Lambda_{c}^{+}(2S)-pole$   &$9.54$   &$\Sigma_{c}^{+}(1^{2}P_{\lambda})-pole$   &$0$\\
&$\Lambda_{c}^{+}(2S)-pole$   &$-9.54$   &$\Sigma_{c}^{+}(1^{2}P_{\lambda})-pole$    &$0$   &   &$\Sigma_{c}^{+}(1S)-pole$   &$\times~(\text{w})$   &$\Sigma_{c}^{+}(1^{4}P_{\lambda})-pole$   &$0$\\
&$\Sigma_{c}^{+}(1S)-pole$    &$\times~(\text{w})$   &$\Sigma_{c}^{+}(1^{4}P_{\lambda})-pole$    &$0$   &   &$\Sigma_{c}^{+}(2S)-pole$   &$\times~(\text{w})$   &   &\\
&$\Sigma_{c}^{+}(2S)-pole$    &$\times~(\text{w})$   &   &                                             &  &$\text{total}$     &$27.63$   &$\text{total}$   &$-0.27$\\
&$\text{total}$   &$-37.45$   &$\text{total}$   &$3.77$  &  &   &   &   &\\
\specialrule{0em}{3pt}{3pt}
\multirow{1}*{\shortstack{$\Xi_{cc}^{+}\to\Xi_{c}^{0}K^{+}$}}   &$T-diagram$   &$-6.59$   &$T-diagram$   &$16.06$
&\multirow{1}*{\shortstack{$\Xi_{cc}^{+}\to\Xi_{c}^{\prime0}K^{+}$}}   &$T-diagram$   &$17.52$   &$T-diagram$   &$-9.70$\\
&$\Lambda_{c}^{+}(1S)-pole$   &$\times~(\text{s})$   &$\Lambda_{c}^{+}(1^{2}P_{\lambda})-pole$   &$\times~(\text{s})$  &  &$\Lambda_{c}^{+}(1S)-pole$   &$-13.53$   &$\Lambda_{c}^{+}(1^{2}P_{\lambda})-pole$   &$0.16$\\
&$\Lambda_{c}^{+}(2S)-pole$   &$\times~(\text{s})$   &$\Sigma_{c}^{+}(1^{2}P_{\lambda})-pole$   &$0$  &  &$\Lambda_{c}^{+}(2S)-pole$   &$-5.45$   &$\Sigma_{c}^{+}(1^{2}P_{\lambda})-pole$   &$0$\\
&$\Sigma_{c}^{+}(1S)-pole$   &$\times~(\text{w})$   &$\Sigma_{c}^{+}(1^{4}P_{\lambda})-pole$   &$0$  &  &$\Sigma_{c}^{+}(1S)-pole$   &$\times~(\text{w})$   &$\Sigma_{c}^{+}(1^{4}P_{\lambda})-pole$   &$0$\\
&$\Sigma_{c}^{+}(2S)-pole$   &$\times~(\text{w})$   &   &                                           &  &$\Sigma_{c}^{+}(2S)-pole$   &$\times~(\text{w})$   &   &\\
&$\text{total}$   &$-6.59$   &$\text{total}$   &$16.06$  &  &$\text{total}$   &$-1.46$   &$\text{total}$   &$-9.54$\\
\specialrule{0em}{3pt}{3pt}
\multirow{1}*{\shortstack{$\Xi_{cc}^{+}\to\Lambda_{c}^{+}\pi^{0}$}}   &$C-diagram$   &$-1.35$   &$C-diagram$   &$2.05$
&\multirow{1}*{\shortstack{$\Xi_{cc}^{+}\to\Sigma_{c}^{+}\pi^{0}$}}   &$C-diagram$   &$-3.42$   &$C-diagram$   &$1.22$\\
&$C^{\prime}-diagram$   &$-0.36$   &$C^{\prime}-diagram$   &$1.21$
& &$C^{\prime}-diagram$   &$0$   &$C^{\prime}-diagram$   &$0$\\
&$\Xi_{cc}^{+}(1S)-pole$   &$7.11$   &$\Xi_{cc}^{+}(1^{2}P_{\rho})-pole$   &$\times~(\text{s})$  &  &$\Xi_{cc}^{+}(1S)-pole$   &$\times~(\text{w})$   &$\Xi_{cc}^{+}(1^{2}P_{\rho})-pole$   &$\times~(\text{s})$\\
&$\Xi_{cc}^{+}(2S)-pole$   &$0.35$   &$\Xi_{cc}^{+}(1^{2}P_{\lambda})-pole$   &$0.31$  &  &$\Xi_{cc}^{+}(2S)-pole$   &$\times~(\text{w})$   &$\Xi_{cc}^{+}(1^{2}P_{\lambda})-pole$   &$0$\\
&$\Lambda_{c}^{+}(1S)-pole$   &$\times~(\text{s})$   &$\Xi_{cc}^{+}(1^{4}P_{\lambda})-pole$   &$-20.97$  &  &$\Lambda_{c}^{+}(1S)-pole$   &$-18.09$   &$\Xi_{cc}^{+}(1^{4}P_{\lambda})-pole$   &$\times~(\text{w})$\\
&$\Lambda_{c}^{+}(2S)-pole$   &$\times~(\text{s})$   &$\Lambda_{c}^{+}(1^{2}P_{\lambda})-pole$   &$\times~(\text{s})$  &  &$\Lambda_{c}^{+}(2S)-pole$   &$-9.54$   &$\Lambda_{c}^{+}(1^{2}P_{\lambda})-pole$   &$0.27$\\
&$\Sigma_{c}^{+}(1S)-pole$   &$\times~(\text{w})$   &$\Sigma_{c}^{+}(1^{2}P_{\lambda})-pole$   &$0$  &  &$\Sigma_{c}^{+}(1S)-pole$   &$\times~(\text{w,s})$   &$\Sigma_{c}^{+}(1^{2}P_{\lambda})-pole$   &$\times~(\text{s})$\\
&$\Sigma_{c}^{+}(2S)-pole$   &$\times~(\text{w})$   &$\Sigma_{c}^{+}(1^{4}P_{\lambda})-pole$   &$0$  &  &$\Sigma_{c}^{+}(2S)-pole$   &$\times~(\text{w,s})$   &$\Sigma_{c}^{+}(1^{4}P_{\lambda})-pole$   &$\times~(\text{s})$\\
&$\text{total}$   &$5.75$   &$\text{total}$   &$-17.40$  &  &$\text{total}$   &$-31.05$   &$\text{total}$   &$1.49$\\
\specialrule{0em}{3pt}{3pt}
\multirow{1}*{\shortstack{$\Xi_{cc}^{+}\to\Lambda_{c}^{+}\eta$}}   &$C-diagram$   &$3.42$   &$C-diagram$   &$-6.72$
&\multirow{1}*{\shortstack{$\Xi_{cc}^{+}\to\Sigma_{c}^{+}\eta$}}   &$C-diagram$   &$8.56$   &$C-diagram$   &$-4.09$\\
&$C^{\prime}-diagram$   &$0.39$   &$C^{\prime}-diagram$   &$-1.32$
& &$C^{\prime}-diagram$   &$0$   &$C^{\prime}-diagram$   &$0$\\
&$\Xi_{cc}^{+}(1S)-pole$   &$-5.72$   &$\Xi_{cc}^{+}(1^{2}P_{\rho})-pole$   &$\times~(\text{s})$  &  &$\Xi_{cc}^{+}(1S)-pole$   &$\times~(\text{w})$   &$\Xi_{cc}^{+}(1^{2}P_{\rho})-pole$   &$\times~(\text{s})$\\
&$\Xi_{cc}^{+}(2S)-pole$   &$-0.24$   &$\Xi_{cc}^{+}(1^{2}P_{\lambda})-pole$   &$-0.25$  &  &$\Xi_{cc}^{+}(2S)-pole$   &$\times~(\text{w})$   &$\Xi_{cc}^{+}(1^{2}P_{\lambda})-pole$   &$0$\\
&$\Lambda_{c}^{+}(1S)-pole$   &$\times~(\text{s})$   &$\Xi_{cc}^{+}(1^{4}P_{\lambda})-pole$   &$-15.32$  &  &$\Lambda_{c}^{+}(1S)-pole$   &$\times~(\text{s})$   &$\Xi_{cc}^{+}(1^{4}P_{\lambda})-pole$   &$\times~(\text{w})$\\
&$\Lambda_{c}^{+}(2S)-pole$   &$\times~(\text{s})$   &$\Lambda_{c}^{+}(1^{2}P_{\lambda})-pole$   &$\times~(\text{s})$  &  &$\Lambda_{c}^{+}(2S)-pole$   &$\times~(\text{s})$   &$\Lambda_{c}^{+}(1^{2}P_{\lambda})-pole$   &$\times~(\text{s})$\\
&$\Sigma_{c}^{+}(1S)-pole$   &$\times~(\text{w,s})$   &$\Sigma_{c}^{+}(1^{2}P_{\lambda})-pole$   &$\times~(\text{s})$  &  &$\Sigma_{c}^{+}(1S)-pole$   &$\times~(\text{w})$   &$\Sigma_{c}^{+}(1^{2}P_{\lambda})-pole$   &$0$\\
&$\Sigma_{c}^{+}(2S)-pole$   &$\times~(\text{w,s})$   &$\Sigma_{c}^{+}(1^{4}P_{\lambda})-pole$   &$\times~(\text{s})$  &  &$\Sigma_{c}^{+}(2S)-pole$   &$\times~(\text{w})$   &$\Sigma_{c}^{+}(1^{4}P_{\lambda})-pole$   &$0$\\
&$\text{total}$   &$-2.15$   &$\text{total}$   &$-23.61$  &  &$\text{total}$   &$8.56$   &$\text{total}$   &$-4.09$\\
\specialrule{0em}{3pt}{3pt}
\multirow{1}*{\shortstack{$\Xi_{cc}^{+}\to\Lambda_{c}^{+}\eta^{\prime}$}}   &$C-diagram$   &$-1.27$   &$C-diagram$   &$8.52$
&\multirow{1}*{\shortstack{$\Xi_{cc}^{+}\to\Sigma_{c}^{+}\eta^{\prime}$}}   &$C-diagram$   &$-2.60$   &$C-diagram$   &$5.52$\\
&$C^{\prime}-diagram$   &$0.27$   &$C^{\prime}-diagram$   &$-1.06$
& &$C^{\prime}-diagram$   &$0$   &$C^{\prime}-diagram$   &$0$\\
&$\Xi_{cc}^{+}(1S)-pole$   &$-5.60$   &$\Xi_{cc}^{+}(1^{2}P_{\rho})-pole$   &$\times~(\text{s})$  &  &$\Xi_{cc}^{+}(1S)-pole$   &$\times~(\text{w})$   &$\Xi_{cc}^{+}(1^{2}P_{\rho})-pole$   &$\times~(\text{s})$\\
&$\Xi_{cc}^{+}(2S)-pole$   &$-0.13$   &$\Xi_{cc}^{+}(1^{2}P_{\lambda})-pole$   &$-0.23$  &  &$\Xi_{cc}^{+}(2S)-pole$   &$\times~(\text{w})$   &$\Xi_{cc}^{+}(1^{2}P_{\lambda})-pole$   &$0$\\
&$\Lambda_{c}^{+}(1S)-pole$   &$\times~(\text{s})$   &$\Xi_{cc}^{+}(1^{4}P_{\lambda})-pole$   &$-11.31$  &  &$\Lambda_{c}^{+}(1S)-pole$   &$\times~(\text{s})$   &$\Xi_{cc}^{+}(1^{4}P_{\lambda})-pole$   &$\times~(\text{w})$\\
&$\Lambda_{c}^{+}(2S)-pole$   &$\times~(\text{s})$   &$\Lambda_{c}^{+}(1^{2}P_{\lambda})-pole$   &$\times~(\text{s})$  &  &$\Lambda_{c}^{+}(2S)-pole$   &$\times~(\text{s})$   &$\Lambda_{c}^{+}(1^{2}P_{\lambda})-pole$   &$\times~(\text{s})$\\
&$\Sigma_{c}^{+}(1S)-pole$   &$\times~(\text{w,s})$   &$\Sigma_{c}^{+}(1^{2}P_{\lambda})-pole$   &$\times~(\text{s})$  &  &$\Sigma_{c}^{+}(1S)-pole$   &$\times~(\text{w})$   &$\Sigma_{c}^{+}(1^{2}P_{\lambda})-pole$   &$0$\\
&$\Sigma_{c}^{+}(2S)-pole$   &$\times~(\text{w,s})$   &$\Sigma_{c}^{+}(1^{4}P_{\lambda})-pole$   &$\times~(\text{s})$  &  &$\Sigma_{c}^{+}(2S)-pole$   &$\times~(\text{w})$   &$\Sigma_{c}^{+}(1^{4}P_{\lambda})-pole$   &$0$\\
&$\text{total}$   &$-6.73$   &$\text{total}$   &$-4.08$  &  &$\text{total}$   &$-2.60$   &$\text{total}$   &$5.52$\\
\specialrule{0em}{3pt}{3pt}
\multirow{1}*{\shortstack{$\Xi_{cc}^{+}\to\Xi_{c}^{+}K^{0}$}}   &$C^{\prime}-diagram$   &$-0.45$   &$C^{\prime}-diagram$   &$2.66$
&\multirow{1}*{\shortstack{$\Xi_{cc}^{+}\to\Xi_{c}^{\prime+}K^{0}$}} &$C^{\prime}-diagram$   &$0$   &$C^{\prime}-diagram$   &$0$\\
&$\Omega_{cc}^{+}(1S)-pole$   &$15.57$   &$\Omega_{cc}^{+}(1^{2}P_{\rho})-pole$   &$\times~(\text{s})$  &  &$\Omega_{cc}^{+}(1S)-pole$   &$\times~(\text{w})$   &$\Omega_{cc}^{+}(1^{2}P_{\rho})-pole$   &$\times~(\text{s})$\\
&$\Omega_{cc}^{+}(2S)-pole$   &$0.78$   &$\Omega_{cc}^{+}(1^{2}P_{\lambda})-pole$   &$0.58$  &  &$\Omega_{cc}^{+}(2S)-pole$   &$\times~(\text{w})$   &$\Omega_{cc}^{+}(1^{2}P_{\lambda})-pole$   &$0$\\
&$\Lambda_{c}^{+}(1S)-pole$   &$\times~(\text{s})$   &$\Omega_{cc}^{+}(1^{4}P_{\lambda})-pole$   &$-11.66$  &  &$\Lambda_{c}^{+}(1S)-pole$   &$13.53$   &$\Omega_{cc}^{+}(1^{4}P_{\lambda})-pole$   &$\times~(\text{w})$\\
&$\Lambda_{c}^{+}(2S)-pole$   &$\times~(\text{s})$   &$\Lambda_{c}^{+}(1^{2}P_{\lambda})-pole$   &$\times~(\text{s})$  &  &$\Lambda_{c}^{+}(2S)-pole$   &$5.45$   &$\Lambda_{c}^{+}(1^{2}P_{\lambda})-pole$   &$-0.16$\\
&$\Sigma_{c}^{+}(1S)-pole$   &$\times~(\text{w})$   &$\Sigma_{c}^{+}(1^{2}P_{\lambda})-pole$   &$0$  &  &$\Sigma_{c}^{+}(1S)-pole$   &$\times~(\text{w})$   &$\Sigma_{c}^{+}(1^{2}P_{\lambda})-pole$   &$0$\\
&$\Sigma_{c}^{+}(2S)-pole$   &$\times~(\text{w})$   &$\Sigma_{c}^{+}(1^{4}P_{\lambda})-pole$   &$0$  &  &$\Sigma_{c}^{+}(2S)-pole$   &$\times~(\text{w})$   &$\Sigma_{c}^{+}(1^{4}P_{\lambda})-pole$   &$0$\\
&$\text{total}$   &$15.90$   &$\text{total}$   &$-8.42$  &  &$\text{total}$   &$18.98$   &$\text{total}$   &$-0.16$\\
\bottomrule[0.5pt]
\bottomrule[1pt]
\end{tabular*}
\end{table*}

\begin{table*}[htbp]\centering
\caption{Same as Table~\ref{tab:CFamplitudes} but for DCS decays.}
\label{tab:DCSamplitudes}
\renewcommand\arraystretch{1.05}
\begin{tabular*}{120mm}{l@{\extracolsep{\fill}}llll}
\toprule[1pt]
\toprule[0.5pt]
\specialrule{0em}{1pt}{1pt}
&\multicolumn{2}{c}{{\normalsize $\mathcal{M}_{\text{PC}}^{-1/2,-1/2}$}}  &\multicolumn{2}{c}{{\normalsize $\mathcal{M}_{\text{PV}}^{-1/2,-1/2}$}}\\
\specialrule{0em}{1pt}{1pt}
\midrule[0.5pt]
\multirow{1}*{\shortstack{$\Xi_{cc}^{+}\to\Sigma_{c}^{0}K^{+}$}}   &$T-diagram$   &$-4.78$  &$T-diagram$   &$2.20$\\
\specialrule{0em}{3pt}{3pt}
\multirow{1}*{\shortstack{$\Xi_{cc}^{+}\to\Lambda_{c}^{+}K^{0}$}}   &$C-diagram$   &$0.94$   &$C-diagram$   &$-1.79$  \\
                                                                    &$C^{\prime}-diagram$   &$0.10$   &$C^{\prime}-diagram$   &$-0.55$  \\
                                                                    &$\Omega_{cc}^{+}(1S)-pole$   &$-3.11$   &$\Omega_{cc}^{+}(1^{2}P_{\rho})-pole$   &$\times~(\text{s})$ \\
                                                                    &$\Omega_{cc}^{+}(2S)-pole$   &$-0.18$   &$\Omega_{cc}^{+}(1^{2}P_{\lambda})-pole$   &$-0.13$ \\
                                                                    &   &                               &$\Omega_{cc}^{+}(1^{4}P_{\lambda})-pole$   &$2.61$\\
                                                                    &$\text{total}$   &$-2.25$   &$\text{total}$   &$0.14$  \\
\specialrule{0em}{3pt}{3pt}
\multirow{1}*{\shortstack{$\Xi_{cc}^{+}\to\Sigma_{c}^{+}K^{0}$}}   &$C-diagram$   &$2.35$   &$C-diagram$   &$-1.08$  \\
                                                                   &$C^{\prime}-diagram$   &$0$   &$C^{\prime}-diagram$   &$0$  \\
                                                                   &$\Omega_{cc}^{+}(1S)-pole$   &$\times~(\text{w})$   &$\Omega_{cc}^{+}(1^{2}P_{\rho})-pole$   &$\times~(\text{s})$ \\
                                                                   &$\Omega_{cc}^{+}(2S)-pole$   &$\times~(\text{w})$   &$\Omega_{cc}^{+}(1^{2}P_{\lambda})-pole$   &$0$ \\
                                                                   &   &                               &$\Omega_{cc}^{+}(1^{4}P_{\lambda})-pole$   &$\times~(\text{w})$\\
                                                                   &$\text{total}$   &$2.35$   &$\text{total}$   &$-1.08$  \\
\bottomrule[0.5pt]
\bottomrule[1pt]
\end{tabular*}
\end{table*}

\begin{table*}[htbp]\centering
\caption{The comparison of different theoretical predictions of branching fractions and asymmetry parameters $\alpha$ of SCS and DCS decays of $\Xi_{cc}^{+}$ baryon.}
\label{tab:SCSDCS}
\renewcommand\arraystretch{1.15}
\begin{tabular*}{162mm}{l@{\extracolsep{\fill}}cccc}
\toprule[1pt]
\toprule[0.5pt]
Modes   &\multicolumn{2}{c}{This work}   &\multicolumn{2}{c}{Hu, FSI~\cite{Hu:2024uia}}\\
        &$\mathcal{B}$   &$\alpha$       &$\mathcal{B}$   &$\alpha$\\
\midrule[0.5pt]
SCS decays  &&&&\\
\quad $\Xi_{cc}^{+}\to\Sigma_{c}^{0}\pi^{+}$    &$7.54\times10^{-3}$  &$-0.20$  &$(0.28^{+0.14}_{-0.11})\times10^{-3}$ &$-0.89^{+0.12}_{-0.04}$\\
\quad $\Xi_{cc}^{+}\to\Sigma_{c}^{++}\pi^{-}$    &$4.06\times10^{-3}$  &$-0.02$  &$(0.18^{+0.21}_{-0.11})\times10^{-4}$ &$0.73^{+0.10}_{-0.24}$\\
\quad $\Xi_{cc}^{+}\to\Xi_{c}^{0}K^{+}$    &$1.46\times10^{-3}$  &$-0.70$  &$(0.85^{+0.01}_{-0.01})\times10^{-3}$ &$-0.55^{+0.03}_{-0.03}$\\
\quad $\Xi_{cc}^{+}\to\Xi_{c}^{\prime0}K^{+}$    &$0.38\times10^{-3}$  &$0.30$  &$(0.41^{+0.01}_{-0.01})\times10^{-3}$ &$-0.69^{+0.26}_{-0.18}$\\
\quad $\Xi_{cc}^{+}\to\Lambda_{c}^{+}\pi^{0}$    &$2.12\times10^{-3}$  &$-0.60$  &$(0.93^{+1.07}_{-0.61})\times10^{-4}$ &$-0.62^{+0.06}_{-0.05}$\\
\quad $\Xi_{cc}^{+}\to\Sigma_{c}^{+}\pi^{0}$    &$5.14\times10^{-3}$  &$-0.10$  &$(0.28^{+0.31}_{-0.17})\times10^{-4}$ &$-0.56^{+0.14}_{-0.08}$\\
\quad $\Xi_{cc}^{+}\to\Lambda_{c}^{+}\eta$    &$3.31\times10^{-3}$  &$0.18$  &$-$ &$-$\\
\quad $\Xi_{cc}^{+}\to\Sigma_{c}^{+}\eta$    &$0.43\times10^{-3}$  &$-0.78$  &$-$ &$-$\\
\quad $\Xi_{cc}^{+}\to\Lambda_{c}^{+}\eta^{\prime}$    &$0.29\times10^{-3}$  &$0.89$  &$-$ &$-$\\
\quad $\Xi_{cc}^{+}\to\Sigma_{c}^{+}\eta^{\prime}$    &$0.12\times10^{-3}$  &$-0.77$  &$-$ &$-$\\
\quad $\Xi_{cc}^{+}\to\Xi_{c}^{+}K^{0}$    &$1.56\times10^{-3}$  &$-0.83$  &$(0.44^{+0.47}_{-0.27})\times10^{-4}$ &$-0.19^{+0.14}_{-0.16}$\\
\quad $\Xi_{cc}^{+}\to\Xi_{c}^{\prime+}K^{0}$    &$1.49\times10^{-3}$  &$-0.02$  &$(0.32^{+0.35}_{-0.21})\times10^{-4}$ &$0.96^{+0.01}_{-0.02}$\\
\midrule[0.5pt]
DCS decays  &&&&\\
\quad $\Xi_{cc}^{+}\to\Sigma_{c}^{0}K^{+}$    &$1.37\times10^{-4}$  &$-0.76$  &$(1.00^{+0.04}_{-0.04})\times10^{-3}$ &$-0.98$\\
\quad $\Xi_{cc}^{+}\to\Lambda_{c}^{+}K^{0}$    &$0.30\times10^{-4}$  &$-0.12$  &$(0.50^{+0.61}_{-0.34})\times10^{-5}$ &$-0.55^{+0.05}_{-0.06}$\\
\quad $\Xi_{cc}^{+}\to\Sigma_{c}^{+}K^{0}$    &$0.33\times10^{-4}$  &$-0.76$  &$(0.09^{+0.09}_{-0.05})\times10^{-5}$ &$0.05^{+0.19}_{-0.15}$\\
\bottomrule[0.5pt]
\bottomrule[1pt]
\end{tabular*}
\end{table*}

Among the SCS and DCS decay modes, the $\Xi_{cc}^{+}\to\Xi_{c}^{0}K^{+}$ and $\Xi_{cc}^{+}\to\Sigma_{c}^{0}K^{+}$ decay modes receive contributions from only factorizable $T$ diagram, whereas the $\Xi_{cc}^{+}\to\Sigma_{c}^{+}\eta^{(\prime)}$ and $\Xi_{cc}^{+}\to\Sigma_{c}^{+}K^{0}$ receive contributions from only factorizable $C$ diagram. In contrast, the decay modes $\Xi_{cc}^{+}\to\Sigma_{c}^{++}\pi^{-}$ and $\Xi_{cc}^{+}\to\Xi_{c}^{\prime+}K^{0}$ are mediated by the nonfactorizable $\Lambda_{c}^{+}$ pole contributions.

Besides, for the $\Xi_{cc}^{+}\to\Sigma_{c}^{0}\pi^{+}$ and $\Xi_{cc}^{+}\to\Sigma_{c}^{+}\pi^{0}$ decay modes, the factorizable and nonfactorizable amplitudes interfere constructively, leading to relatively large branching fractions of $7.54\times10^{-3}$ and $5.14\times10^{-3}$, respectively. By contrast, a strong cancellation between the factorizable and nonfactorizable PC amplitudes occurs in the $\Xi_{cc}^{+}\to\Xi_{c}^{\prime0}K^{+}$ decay mode, contributing to a small branching fraction as $\mathcal{B}(\Xi_{cc}^{+}\to\Xi_{c}^{\prime0}K^{+})=0.38\times10^{-3}$.

\section{Conclusion}
\label{sec5}

The nonleptonic decays of double-charmed baryons offer valuable insights into the internal structures of double-charmed baryons and underlying weak decay dynamics. Very recently, the LHCb Collaboration reported the observation of double-charmed baryon $\Xi_{cc}^{+}$. This provides important experimental motivation for a systematic theoretical investigation of its nonleptonic decays. In this paper, we systematically investigate the two-body nonleptonic decays $\Xi_{cc}^{+}\to\mathcal{B}_{c}+P$, where $\mathcal{B}_{c}$ and $P$ denote a single-charmed baryon and a light pseudoscalar meson, respectively. 

In our calculation, the relevant decay amplitudes are evaluated within the nonrelativistic quark model. In particular, the nonfactorizable contributions from $W$-exchange diagrams are calculated using the pole model, in which the low-lying intermediate baryon poles with $J^{P}=1/2^{+}$ and $1/2^{-}$ are included to evaluated the PC and PV decay amplitudes, respectively. Moreover, to reduce the arbitrariness associated with the choice of baryon spatial wave functions, we determine the charmed-baryon wave functions by solving the three-quark Schr\"{o}dinger equation with a nonrelativistic quark potential model using the Gaussian expansion method, rather than relying on phenomenological parametrizations. This provides a dynamically determination of the baryon wave functions for evaluating the decay amplitudes.

Taking the dynamically determined charmed-baryon wave functions as inputs, we numerically calculate both the factorizable and nonfactorizable contributions to the decay amplitudes. Our results reveal that the nonfactorizable amplitudes, particularly those arising from the $W$-exchange diagrams mediated by intermediate charmed-baryon poles, are comparable to or even dominate the factorizable decay amplitudes.

With the obtained decay amplitudes, we present systematic predictions for the branching fractions of the concerned CF, SCS, and DCS decay modes. Assuming a lifetime of $\tau_{\Xi_{cc}^{+}}=47~\text{fs}$, several CF decay modes are predicted to have large branching fractions, i.e., $\mathcal{B}(\Xi_{cc}^{+}\to\Xi_{c}^{\prime0}\pi^{+})=8.90\%$, $\mathcal{B}(\Xi_{cc}^{+}\to\Omega_{c}^{0}K^{+})=7.09\%$, $\mathcal{B}(\Xi_{cc}^{+}\to\Sigma_{c}^{+}\bar{K}^{0})=6.92\%$, $\mathcal{B}(\Xi_{cc}^{+}\to\Xi_{c}^{+}\pi^{0})=5.28\%$, and $\mathcal{B}(\Xi_{cc}^{+}\to\Sigma_{c}^{++}K^{-})=5.17\%$. These channels are promising candidates for future measurements by the LHCb Collaboration. Besides, we also calculate the corresponding decay asymmetry parameters $\alpha$. Future experimental measurements of the decay asymmetry parameters would provide valuable information on the relative magnitudes and phases of the PC and PV decay amplitudes, thereby helping to elucidate the underlying weak decay dynamics.

\section*{ACKNOWLEDGMENTS}

This work is supported by the National Natural Science Foundation of China under Grant No. 12447155 and 12447172, and the Fundamental Research Funds for the Central Universities under Grant No. 26CX06073A.

\appendix

\section{Spin-flavor matrix elements for $W$-emission processes}
\label{app01}

\setcounter{equation}{0}
\renewcommand{\theequation}{A.\arabic{equation}}

\setcounter{table}{0}
\renewcommand{\thetable}{A\Roman{table}}

In this appendix, the spin-flavor matrix elements for $W$-emission processes are presented. We first specify the spin-flavor wave functions for the relevant charmed baryons and pseudoscalar mesons. The $S$-wave double-charmed baryons form a flavor triplet state, for which the spin-flavor wave functions are taken as~\cite{Perez-Marcial:1989sch,Li:2025noz}
\begin{equation}
\begin{split}
\vert\Omega_{cc}^{+}\rangle=&-\frac{1}{\sqrt{3}}\big{[}ccs\chi_{S}+(13)+(23)\big{]},\\
\vert\Xi_{cc}^{++}\rangle=&-\frac{1}{\sqrt{3}}\big{[}ccu\chi_{S}+(13)+(23)\big{]},\\
\vert\Xi_{cc}^{+}\rangle=&-\frac{1}{\sqrt{3}}\big{[}ccd\chi_{S}+(13)+(23)\big{]}.
\end{split}
\end{equation}
For the $S$-wave single-charmed baryons, the $\Lambda_{c}^{+}$ and $\Xi_{c}^{+,0}$ form a flavor anti-triplet state, whereas the $\Sigma_{c}^{++,+,0}$, $\Xi_{c}^{\prime+,0}$, and $\Omega_{c}^{0}$ form a flavor sextet state. The spin-flavor wave functions of the anti-triplet baryons are taken as
\begin{equation}
\begin{split}
\vert\Lambda_{c}^{+}\rangle=&\frac{1}{\sqrt{6}}\big{[}(udc-duc)\chi_{A}+(13)+(23)\big{]},\\
\vert\Xi_{c}^{+}\rangle=&\frac{1}{\sqrt{6}}\big{[}(usc-suc)\chi_{A}+(13)+(23)\big{]},\\
\vert\Xi_{c}^{0}\rangle=&\frac{1}{\sqrt{6}}\big{[}(dsc-sdc)\chi_{A}+(13)+(23)\big{]},
\end{split}
\end{equation}
while the ones of the sextet baryons are taken as
\begin{equation}
\begin{split}
\vert\Sigma_{c}^{++}\rangle=&\frac{1}{\sqrt{3}}\big{[}uuc\chi_{S}+(13)+(23)\big{]},\\
\vert\Sigma_{c}^{+}\rangle=&\frac{1}{\sqrt{6}}\big{[}(udc+duc)\chi_{S}+(13)+(23)\big{]},\\
\vert\Sigma_{c}^{0}\rangle=&\frac{1}{\sqrt{3}}\big{[}ddc\chi_{S}+(13)+(23)\big{]},\\
\vert\Xi_{c}^{\prime+}\rangle=&\frac{1}{\sqrt{6}}\big{[}(usc+suc)\chi_{S}+(13)+(23)\big{]},\\
\vert\Xi_{c}^{\prime0}\rangle=&\frac{1}{\sqrt{6}}\big{[}(dsc+sdc)\chi_{S}+(13)+(23)\big{]},\\
\vert\Omega_{c}^{0}\rangle=&\frac{1}{\sqrt{3}}\big{[}ssc\chi_{S}+(13)+(23)\big{]}.
\end{split}
\end{equation}

Here, we have
\begin{equation}
\begin{split}
abc\chi_{S}^{\downarrow}=&\frac{1}{\sqrt{6}}(-2a^{\downarrow}b^{\downarrow}c^{\uparrow}+a^{\uparrow}b^{\downarrow}c^{\downarrow}+a^{\downarrow}b^{\uparrow}c^{\downarrow}),\\
abc\chi_{A}^{\downarrow}=&\frac{1}{\sqrt{2}}(a^{\uparrow}b^{\downarrow}c^{\downarrow}-a^{\downarrow}b^{\uparrow}c^{\downarrow}),
\end{split}
\end{equation}
for the baryons of $\vert{S,S_{z}}\rangle=\vert{1/2,-1/2}\rangle$, and
\begin{equation}
\begin{split}
abc\chi_{S}^{\uparrow}=&\frac{1}{\sqrt{6}}(2a^{\uparrow}b^{\uparrow}c^{\downarrow}-a^{\uparrow}b^{\downarrow}c^{\uparrow}-a^{\downarrow}b^{\uparrow}c^{\uparrow}),\\
abc\chi_{A}^{\uparrow}=&\frac{1}{\sqrt{2}}(a^{\uparrow}b^{\downarrow}c^{\uparrow}-a^{\downarrow}b^{\uparrow}c^{\uparrow}),
\end{split}
\end{equation}
for that of $\vert{S,S_{z}}\rangle=\vert{1/2,+1/2}\rangle$. Besides, the spin wave function of pseudoscalar meson is adopted as $\chi_{P}=\frac{1}{\sqrt{2}}(\uparrow\downarrow-\downarrow\uparrow)$. For calculating the spin matrix element, the particle-hole conjugation $\langle{j,-m}\vert\to{(-1)^{j+m}}\vert{j,m}\rangle$ is useful. This relation makes the antiquark spin transform as $\langle\bar{\uparrow}\vert{\to}\vert{\downarrow}\rangle$ and $\langle\bar{\downarrow}\vert\to -\vert{\uparrow}\rangle$.

Using the above spin-flavor conventions, the spin-flavor matrix elements entering $W$-emission processes can be calculated. The corresponding values of $\langle{\mathcal{B}_{f}}{\downarrow}\vert\hat{\mathcal{O}}\vert{\mathcal{B}_{i}}{\downarrow}\rangle$ are summarized in Table~\ref{tab:TCfactor} for $T$ and $C$ diagrams, while the ones for $C^{\prime}$ diagram are summarized in Table~\ref{tab:Cprimfactor}. We emphasize that particularly for the $T$ diagram, the spin matrix elements associated with $\langle{s_{z,5}\bar{s}_{z,4}}\vert{\sigma_{x,y,z}}\vert{0}\rangle$ are all vanishing~\cite{Li:2025mxz}. Besides, that associated with $\langle{s_{z,1}^{\prime}\bar{s}_{z,4}}\vert{\sigma_{x,y,z}}\vert{0}\rangle$ for the $C$ diagram, are also vanishing. This is caused by the spin wave function of pseudoscalar meson.

\begin{table*}[htbp]\centering
\caption{The spin-flavor matrix elements of PC (up panel) and PV (bottom panel) amplitudes for $T$ and $C$ diagrams. For simplicity, the spin and flavor wave functions of pseudoscalar meson is omitted here.}
\label{tab:TCfactor}
\renewcommand\arraystretch{1.05}
\begin{tabular*}{150mm}{c@{\extracolsep{\fill}}{|}ccc{|}cc}
\toprule[1pt]
\toprule[0.5pt]
\specialrule{0em}{1pt}{1pt}
{}   &\multicolumn{3}{c}{$T-diagrams$}   &\multicolumn{2}{c}{$C-diagrams$}\\
\midrule[0.5pt]
\specialrule{0em}{1pt}{1pt}
matrix elements
&$\langle{\Xi_{c}^{0}}{\downarrow}\vert\hat{\mathcal{O}}\vert{\Xi_{cc}^{+}}{\downarrow}\rangle$  &$\langle{\Xi_{c}^{\prime0}}{\downarrow}\vert\hat{\mathcal{O}}\vert{\Xi_{cc}^{+}}{\downarrow}\rangle$   &$\langle{\Sigma_{c}^{0}}{\downarrow}\vert\hat{\mathcal{O}}\vert{\Xi_{cc}^{+}}{\downarrow}\rangle$   
&$\langle{\Lambda_{c}^{+}}{\downarrow}\vert\hat{\mathcal{O}}\vert{\Xi_{cc}^{+}}{\downarrow}\rangle$   &$\langle{\Sigma_{c}^{+}}{\downarrow}\vert\hat{\mathcal{O}}\vert{\Xi_{cc}^{+}}{\downarrow}\rangle$\\
\specialrule{0em}{1pt}{1pt}
\midrule[0.5pt]
$\langle{s_{z,1}^{\prime}}\vert{I}\vert{s_{z,1}}\rangle\langle{s_{z,5}\bar{s}_{z,4}}\vert{\sigma_{z}}\vert{0}\rangle$  &$0$  &$0$  &$0$  &$-\frac{1}{6\sqrt{3}}$  &$\frac{5}{18}$\\
$\langle{s_{z,1}^{\prime}}\vert{\sigma_{z}}\vert{s_{z,1}}\rangle\langle{s_{z,5}\bar{s}_{z,4}}\vert{I}\vert{0}\rangle$  &$\frac{1}{3\sqrt{3}}$  &$\frac{5}{9}$  &$\frac{5\sqrt{2}}{9}$  &$-\frac{1}{6\sqrt{3}}$  &$\frac{5}{18}$\\
$(\langle{s_{z,1}^{\prime}}\vert{\pmb{\sigma}}\vert{s_{z,1}}\rangle\times\langle{s_{z,5}\bar{s}_{z,4}}\vert{\pmb{\sigma}}\vert{0}\rangle)_{z}$  &$0$  &$0$  &$0$  &$\frac{i}{3\sqrt{3}}$  &$-\frac{5i}{9}$\\
\specialrule{0em}{1pt}{1pt}
$\langle{s_{z,5}}\vert{I}\vert{s_{z,1}}\rangle\langle{s_{z,1}^{\prime}\bar{s}_{z,4}}\vert{\sigma_{z}}\vert{0}\rangle$  &$\frac{1}{6\sqrt{3}}$  &$\frac{5}{18}$  &$\frac{5}{9\sqrt{2}}$  &$0$  &$0$\\
$\langle{s_{z,5}}\vert{\sigma_{z}}\vert{s_{z,1}}\rangle\langle{s_{z,1}^{\prime}\bar{s}_{z,4}}\vert{I}\vert{0}\rangle$  &$\frac{1}{6\sqrt{3}}$  &$\frac{5}{18}$  &$\frac{5}{9\sqrt{2}}$  &$-\frac{1}{3\sqrt{3}}$  &$\frac{5}{9}$\\
$(\langle{s_{z,5}}\vert{\pmb{\sigma}}\vert{s_{z,1}}\rangle\times\langle{s_{z,1}^{\prime}\bar{s}_{z,4}}\vert{\pmb{\sigma}}\vert{0}\rangle)_{z}$  &$-\frac{i}{3\sqrt{3}}$  &$-\frac{5i}{9}$  &$-\frac{5\sqrt{2}i}{9}$  &$0$  &$0$\\
\specialrule{0em}{1pt}{1pt}
\midrule[0.5pt]
\specialrule{0em}{1pt}{1pt}
$\langle{s_{z,1}^{\prime}}\vert{I}\vert{s_{z,1}}\rangle\langle{s_{z,5}\bar{s}_{z,4}}\vert{I}\vert{0}\rangle$  &$-\frac{1}{\sqrt{3}}$  &$-\frac{1}{3}$  &$-\frac{\sqrt{2}}{3}$  &$\frac{1}{2\sqrt{3}}$  &$-\frac{1}{6}$\\
$\langle{s_{z,1}^{\prime}}\vert{\sigma_{x}}\vert{s_{z,1}}\rangle\langle{s_{z,5}\bar{s}_{z,4}}\vert{\sigma_{x}}\vert{0}\rangle$  &$0$  &$0$  &$0$  &$\frac{1}{2\sqrt{3}}$  &$-\frac{1}{6}$\\
$\langle{s_{z,1}^{\prime}}\vert{\sigma_{y}}\vert{s_{z,1}}\rangle\langle{s_{z,5}\bar{s}_{z,4}}\vert{\sigma_{y}}\vert{0}\rangle$  &$0$  &$0$  &$0$  &$\frac{1}{2\sqrt{3}}$  &$-\frac{1}{6}$\\
$\langle{s_{z,1}^{\prime}}\vert{\sigma_{z}}\vert{s_{z,1}}\rangle\langle{s_{z,5}\bar{s}_{z,4}}\vert{\sigma_{z}}\vert{0}\rangle$  &$0$  &$0$  &$0$  &$\frac{1}{2\sqrt{3}}$  &$-\frac{1}{6}$\\
\specialrule{0em}{1pt}{1pt}
$\langle{s_{z,5}}\vert{I}\vert{s_{z,1}}\rangle\langle{s_{z,1}^{\prime}\bar{s}_{z,4}}\vert{I}\vert{0}\rangle$  &$-\frac{1}{2\sqrt{3}}$  &$-\frac{1}{6}$  &$-\frac{1}{3\sqrt{2}}$  &$\frac{1}{\sqrt{3}}$  &$-\frac{1}{3}$\\
$\langle{s_{z,5}}\vert{\sigma_{x}}\vert{s_{z,1}}\rangle\langle{s_{z,1}^{\prime}\bar{s}_{z,4}}\vert{\sigma_{x}}\vert{0}\rangle$  &$-\frac{1}{2\sqrt{3}}$  &$-\frac{1}{6}$  &$-\frac{1}{3\sqrt{2}}$  &$0$  &$0$\\
$\langle{s_{z,5}}\vert{\sigma_{y}}\vert{s_{z,1}}\rangle\langle{s_{z,1}^{\prime}\bar{s}_{z,4}}\vert{\sigma_{y}}\vert{0}\rangle$  &$-\frac{1}{2\sqrt{3}}$  &$-\frac{1}{6}$  &$-\frac{1}{3\sqrt{2}}$  &$0$  &$0$\\
$\langle{s_{z,5}}\vert{\sigma_{z}}\vert{s_{z,1}}\rangle\langle{s_{z,1}^{\prime}\bar{s}_{z,4}}\vert{\sigma_{z}}\vert{0}\rangle$  &$-\frac{1}{2\sqrt{3}}$  &$-\frac{1}{6}$  &$-\frac{1}{3\sqrt{2}}$  &$0$  &$0$\\
\bottomrule[0.5pt]
\bottomrule[1pt]
\end{tabular*}
\end{table*}

\begin{table*}[htbp]\centering
\caption{Same as Table~\ref{tab:TCfactor} but for $C^{\prime}$ diagrams.}
\label{tab:Cprimfactor}
\renewcommand\arraystretch{1.05}
\begin{tabular*}{140mm}{c@{\extracolsep{\fill}}|cccc}
\toprule[1pt]
\toprule[0.5pt]
\specialrule{0em}{1pt}{1pt}
matrix elements &$\langle{\Xi_{c}^{+}}{\downarrow}\vert\hat{\mathcal{O}}\vert{\Xi_{cc}^{+}}{\downarrow}\rangle$  &$\langle{\Xi_{c}^{\prime+}}{\downarrow}\vert\hat{\mathcal{O}}\vert{\Xi_{cc}^{+}}{\downarrow}\rangle$  &$\langle{\Lambda_{c}^{+}}{\downarrow}\vert\hat{\mathcal{O}}\vert{\Xi_{cc}^{+}}{\downarrow}\rangle$  &$\langle{\Sigma_{c}^{+}}{\downarrow}\vert\hat{\mathcal{O}}\vert{\Xi_{cc}^{+}}{\downarrow}\rangle$\\
\specialrule{0em}{1pt}{1pt}
\midrule[0.5pt]
$\langle{s_{z,1}^{\prime}}\vert{I}\vert{s_{z,1}}\rangle\langle{s_{z,5}\bar{s}_{z,4}}\vert{\sigma_{z}}\vert{0}\rangle$  &$-\frac{1}{12\sqrt{3}}$  &$-\frac{1}{36}$  &$-\frac{1}{12\sqrt{3}}$  &$-\frac{1}{36}$\\
$\langle{s_{z,1}^{\prime}}\vert{\sigma_{z}}\vert{s_{z,1}}\rangle\langle{s_{z,5}\bar{s}_{z,4}}\vert{I}\vert{0}\rangle$  &$\frac{1}{12\sqrt{3}}$  &$\frac{5}{36}$  &$\frac{1}{12\sqrt{3}}$  &$\frac{5}{36}$\\
$(\langle{s_{z,1}^{\prime}}\vert{\pmb{\sigma}}\vert{s_{z,1}}\rangle\times\langle{s_{z,5}\bar{s}_{z,4}}\vert{\pmb{\sigma}}\vert{0}\rangle)_{z}$  &$\frac{i}{6\sqrt{3}}$  &$-\frac{i}{6}$  &$\frac{i}{6\sqrt{3}}$  &$-\frac{i}{6}$\\
\specialrule{0em}{1pt}{1pt}
$\langle{s_{z,5}}\vert{I}\vert{s_{z,1}}\rangle\langle{s_{z,1}^{\prime}\bar{s}_{z,4}}\vert{\sigma_{z}}\vert{0}\rangle$  &$\frac{1}{12\sqrt{3}}$  &$-\frac{1}{36}$  &$\frac{1}{12\sqrt{3}}$  &$-\frac{1}{36}$\\
$\langle{s_{z,5}}\vert{\sigma_{z}}\vert{s_{z,1}}\rangle\langle{s_{z,1}^{\prime}\bar{s}_{z,4}}\vert{I}\vert{0}\rangle$  &$-\frac{1}{12\sqrt{3}}$  &$\frac{5}{36}$  &$-\frac{1}{12\sqrt{3}}$  &$\frac{5}{36}$\\
$(\langle{s_{z,5}}\vert{\pmb{\sigma}}\vert{s_{z,1}}\rangle\times\langle{s_{z,1}^{\prime}\bar{s}_{z,4}}\vert{\pmb{\sigma}}\vert{0}\rangle)_{z}$  &$-\frac{i}{6\sqrt{3}}$  &$-\frac{i}{6}$  &$-\frac{i}{6\sqrt{3}}$  &$-\frac{i}{6}$\\
\specialrule{0em}{1pt}{1pt}
\midrule[0.5pt]
\specialrule{0em}{1pt}{1pt}
$\langle{s_{z,1}^{\prime}}\vert{I}\vert{s_{z,1}}\rangle\langle{s_{z,5}\bar{s}_{z,4}}\vert{I}\vert{0}\rangle$  &$-\frac{1}{4\sqrt{3}}$  &$-\frac{1}{12}$  &$-\frac{1}{4\sqrt{3}}$  &$-\frac{1}{12}$\\
$\langle{s_{z,1}^{\prime}}\vert{\sigma_{x}}\vert{s_{z,1}}\rangle\langle{s_{z,5}\bar{s}_{z,4}}\vert{\sigma_{x}}\vert{0}\rangle$  &$\frac{1}{4\sqrt{3}}$  &$-\frac{1}{36}$  &$\frac{1}{4\sqrt{3}}$  &$-\frac{1}{36}$\\
$\langle{s_{z,1}^{\prime}}\vert{\sigma_{y}}\vert{s_{z,1}}\rangle\langle{s_{z,5}\bar{s}_{z,4}}\vert{\sigma_{y}}\vert{0}\rangle$  &$\frac{1}{4\sqrt{3}}$  &$-\frac{1}{36}$  &$\frac{1}{4\sqrt{3}}$  &$-\frac{1}{36}$\\
$\langle{s_{z,1}^{\prime}}\vert{\sigma_{z}}\vert{s_{z,1}}\rangle\langle{s_{z,5}\bar{s}_{z,4}}\vert{\sigma_{z}}\vert{0}\rangle$  &$\frac{1}{4\sqrt{3}}$  &$-\frac{1}{36}$  &$\frac{1}{4\sqrt{3}}$  &$-\frac{1}{36}$\\
\specialrule{0em}{1pt}{1pt}
$\langle{s_{z,5}}\vert{I}\vert{s_{z,1}}\rangle\langle{s_{z,1}^{\prime}\bar{s}_{z,4}}\vert{I}\vert{0}\rangle$  &$\frac{1}{4\sqrt{3}}$  &$-\frac{1}{12}$  &$\frac{1}{4\sqrt{3}}$  &$-\frac{1}{12}$\\
$\langle{s_{z,5}}\vert{\sigma_{x}}\vert{s_{z,1}}\rangle\langle{s_{z,1}^{\prime}\bar{s}_{z,4}}\vert{\sigma_{x}}\vert{0}\rangle$  &$-\frac{1}{4\sqrt{3}}$  &$-\frac{1}{36}$  &$-\frac{1}{4\sqrt{3}}$  &$-\frac{1}{36}$\\
$\langle{s_{z,5}}\vert{\sigma_{y}}\vert{s_{z,1}}\rangle\langle{s_{z,1}^{\prime}\bar{s}_{z,4}}\vert{\sigma_{y}}\vert{0}\rangle$  &$-\frac{1}{4\sqrt{3}}$  &$-\frac{1}{36}$  &$-\frac{1}{4\sqrt{3}}$  &$-\frac{1}{36}$\\
$\langle{s_{z,5}}\vert{\sigma_{z}}\vert{s_{z,1}}\rangle\langle{s_{z,1}^{\prime}\bar{s}_{z,4}}\vert{\sigma_{z}}\vert{0}\rangle$  &$-\frac{1}{4\sqrt{3}}$  &$-\frac{1}{36}$  &$-\frac{1}{4\sqrt{3}}$  &$-\frac{1}{36}$\\
\bottomrule[0.5pt]
\bottomrule[1pt]
\end{tabular*}
\end{table*}

Similarly, the spin-flavor matrix elements $\langle{\mathcal{B}_{f}}{\uparrow}\vert\hat{\mathcal{O}}\vert{\mathcal{B}_{i}}{\uparrow}\rangle$ are related to the spin-down ones as
\begin{equation}
\begin{split}
\langle{\mathcal{B}_{f}}{\uparrow}\vert\hat{\mathcal{O}}\vert{\mathcal{B}_{i}}{\uparrow}\rangle
=\left\{
\begin{array}{ll}
-\langle{\mathcal{B}_{f}}{\downarrow}\vert\hat{\mathcal{O}}\vert{\mathcal{B}_{i}}{\downarrow}\rangle, &\text{for PC amplitudes},\\
\langle{\mathcal{B}_{f}}{\downarrow}\vert\hat{\mathcal{O}}\vert{\mathcal{B}_{i}}{\downarrow}\rangle, &\text{for PV amplitudes}.\\
\end{array}
\right.
\label{eq:relation11}
\end{split}
\end{equation}
In addition, the spin-flip matrix elements vanish,
\begin{equation}
\langle{\mathcal{B}_{f}}{\uparrow}\vert\hat{\mathcal{O}}\vert{\mathcal{B}_{i}}{\downarrow}\rangle=\langle{\mathcal{B}_{f}}{\downarrow}\vert\hat{\mathcal{O}}\vert{\mathcal{B}_{i}}{\uparrow}\rangle=0.
\end{equation}

\section{Spin-flavor matrix elements for $W$-exchange processes}
\label{app02}

\setcounter{equation}{0}
\renewcommand{\theequation}{B.\arabic{equation}}

\setcounter{table}{0}
\renewcommand{\thetable}{B\Roman{table}}

In this appendix, the spin-flavor matrix elements for $W$-exchange processes are presented. 

For the evaluation of these spin-flavor matrix elements, we employed the following spin-flavor wave functions for excited baryons. For $\lambda$-mode excited charmed baryons with total spin $\vert{S,S_{z}}\rangle=\vert{1/2,\pm1/2}\rangle$, the spin-flavor wave functions $\vert{\mathcal{B}}\rangle_{{}^{2}{\lambda}}$ are taken to be the same as those of the $S$-wave states. In addition, for $\lambda$-mode excited double-charmed baryons with total spin $\vert{S,S_{z}}\rangle=\vert{3/2,\pm1/2}\rangle$, we adopted as~\cite{Li:2025noz}
\begin{equation}
\begin{split}
\vert\Omega_{cc}^{+}\rangle_{{}^{4}\lambda}=&-\frac{1}{\sqrt{3}}\big{[}ccs\chi_{s}+(13)+(23)\big{]},\\
\vert\Xi_{cc}^{++}\rangle_{{}^{4}\lambda}=&-\frac{1}{\sqrt{3}}\big{[}ccu\chi_{s}+(13)+(23)\big{]},\\
\vert\Xi_{cc}^{+}\rangle_{{}^{4}\lambda}=&-\frac{1}{\sqrt{3}}\big{[}ccd\chi_{s}+(13)+(23)\big{]},
\end{split}
\end{equation}
while for $\lambda$-mode excited single-charmed baryons we adopted as~\cite{Li:2025noz}
\begin{equation}
\begin{split}
\vert\Omega_{c}^{0}\rangle_{{}^{4}\lambda}=&\frac{1}{\sqrt{3}}\big{[}ssc\chi_{s}+(13)+(23)\big{]},\\
\vert\Sigma_{c}^{++}\rangle_{{}^{4}\lambda}=&\frac{1}{\sqrt{3}}\big{[}uuc\chi_{s}+(13)+(23)\big{]}\\
\vert\Sigma_{c}^{+}\rangle_{{}^{4}\lambda}=&\frac{1}{\sqrt{6}}\big{[}(udc+duc)\chi_{s}+(13)+(23)\big{]},\\
\vert\Sigma_{c}^{0}\rangle_{{}^{4}\lambda}=&\frac{1}{\sqrt{3}}\big{[}ddc\chi_{s}+(13)+(23)\big{]}\\
\vert\Xi_{c}^{\prime+}\rangle_{{}^{4}\lambda}=&\frac{1}{\sqrt{6}}\big{[}(usc+suc)\chi_{s}+(13)+(23)\big{]},\\
\vert\Xi_{c}^{\prime0}\rangle_{{}^{4}\lambda}=&\frac{1}{\sqrt{6}}\big{[}(dsc+sdc)\chi_{s}+(13)+(23)\big{]},
\end{split}
\end{equation}
where
\begin{equation}
\begin{split}
abc\chi_{s}^{\downarrow}=&\frac{1}{\sqrt{3}}(a^{\downarrow}b^{\downarrow}c^{\uparrow}+a^{\uparrow}b^{\downarrow}c^{\downarrow}+a^{\downarrow}b^{\uparrow}c^{\downarrow}),\\
abc\chi_{s}^{\uparrow}=&\frac{1}{\sqrt{3}}(a^{\downarrow}b^{\uparrow}c^{\uparrow}+a^{\uparrow}b^{\uparrow}c^{\downarrow}+a^{\uparrow}b^{\downarrow}c^{\uparrow}).
\end{split}
\end{equation}
Besides, the spin-flavor wave functions of $\rho$-mode excited double-charmed baryons with total spin $\vert{S,S_{z}}\rangle=\vert{1/2,\pm1/2}\rangle$ can be written as~\cite{Li:2025noz}
\begin{equation}
\begin{split}
\vert\Omega_{cc}^{+}\rangle_{{}^{2}\rho}=&-\frac{1}{\sqrt{3}}\big{[}ccs\chi_{A}+(13)+(23)\big{]},\\
\vert\Xi_{cc}^{++}\rangle_{{}^{2}\rho}=&-\frac{1}{\sqrt{3}}\big{[}ccu\chi_{A}+(13)+(23)\big{]},\\
\vert\Xi_{cc}^{+}\rangle_{{}^{2}\rho}=&-\frac{1}{\sqrt{3}}\big{[}ccd\chi_{A}+(13)+(23)\big{]}.
\end{split}
\end{equation}

Using the spin-flavor conventions specified above, the weak-transition spin-flavor matrix elements entering the $W$-exchange processes can be calculated. Our results for the spin-flavor matrix elements of PC and PV amplitudes are summarized in Tables~\ref{tab:polefactorPC} and \ref{tab:polefactorPV}, respectively. For compactness, the following shorthand notations
\begin{equation}
\begin{split}
\pmb{\sigma}_{1}\cdot\pmb{\sigma}_{2}\equiv&\langle{s_{z,1}^{\prime}}\vert{\pmb{\sigma}_{1}}\vert{s_{z,1}}\rangle\cdot\langle{s_{z,2}^{\prime}}\vert{\pmb{\sigma}_{2}}\vert{s_{z,2}}\rangle,\\
\tilde{\pmb{\sigma}}_{1}\cdot\tilde{\pmb{\sigma}}_{2}\equiv&\langle{s_{z,2}^{\prime}}\vert{\pmb{\sigma}_{1}}\vert{s_{z,1}}\rangle\cdot\langle{s_{z,1}^{\prime}}\vert{\pmb{\sigma}_{2}}\vert{s_{z,2}}\rangle,
\end{split}
\end{equation}
are adopted. Analogous shorthand notation is used for the $\pmb{\sigma}_{1}-\pmb{\sigma}_{2}$ and $\pmb{\sigma}_{1}\times\pmb{\sigma}_{2}$.

\begin{table*}[htbp]\centering
\caption{The spin-flavor matrix elements for PC amplitudes of $W$-exchange diagrams, in which $\mathcal{B}_{cc}^{+}=(\Xi_{cc}^{+},\Omega_{cc}^{+})$.}
\label{tab:polefactorPC}
\renewcommand\arraystretch{1.05}
\begin{tabular*}{120mm}{c@{\extracolsep{\fill}}cccc}
\toprule[1pt]
\toprule[0.5pt]
\specialrule{0em}{1pt}{1pt}
matrix elements &$\langle{\Xi_{c}^{+}}{\downarrow}\vert\hat{\mathcal{O}}\vert{\mathcal{B}_{cc}^{+}}{\downarrow}\rangle$  &$\langle{\Xi_{c}^{\prime+}}{\downarrow}\vert\hat{\mathcal{O}}\vert{\mathcal{B}_{cc}^{+}}{\downarrow}\rangle$  &$\langle{\Lambda_{c}^{+}}{\downarrow}\vert\hat{\mathcal{O}}\vert{\mathcal{B}_{cc}^{+}}{\downarrow}\rangle$  &$\langle{\Sigma_{c}^{+}}{\downarrow}\vert\hat{\mathcal{O}}\vert{\mathcal{B}_{cc}^{+}}{\downarrow}\rangle$\\
\specialrule{0em}{1pt}{1pt}
\midrule[0.5pt]
$\hat{\alpha}_{1}^{(-)}\hat{\beta}_{2}^{(+)}(1-\pmb{\sigma}_{1}\cdot\pmb{\sigma}_{2})$  &$\sqrt{\frac{2}{3}}$  &$0$  &$\sqrt{\frac{2}{3}}$  &$0$\\
$\hat{\alpha}_{1}^{(-)}\hat{\beta}_{2}^{(+)}(1-\tilde{\pmb{\sigma}}_{1}\cdot\tilde{\pmb{\sigma}}_{2})$  &$\sqrt{\frac{2}{3}}$  &$0$  &$\sqrt{\frac{2}{3}}$  &$0$\\
\bottomrule[0.5pt]
\bottomrule[1pt]
\end{tabular*}
\end{table*}

\begin{table*}[htbp]\centering
\caption{The spin-flavor matrix elements for PV amplitudes of $W$-exchange diagrams, in which $\mathcal{B}_{cc}^{+}=(\Xi_{cc}^{+},\Omega_{cc}^{+})$, and the subscripts $\rho$ and $\lambda$ represent $\rho$-mode and $\lambda$-mode excitations respectively.}
\label{tab:polefactorPV}
\renewcommand\arraystretch{1.05}
\begin{tabular*}{160mm}{c@{\extracolsep{\fill}}cccccc}
\toprule[1pt]
\specialrule{0em}{1pt}{1pt}
\midrule[0.5pt]
\specialrule{0em}{1pt}{1pt}
matrix elements
&$\langle{\Xi_{c}^{+}}{\downarrow}\vert\hat{\mathcal{O}}\vert{\mathcal{B}_{cc}^{+}}{\downarrow}\rangle_{{}^{2}\lambda}$  &$\langle{\Xi_{c}^{+}}{\downarrow}\vert\hat{\mathcal{O}}\vert{\mathcal{B}_{cc}^{+}}{\downarrow}\rangle_{{}^{2}\rho}$
&$\langle{\Xi_{c}^{+}}{\downarrow}\vert\hat{\mathcal{O}}\vert{\mathcal{B}_{cc}^{+}}{\downarrow}\rangle_{{}^{4}\lambda}$ &$\langle{\Xi_{c}^{\prime+}}{\downarrow}\vert\hat{\mathcal{O}}\vert{\mathcal{B}_{cc}^{+}}{\downarrow}\rangle_{{}^{2}\lambda}$  &$\langle{\Xi_{c}^{\prime+}}{\downarrow}\vert\hat{\mathcal{O}}\vert{\mathcal{B}_{cc}^{+}}{\downarrow}\rangle_{{}^{2}\rho}$
&$\langle{\Xi_{c}^{\prime+}}{\downarrow}\vert\hat{\mathcal{O}}\vert{\mathcal{B}_{cc}^{+}}{\downarrow}\rangle_{{}^{4}\lambda}$\\
\specialrule{0em}{1pt}{1pt}
$-\hat{\alpha}_{1}^{(-)}\hat{\beta}_{2}^{(+)}(\pmb{\sigma}_{1}-\pmb{\sigma}_{2})_{z}$  &$\frac{1}{3\sqrt{6}}$  &$-\frac{1}{3\sqrt{2}}$  &$-\frac{2}{3\sqrt{3}}$
&$\frac{1}{3\sqrt{2}}$  &$\frac{1}{3\sqrt{6}}$   &$0$\\
$\hat{\alpha}_{1}^{(-)}\hat{\beta}_{2}^{(+)}(\pmb{\sigma}_{1}\times\pmb{\sigma}_{2})_{z}$  &$-\frac{i}{3\sqrt{6}}$  &$\frac{i}{3\sqrt{2}}$  &$\frac{2i}{3\sqrt{3}}$
&$\frac{i}{3\sqrt{2}}$  &$\frac{i}{3\sqrt{6}}$   &$0$\\
$-\hat{\alpha}_{1}^{(-)}\hat{\beta}_{2}^{(+)}(\tilde{\pmb{\sigma}}_{1}-\tilde{\pmb{\sigma}}_{2})_{z}$  &$\frac{1}{3\sqrt{6}}$  &$-\frac{1}{3\sqrt{2}}$  &$-\frac{2}{3\sqrt{3}}$
&$\frac{1}{3\sqrt{2}}$  &$\frac{1}{3\sqrt{6}}$   &$0$\\
$\hat{\alpha}_{1}^{(-)}\hat{\beta}_{2}^{(+)}(\tilde{\pmb{\sigma}}_{1}\times\tilde{\pmb{\sigma}}_{2})_{z}$  &$-\frac{i}{3\sqrt{6}}$  &$\frac{i}{3\sqrt{2}}$  &$\frac{2i}{3\sqrt{3}}$
&$\frac{i}{3\sqrt{2}}$  &$\frac{i}{3\sqrt{6}}$   &$0$\\
\specialrule{0em}{1pt}{1pt}
\midrule[0.5pt]
\specialrule{0em}{1pt}{1pt}
matrix elements
&$\langle{\Lambda_{c}^{+}}{\downarrow}\vert\hat{\mathcal{O}}\vert{\mathcal{B}_{cc}^{+}}{\downarrow}\rangle_{{}^{2}\lambda}$  &$\langle{\Lambda_{c}^{+}}{\downarrow}\vert\hat{\mathcal{O}}\vert{\mathcal{B}_{cc}^{+}}{\downarrow}\rangle_{{}^{2}\rho}$
&$\langle{\Lambda_{c}^{+}}{\downarrow}\vert\hat{\mathcal{O}}\vert{\mathcal{B}_{cc}^{+}}{\downarrow}\rangle_{{}^{4}\lambda}$ &$\langle{\Sigma_{c}^{+}}{\downarrow}\vert\hat{\mathcal{O}}\vert{\mathcal{B}_{cc}^{+}}{\downarrow}\rangle_{{}^{2}\lambda}$  &$\langle{\Sigma_{c}^{+}}{\downarrow}\vert\hat{\mathcal{O}}\vert{\mathcal{B}_{cc}^{+}}{\downarrow}\rangle_{{}^{2}\rho}$
&$\langle{\Sigma_{c}^{+}}{\downarrow}\vert\hat{\mathcal{O}}\vert{\mathcal{B}_{cc}^{+}}{\downarrow}\rangle_{{}^{4}\lambda}$\\
$-\hat{\alpha}_{1}^{(-)}\hat{\beta}_{2}^{(+)}(\pmb{\sigma}_{1}-\pmb{\sigma}_{2})_{z}$  &$\frac{1}{3\sqrt{6}}$  &$-\frac{1}{3\sqrt{2}}$  &$-\frac{2}{3\sqrt{3}}$
&$\frac{1}{3\sqrt{2}}$  &$\frac{1}{3\sqrt{6}}$   &$0$\\
$\hat{\alpha}_{1}^{(-)}\hat{\beta}_{2}^{(+)}(\pmb{\sigma}_{1}\times\pmb{\sigma}_{2})_{z}$  &$-\frac{i}{3\sqrt{6}}$  &$\frac{i}{3\sqrt{2}}$  &$\frac{2i}{3\sqrt{3}}$
&$\frac{i}{3\sqrt{2}}$  &$\frac{i}{3\sqrt{6}}$   &$0$\\
$-\hat{\alpha}_{1}^{(-)}\hat{\beta}_{2}^{(+)}(\tilde{\pmb{\sigma}}_{1}-\tilde{\pmb{\sigma}}_{2})_{z}$  &$\frac{1}{3\sqrt{6}}$  &$-\frac{1}{3\sqrt{2}}$  &$-\frac{2}{3\sqrt{3}}$
&$\frac{1}{3\sqrt{2}}$  &$\frac{1}{3\sqrt{6}}$   &$0$\\
$\hat{\alpha}_{1}^{(-)}\hat{\beta}_{2}^{(+)}(\tilde{\pmb{\sigma}}_{1}\times\tilde{\pmb{\sigma}}_{2})_{z}$  &$-\frac{i}{3\sqrt{6}}$  &$\frac{i}{3\sqrt{2}}$  &$\frac{2i}{3\sqrt{3}}$
&$\frac{i}{3\sqrt{2}}$  &$\frac{i}{3\sqrt{6}}$   &$0$\\
\specialrule{0em}{1pt}{1pt}
\midrule[0.5pt]
\specialrule{0em}{1pt}{1pt}
matrix elements
&${}_{{}^{2}\lambda}\langle{\Xi_{c}^{+}}{\downarrow}\vert\hat{\mathcal{O}}\vert{\Xi_{cc}^{+}}{\downarrow}\rangle$  &${}_{{}^{2}\lambda}\langle{\Xi_{c}^{\prime+}}{\downarrow}\vert\hat{\mathcal{O}}\vert{\Xi_{cc}^{+}}{\downarrow}\rangle$
&${}_{{}^{4}\lambda}\langle{\Xi_{c}^{\prime+}}{\downarrow}\vert\hat{\mathcal{O}}\vert{\Xi_{cc}^{+}}{\downarrow}\rangle$   &${}_{{}^{2}\lambda}\langle{\Lambda_{c}^{+}}{\downarrow}\vert\hat{\mathcal{O}}\vert{\Xi_{cc}^{+}}{\downarrow}\rangle$  &${}_{{}^{2}\lambda}\langle{\Sigma_{c}^{+}}{\downarrow}\vert\hat{\mathcal{O}}\vert{\Xi_{cc}^{+}}{\downarrow}\rangle$
&${}_{{}^{4}\lambda}\langle{\Sigma_{c}^{+}}{\downarrow}\vert\hat{\mathcal{O}}\vert{\Xi_{cc}^{+}}{\downarrow}\rangle$\\
$-\hat{\alpha}_{1}^{(-)}\hat{\beta}_{2}^{(+)}(\pmb{\sigma}_{1}-\pmb{\sigma}_{2})_{z}$  &$\frac{1}{3\sqrt{6}}$  &$\frac{1}{3\sqrt{2}}$  &$\frac{1}{3}$
&$\frac{1}{3\sqrt{6}}$  &$\frac{1}{3\sqrt{2}}$   &$\frac{1}{3}$\\
$\hat{\alpha}_{1}^{(-)}\hat{\beta}_{2}^{(+)}(\pmb{\sigma}_{1}\times\pmb{\sigma}_{2})_{z}$  &$-\frac{i}{3\sqrt{6}}$  &$\frac{i}{3\sqrt{2}}$  &$\frac{i}{3}$
&$-\frac{i}{3\sqrt{6}}$  &$\frac{i}{3\sqrt{2}}$   &$\frac{i}{3}$\\
$-\hat{\alpha}_{1}^{(-)}\hat{\beta}_{2}^{(+)}(\tilde{\pmb{\sigma}}_{1}-\tilde{\pmb{\sigma}}_{2})_{z}$  &$\frac{1}{3\sqrt{6}}$  &$\frac{1}{3\sqrt{2}}$  &$\frac{1}{3}$
&$\frac{1}{3\sqrt{6}}$  &$\frac{1}{3\sqrt{2}}$   &$\frac{1}{3}$\\
$\hat{\alpha}_{1}^{(-)}\hat{\beta}_{2}^{(+)}(\tilde{\pmb{\sigma}}_{1}\times\tilde{\pmb{\sigma}}_{2})_{z}$  &$-\frac{i}{3\sqrt{6}}$  &$\frac{i}{3\sqrt{2}}$  &$\frac{i}{3}$
&$-\frac{i}{3\sqrt{6}}$  &$\frac{i}{3\sqrt{2}}$   &$\frac{i}{3}$\\
\bottomrule[0.5pt]
\bottomrule[1pt]
\end{tabular*}
\end{table*}

In addition, our results of the spin-flavor matrix elements $\langle{\mathcal{B}^{\prime}{\downarrow}}\vert\hat{I}_{P}\sigma_{z}\vert{\mathcal{B}{\downarrow}}\rangle$ associated with the baryon-pseudoscalar meson vertices are summarized in Table~\ref{tab:gA}. The corresponding spin-up matrix elements can be evaluated analogously, while the spin-flip matrix elements vanish,
\begin{equation}
\langle{\mathcal{B}^{\prime}{\uparrow}}\vert\hat{I}_{P}\sigma_{z}\vert{\mathcal{B}{\downarrow}}\rangle=\langle{\mathcal{B}^{\prime}{\downarrow}}\vert\hat{I}_{P}\sigma_{z}\vert{\mathcal{B}{\uparrow}}\rangle=0.
\end{equation}
It is also found that the spin-flavor matrix elements between two antitriplet baryons vanish,
\begin{equation}
\langle{\mathcal{B}_{\bar{3}}^{\prime}}\vert\hat{I}_{P}\sigma_{z}\vert{\mathcal{B}_{\bar{3}}}\rangle=0.
\end{equation}

\begin{sidewaystable}[h!]
\centering
\caption{Spin-flavor matrix elements $\langle{\mathcal{B}^{\prime}{\downarrow}}\vert\hat{I}_{P}\sigma_{z}\vert{\mathcal{B}{\downarrow}}\rangle$, The first and second parts correspond to $E^{\prime}$ and $E_{1,2}$ diagrams, respectively. For the $\Xi_{c}^{(\prime)+}\to\Xi_{c}^{(\prime)+}\eta^{(\prime)}$ transitions, the two entries in braces correspond to the $\eta_{q}$ and $\eta_{s}$ components, respectively; all other $\eta^{(\prime)}$ transitions involve only the $\eta_{q}$ component.}\label{tab:gA}
\hspace*{-0.5cm}
\vspace*{-8cm}
\renewcommand\arraystretch{1.2}
\renewcommand\tabcolsep{0.65 pt}
\begin{tabular}{c@{\extracolsep{\fill}}ccccccccc}
\toprule[1.50pt]
\toprule[0.50pt]
matrix elements &values    &matrix elements &values    &matrix elements &values    &matrix elements &values    &matrix elements &values\\
\midrule[0.5pt]
$\langle{\Xi_{cc}^{+}{\downarrow}}\vert\hat{I}_{\pi^{0}}\sigma_{z}\vert{\Xi_{cc}^{+}{\downarrow}}\rangle$  &$-\frac{1}{3\sqrt{2}}$
&${}_{^{2}\lambda}\langle{\Xi_{cc}^{+}{\downarrow}}\vert\hat{I}_{\pi^{0}}\sigma_{z}\vert{\Xi_{cc}^{+}{\downarrow}}\rangle$  &$-\frac{1}{3\sqrt{2}}$
&${}_{^{2}\rho}\langle{\Xi_{cc}^{+}{\downarrow}}\vert\hat{I}_{\pi^{0}}\sigma_{z}\vert{\Xi_{cc}^{+}{\downarrow}}\rangle$  &$0$  &${}_{^{4}\lambda}\langle{\Xi_{cc}^{+}{\downarrow}}\vert\hat{I}_{\pi^{0}}\sigma_{z}\vert{\Xi_{cc}^{+}{\downarrow}}\rangle$  &$-\frac{2}{3}$  &&\\
$\langle{\Omega_{cc}^{+}{\downarrow}}\vert\hat{I}_{K^{0}}\sigma_{z}\vert{\Xi_{cc}^{+}{\downarrow}}\rangle$  &$\frac{1}{3}$
&${}_{^{2}\lambda}\langle{\Omega_{cc}^{+}{\downarrow}}\vert\hat{I}_{K^{0}}\sigma_{z}\vert{\Xi_{cc}^{+}{\downarrow}}\rangle$  &$\frac{1}{3}$
&${}_{^{2}\rho}\langle{\Omega_{cc}^{+}{\downarrow}}\vert\hat{I}_{K^{0}}\sigma_{z}\vert{\Xi_{cc}^{+}{\downarrow}}\rangle$  &$0$  &${}_{^{4}\lambda}\langle{\Omega_{cc}^{+}{\downarrow}}\vert\hat{I}_{K^{0}}\sigma_{z}\vert{\Xi_{cc}^{+}{\downarrow}}\rangle$  &$\frac{2\sqrt{2}}{3}$  &&\\
$\langle{\Xi_{cc}^{+}{\downarrow}}\vert\hat{I}_{\eta}\sigma_{z}\vert{\Xi_{cc}^{+}{\downarrow}}\rangle$  &$\frac{1}{3\sqrt{2}}\cos\zeta$
&${}_{^{2}{\lambda}}\langle{\Xi_{cc}^{+}{\downarrow}}\vert\hat{I}_{\eta}\sigma_{z}\vert{\Xi_{cc}^{+}{\downarrow}}\rangle$  &$\frac{1}{3\sqrt{2}}\cos\zeta$  &${}_{^{2}{\rho}}\langle{\Xi_{cc}^{+}{\downarrow}}\vert\hat{I}_{\eta}\sigma_{z}\vert{\Xi_{cc}^{+}{\downarrow}}\rangle$  &$0$ &${}_{^{4}{\lambda}}\langle{\Xi_{cc}^{+}{\downarrow}}\vert\hat{I}_{\eta}\sigma_{z}\vert{\Xi_{cc}^{+}{\downarrow}}\rangle$  &$-\frac{2}{3}\cos\zeta$  &  &\\
$\langle{\Xi_{cc}^{+}{\downarrow}}\vert\hat{I}_{\eta^{\prime}}\sigma_{z}\vert{\Xi_{cc}^{+}{\downarrow}}\rangle$  &$\frac{1}{3\sqrt{2}}\sin\zeta$  
&${}_{^{2}{\lambda}}\langle{\Xi_{cc}^{+}{\downarrow}}\vert\hat{I}_{\eta^{\prime}}\sigma_{z}\vert{\Xi_{cc}^{+}{\downarrow}}\rangle$ &$\frac{1}{3\sqrt{2}}\sin\zeta$
&${}_{^{2}{\rho}}\langle{\Xi_{cc}^{+}{\downarrow}}\vert\hat{I}_{\eta^{\prime}}\sigma_{z}\vert{\Xi_{cc}^{+}{\downarrow}}\rangle$  &$0$
&${}_{^{4}{\lambda}}\langle{\Xi_{cc}^{+}{\downarrow}}\vert\hat{I}_{\eta^{\prime}}\sigma_{z}\vert{\Xi_{cc}^{+}{\downarrow}}\rangle$  &$-\frac{2}{3}\sin\zeta$  &  &\\
\midrule[0.5pt]
$\langle{\Xi_{c}^{+}{\downarrow}}\vert\hat{I}_{\pi^{0}}\sigma_{z}\vert{\Xi_{c}^{+}{\downarrow}}\rangle$  &$0$
&$\langle{\Xi_{c}^{+}{\downarrow}}\vert\hat{I}_{\pi^{0}}\sigma_{z}\vert{\Xi_{c}^{+}{\downarrow}}\rangle_{^{2}\lambda}$  &$0$
&$\langle{\Xi_{c}^{+}{\downarrow}}\vert\hat{I}_{\pi^{0}}\sigma_{z}\vert{\Xi_{c}^{\prime+}{\downarrow}}\rangle$  &$\frac{1}{\sqrt{6}}$  &$\langle{\Xi_{c}^{+}{\downarrow}}\vert\hat{I}_{\pi^{0}}\sigma_{z}\vert{\Xi_{c}^{\prime+}{\downarrow}}\rangle_{^{2}\lambda}$  &$\frac{1}{\sqrt{6}}$  &$\langle{\Xi_{c}^{+}{\downarrow}}\vert\hat{I}_{\pi^{0}}\sigma_{z}\vert{\Xi_{c}^{\prime+}{\downarrow}}\rangle_{^{4}\lambda}$  &$\frac{1}{\sqrt{3}}$\\
$\langle{\Xi_{c}^{\prime+}{\downarrow}}\vert\hat{I}_{\pi^{0}}\sigma_{z}\vert{\Xi_{c}^{+}{\downarrow}}\rangle$  &$\frac{1}{\sqrt{6}}$
&$\langle{\Xi_{c}^{\prime+}{\downarrow}}\vert\hat{I}_{\pi^{0}}\sigma_{z}\vert{\Xi_{c}^{+}{\downarrow}}\rangle_{^{2}\lambda}$  &$\frac{1}{\sqrt{6}}$
&$\langle{\Xi_{c}^{\prime+}{\downarrow}}\vert\hat{I}_{\pi^{0}}\sigma_{z}\vert{\Xi_{c}^{\prime+}{\downarrow}}\rangle$  &$-\frac{\sqrt{2}}{3}$  &$\langle{\Xi_{c}^{\prime+}{\downarrow}}\vert\hat{I}_{\pi^{0}}\sigma_{z}\vert{\Xi_{c}^{\prime+}{\downarrow}}\rangle_{^{2}\lambda}$  &$-\frac{\sqrt{2}}{3}$  &$\langle{\Xi_{c}^{\prime+}{\downarrow}}\vert\hat{I}_{\pi^{0}}\sigma_{z}\vert{\Xi_{c}^{\prime+}{\downarrow}}\rangle_{^{4}\lambda}$  &$\frac{1}{3}$\\
$\langle{\Xi_{c}^{0}{\downarrow}}\vert\hat{I}_{\pi^{+}}\sigma_{z}\vert{\Xi_{c}^{+}{\downarrow}}\rangle$  &$0$
&$\langle{\Xi_{c}^{0}{\downarrow}}\vert\hat{I}_{\pi^{+}}\sigma_{z}\vert{\Xi_{c}^{+}{\downarrow}}\rangle_{^{2}\lambda}$  &$0$
&$\langle{\Xi_{c}^{0}{\downarrow}}\vert\hat{I}_{\pi^{+}}\sigma_{z}\vert{\Xi_{c}^{\prime+}{\downarrow}}\rangle$  &$\frac{1}{\sqrt{3}}$  &$\langle{\Xi_{c}^{0}{\downarrow}}\vert\hat{I}_{\pi^{+}}\sigma_{z}\vert{\Xi_{c}^{\prime+}{\downarrow}}\rangle_{^{2}\lambda}$  &$\frac{1}{\sqrt{3}}$  &$\langle{\Xi_{c}^{0}{\downarrow}}\vert\hat{I}_{\pi^{+}}\sigma_{z}\vert{\Xi_{c}^{\prime+}{\downarrow}}\rangle_{^{4}\lambda}$  &$\sqrt{\frac{2}{3}}$\\
$\langle{\Xi_{c}^{\prime0}{\downarrow}}\vert\hat{I}_{\pi^{+}}\sigma_{z}\vert{\Xi_{c}^{+}{\downarrow}}\rangle$  &$\frac{1}{\sqrt{3}}$
&$\langle{\Xi_{c}^{\prime0}{\downarrow}}\vert\hat{I}_{\pi^{+}}\sigma_{z}\vert{\Xi_{c}^{+}{\downarrow}}\rangle_{^{2}\lambda}$  &$\frac{1}{\sqrt{3}}$
&$\langle{\Xi_{c}^{\prime0}{\downarrow}}\vert\hat{I}_{\pi^{+}}\sigma_{z}\vert{\Xi_{c}^{\prime+}{\downarrow}}\rangle$  &$-\frac{2}{3}$  &$\langle{\Xi_{c}^{\prime0}{\downarrow}}\vert\hat{I}_{\pi^{+}}\sigma_{z}\vert{\Xi_{c}^{\prime+}{\downarrow}}\rangle_{^{2}\lambda}$  &$-\frac{2}{3}$  &$\langle{\Xi_{c}^{\prime0}{\downarrow}}\vert\hat{I}_{\pi^{+}}\sigma_{z}\vert{\Xi_{c}^{\prime+}{\downarrow}}\rangle_{^{4}\lambda}$  &$\frac{\sqrt{2}}{3}$\\
$\langle{\Omega_{c}^{0}{\downarrow}}\vert\hat{I}_{K^{+}}\sigma_{z}\vert{\Xi_{c}^{+}{\downarrow}}\rangle$  &$\sqrt{\frac{2}{3}}$
&$\langle{\Omega_{c}^{0}{\downarrow}}\vert\hat{I}_{K^{+}}\sigma_{z}\vert{\Xi_{c}^{+}{\downarrow}}\rangle_{^{2}\lambda}$  &$\sqrt{\frac{2}{3}}$
&$\langle{\Omega_{c}^{0}{\downarrow}}\vert\hat{I}_{K^{+}}\sigma_{z}\vert{\Xi_{c}^{\prime+}{\downarrow}}\rangle$  &$-\frac{2\sqrt{2}}{3}$  &$\langle{\Omega_{c}^{0}{\downarrow}}\vert\hat{I}_{K^{+}}\sigma_{z}\vert{\Xi_{c}^{\prime+}{\downarrow}}\rangle_{^{2}\lambda}$  &$-\frac{2\sqrt{2}}{3}$  &$\langle{\Omega_{c}^{0}{\downarrow}}\vert\hat{I}_{K^{+}}\sigma_{z}\vert{\Xi_{c}^{\prime+}{\downarrow}}\rangle_{^{4}\lambda}$  &$\frac{2}{3}$\\
$\langle{\Sigma_{c}^{++}{\downarrow}}\vert\hat{I}_{K^{-}}\sigma_{z}\vert{\Xi_{c}^{+}{\downarrow}}\rangle$  &$-\sqrt{\frac{2}{3}}$
&$\langle{\Sigma_{c}^{++}{\downarrow}}\vert\hat{I}_{K^{-}}\sigma_{z}\vert{\Xi_{c}^{+}{\downarrow}}\rangle_{^{2}\lambda}$  &$-\sqrt{\frac{2}{3}}$
&$\langle{\Sigma_{c}^{++}{\downarrow}}\vert\hat{I}_{K^{-}}\sigma_{z}\vert{\Xi_{c}^{\prime+}{\downarrow}}\rangle$  &$-\frac{2\sqrt{2}}{3}$  &$\langle{\Sigma_{c}^{++}{\downarrow}}\vert\hat{I}_{K^{-}}\sigma_{z}\vert{\Xi_{c}^{\prime+}{\downarrow}}\rangle_{^{2}\lambda}$  &$-\frac{2\sqrt{2}}{3}$  &$\langle{\Sigma_{c}^{++}{\downarrow}}\vert\hat{I}_{K^{-}}\sigma_{z}\vert{\Xi_{c}^{\prime+}{\downarrow}}\rangle_{^{4}\lambda}$  &$\frac{2}{3}$\\
$\langle{\Lambda_{c}^{+}{\downarrow}}\vert\hat{I}_{\bar{K}^{0}}\sigma_{z}\vert{\Xi_{c}^{+}{\downarrow}}\rangle$  &$0$
&$\langle{\Lambda_{c}^{+}{\downarrow}}\vert\hat{I}_{\bar{K}^{0}}\sigma_{z}\vert{\Xi_{c}^{+}{\downarrow}}\rangle_{^{2}\lambda}$  &$0$
&$\langle{\Lambda_{c}^{+}{\downarrow}}\vert\hat{I}_{\bar{K}^{0}}\sigma_{z}\vert{\Xi_{c}^{\prime+}{\downarrow}}\rangle$  &$-\frac{1}{\sqrt{3}}$  &$\langle{\Lambda_{c}^{+}{\downarrow}}\vert\hat{I}_{\bar{K}^{0}}\sigma_{z}\vert{\Xi_{c}^{\prime+}{\downarrow}}\rangle_{^{2}\lambda}$  &$-\frac{1}{\sqrt{3}}$  &$\langle{\Lambda_{c}^{+}{\downarrow}}\vert\hat{I}_{\bar{K}^{0}}\sigma_{z}\vert{\Xi_{c}^{\prime+}{\downarrow}}\rangle_{^{4}\lambda}$  &$-\sqrt{\frac{2}{3}}$\\
$\langle{\Sigma_{c}^{+}{\downarrow}}\vert\hat{I}_{\bar{K}^{0}}\sigma_{z}\vert{\Xi_{c}^{+}{\downarrow}}\rangle$  &$-\frac{1}{\sqrt{3}}$
&$\langle{\Sigma_{c}^{+}{\downarrow}}\vert\hat{I}_{\bar{K}^{0}}\sigma_{z}\vert{\Xi_{c}^{+}{\downarrow}}\rangle_{^{2}\lambda}$  &$-\frac{1}{\sqrt{3}}$
&$\langle{\Sigma_{c}^{+}{\downarrow}}\vert\hat{I}_{\bar{K}^{0}}\sigma_{z}\vert{\Xi_{c}^{\prime+}{\downarrow}}\rangle$  &$-\frac{2}{3}$  &$\langle{\Sigma_{c}^{+}{\downarrow}}\vert\hat{I}_{\bar{K}^{0}}\sigma_{z}\vert{\Xi_{c}^{\prime+}{\downarrow}}\rangle_{^{2}\lambda}$  &$-\frac{2}{3}$  &$\langle{\Sigma_{c}^{+}{\downarrow}}\vert\hat{I}_{\bar{K}^{0}}\sigma_{z}\vert{\Xi_{c}^{\prime+}{\downarrow}}\rangle_{^{4}\lambda}$  &$\frac{\sqrt{2}}{3}$\\
$\langle{\Lambda_{c}^{+}{\downarrow}}\vert\hat{I}_{\pi^{0}}\sigma_{z}\vert{\Lambda_{c}^{+}{\downarrow}}\rangle$  &$0$
&$\langle{\Lambda_{c}^{+}{\downarrow}}\vert\hat{I}_{\pi^{0}}\sigma_{z}\vert{\Lambda_{c}^{+}{\downarrow}}\rangle_{^{2}\lambda}$  &$0$
&$\langle{\Lambda_{c}^{+}{\downarrow}}\vert\hat{I}_{\pi^{0}}\sigma_{z}\vert{\Sigma_{c}^{+}{\downarrow}}\rangle$  &$\sqrt{\frac{2}{3}}$  &$\langle{\Lambda_{c}^{+}{\downarrow}}\vert\hat{I}_{\pi^{0}}\sigma_{z}\vert{\Sigma_{c}^{+}{\downarrow}}\rangle_{^{2}\lambda}$  &$\sqrt{\frac{2}{3}}$  &$\langle{\Lambda_{c}^{+}{\downarrow}}\vert\hat{I}_{\pi^{0}}\sigma_{z}\vert{\Sigma_{c}^{+}{\downarrow}}\rangle_{^{4}\lambda}$  &$\frac{2}{\sqrt{3}}$\\
$\langle{\Sigma_{c}^{+}{\downarrow}}\vert\hat{I}_{\pi^{0}}\sigma_{z}\vert{\Lambda_{c}^{+}{\downarrow}}\rangle$  &$\sqrt{\frac{2}{3}}$
&$\langle{\Sigma_{c}^{+}{\downarrow}}\vert\hat{I}_{\pi^{0}}\sigma_{z}\vert{\Lambda_{c}^{+}{\downarrow}}\rangle_{^{2}\lambda}$  &$\sqrt{\frac{2}{3}}$
&$\langle{\Sigma_{c}^{+}{\downarrow}}\vert\hat{I}_{\pi^{0}}\sigma_{z}\vert{\Sigma_{c}^{+}{\downarrow}}\rangle$  &$0$  &$\langle{\Sigma_{c}^{+}{\downarrow}}\vert\hat{I}_{\pi^{0}}\sigma_{z}\vert{\Sigma_{c}^{+}{\downarrow}}\rangle_{^{2}\lambda}$  &$0$  &$\langle{\Sigma_{c}^{+}{\downarrow}}\vert\hat{I}_{\pi^{0}}\sigma_{z}\vert{\Sigma_{c}^{+}{\downarrow}}\rangle_{^{4}\lambda}$  &$0$\\
$\langle{\Sigma_{c}^{0}{\downarrow}}\vert\hat{I}_{\pi^{+}}\sigma_{z}\vert{\Lambda_{c}^{+}{\downarrow}}\rangle$  &$\sqrt{\frac{2}{3}}$
&$\langle{\Sigma_{c}^{0}{\downarrow}}\vert\hat{I}_{\pi^{+}}\sigma_{z}\vert{\Lambda_{c}^{+}{\downarrow}}\rangle_{^{2}\lambda}$  &$\sqrt{\frac{2}{3}}$
&$\langle{\Sigma_{c}^{0}{\downarrow}}\vert\hat{I}_{\pi^{+}}\sigma_{z}\vert{\Sigma_{c}^{+}{\downarrow}}\rangle$  &$-\frac{2\sqrt{2}}{3}$  &$\langle{\Sigma_{c}^{0}{\downarrow}}\vert\hat{I}_{\pi^{+}}\sigma_{z}\vert{\Sigma_{c}^{+}{\downarrow}}\rangle_{^{2}\lambda}$  &$-\frac{2\sqrt{2}}{3}$  &$\langle{\Sigma_{c}^{0}{\downarrow}}\vert\hat{I}_{\pi^{+}}\sigma_{z}\vert{\Sigma_{c}^{+}{\downarrow}}\rangle_{^{4}\lambda}$  &$\frac{2}{3}$\\
$\langle{\Xi_{c}^{0}{\downarrow}}\vert\hat{I}_{K^{+}}\sigma_{z}\vert{\Lambda_{c}^{+}{\downarrow}}\rangle$  &$0$
&$\langle{\Xi_{c}^{0}{\downarrow}}\vert\hat{I}_{K^{+}}\sigma_{z}\vert{\Lambda_{c}^{+}{\downarrow}}\rangle_{^{2}\lambda}$  &$0$
&$\langle{\Xi_{c}^{0}{\downarrow}}\vert\hat{I}_{K^{+}}\sigma_{z}\vert{\Sigma_{c}^{+}{\downarrow}}\rangle$  &$-\frac{1}{\sqrt{3}}$  &$\langle{\Xi_{c}^{0}{\downarrow}}\vert\hat{I}_{K^{+}}\sigma_{z}\vert{\Sigma_{c}^{+}{\downarrow}}\rangle_{^{2}\lambda}$  &$-\frac{1}{\sqrt{3}}$  &$\langle{\Xi_{c}^{0}{\downarrow}}\vert\hat{I}_{K^{+}}\sigma_{z}\vert{\Sigma_{c}^{+}{\downarrow}}\rangle_{^{4}\lambda}$  &$-\sqrt{\frac{2}{3}}$\\
$\langle{\Xi_{c}^{\prime0}{\downarrow}}\vert\hat{I}_{K^{+}}\sigma_{z}\vert{\Lambda_{c}^{+}{\downarrow}}\rangle$  &$\frac{1}{\sqrt{3}}$
&$\langle{\Xi_{c}^{\prime0}{\downarrow}}\vert\hat{I}_{K^{+}}\sigma_{z}\vert{\Lambda_{c}^{+}{\downarrow}}\rangle_{^{2}\lambda}$  &$\frac{1}{\sqrt{3}}$
&$\langle{\Xi_{c}^{\prime0}{\downarrow}}\vert\hat{I}_{K^{+}}\sigma_{z}\vert{\Sigma_{c}^{+}{\downarrow}}\rangle$  &$-\frac{2}{3}$  &$\langle{\Xi_{c}^{\prime0}{\downarrow}}\vert\hat{I}_{K^{+}}\sigma_{z}\vert{\Sigma_{c}^{+}{\downarrow}}\rangle_{^{2}\lambda}$  &$-\frac{2}{3}$  &$\langle{\Xi_{c}^{\prime0}{\downarrow}}\vert\hat{I}_{K^{+}}\sigma_{z}\vert{\Sigma_{c}^{+}{\downarrow}}\rangle_{^{4}\lambda}$  &$\frac{\sqrt{2}}{3}$\\
$\langle{\Sigma_{c}^{++}{\downarrow}}\vert\hat{I}_{\pi^{-}}\sigma_{z}\vert{\Lambda_{c}^{+}{\downarrow}}\rangle$  &$-\sqrt{\frac{2}{3}}$
&$\langle{\Sigma_{c}^{++}{\downarrow}}\vert\hat{I}_{\pi^{-}}\sigma_{z}\vert{\Lambda_{c}^{+}{\downarrow}}\rangle_{^{2}\lambda}$  &$-\sqrt{\frac{2}{3}}$
&$\langle{\Sigma_{c}^{++}{\downarrow}}\vert\hat{I}_{\pi^{-}}\sigma_{z}\vert{\Sigma_{c}^{+}{\downarrow}}\rangle$  &$-\frac{2\sqrt{2}}{3}$  &$\langle{\Sigma_{c}^{++}{\downarrow}}\vert\hat{I}_{\pi^{-}}\sigma_{z}\vert{\Sigma_{c}^{+}{\downarrow}}\rangle_{^{2}\lambda}$  &$-\frac{2\sqrt{2}}{3}$  &$\langle{\Sigma_{c}^{++}{\downarrow}}\vert\hat{I}_{\pi^{-}}\sigma_{z}\vert{\Sigma_{c}^{+}{\downarrow}}\rangle_{^{4}\lambda}$  &$\frac{2}{3}$\\
$\langle{\Xi_{c}^{+}{\downarrow}}\vert\hat{I}_{K^{0}}\sigma_{z}\vert{\Lambda_{c}^{+}{\downarrow}}\rangle$  &$0$
&$\langle{\Xi_{c}^{+}{\downarrow}}\vert\hat{I}_{K^{0}}\sigma_{z}\vert{\Lambda_{c}^{+}{\downarrow}}\rangle_{^{2}\lambda}$  &$0$
&$\langle{\Xi_{c}^{+}{\downarrow}}\vert\hat{I}_{K^{0}}\sigma_{z}\vert{\Sigma_{c}^{+}{\downarrow}}\rangle$  &$-\frac{1}{\sqrt{3}}$  &$\langle{\Xi_{c}^{+}{\downarrow}}\vert\hat{I}_{K^{0}}\sigma_{z}\vert{\Sigma_{c}^{+}{\downarrow}}\rangle_{^{2}\lambda}$  &$-\frac{1}{\sqrt{3}}$  &$\langle{\Xi_{c}^{+}{\downarrow}}\vert\hat{I}_{K^{0}}\sigma_{z}\vert{\Sigma_{c}^{+}{\downarrow}}\rangle_{^{4}\lambda}$  &$-\sqrt{\frac{2}{3}}$\\
$\langle{\Xi_{c}^{\prime+}{\downarrow}}\vert\hat{I}_{K^{0}}\sigma_{z}\vert{\Lambda_{c}^{+}{\downarrow}}\rangle$  &$-\frac{1}{\sqrt{3}}$
&$\langle{\Xi_{c}^{\prime+}{\downarrow}}\vert\hat{I}_{K^{0}}\sigma_{z}\vert{\Lambda_{c}^{+}{\downarrow}}\rangle_{^{2}\lambda}$  &$-\frac{1}{\sqrt{3}}$
&$\langle{\Xi_{c}^{\prime+}{\downarrow}}\vert\hat{I}_{K^{0}}\sigma_{z}\vert{\Sigma_{c}^{+}{\downarrow}}\rangle$  &$-\frac{2}{3}$  &$\langle{\Xi_{c}^{\prime+}{\downarrow}}\vert\hat{I}_{K^{0}}\sigma_{z}\vert{\Sigma_{c}^{+}{\downarrow}}\rangle_{^{2}\lambda}$  &$-\frac{2}{3}$  &$\langle{\Xi_{c}^{\prime+}{\downarrow}}\vert\hat{I}_{K^{0}}\sigma_{z}\vert{\Sigma_{c}^{+}{\downarrow}}\rangle_{^{4}\lambda}$  &$\frac{\sqrt{2}}{3}$\\
$\langle{\Xi_{c}^{+}{\downarrow}}\vert\hat{I}_{\eta}\sigma_{z}\vert{\Xi_{c}^{+}{\downarrow}}\rangle$  &$\{0,0\}$  &$\langle{\Xi_{c}^{+}{\downarrow}}\vert\hat{I}_{\eta}\sigma_{z}\vert{\Xi_{c}^{+}{\downarrow}}\rangle_{^{2}{\lambda}}$  &$\{0,0\}$
&$\langle{\Xi_{c}^{+}{\downarrow}}\vert\hat{I}_{\eta}\sigma_{z}\vert{\Xi_{c}^{\prime+}{\downarrow}}\rangle$  &$\{\frac{\cos\zeta}{\sqrt{6}},\frac{\sin\zeta}{\sqrt{3}}\}$  &$\langle{\Xi_{c}^{+}{\downarrow}}\vert\hat{I}_{\eta}\sigma_{z}\vert{\Xi_{c}^{\prime+}{\downarrow}}\rangle_{^{2}{\lambda}}$  &$\{\frac{\cos\zeta}{\sqrt{6}},\frac{\sin\zeta}{\sqrt{3}}\}$  &$\langle{\Xi_{c}^{+}{\downarrow}}\vert\hat{I}_{\eta}\sigma_{z}\vert{\Xi_{c}^{\prime+}{\downarrow}}\rangle_{^{4}{\lambda}}$  &$\{\frac{\cos\zeta}{\sqrt{3}},\sqrt{\frac{2\sin\zeta}{3}}\}$\\
$\langle{\Xi_{c}^{\prime+}{\downarrow}}\vert\hat{I}_{\eta}\sigma_{z}\vert{\Xi_{c}^{+}{\downarrow}}\rangle$  &$\{\frac{\cos\zeta}{\sqrt{6}},\frac{\sin\zeta}{\sqrt{3}}\}$  &$\langle{\Xi_{c}^{\prime+}{\downarrow}}\vert\hat{I}_{\eta}\sigma_{z}\vert{\Xi_{c}^{+}{\downarrow}}\rangle_{^{2}{\lambda}}$  &$\{\frac{\cos\zeta}{\sqrt{6}},\frac{\sin\zeta}{\sqrt{3}}\}$
&$\langle{\Xi_{c}^{\prime+}{\downarrow}}\vert\hat{I}_{\eta}\sigma_{z}\vert{\Xi_{c}^{\prime+}{\downarrow}}\rangle$  &$\{-\frac{\sqrt{2}\cos\zeta}{3},\frac{2\sin\zeta}{3}\}$  &$\langle{\Xi_{c}^{\prime+}{\downarrow}}\vert\hat{I}_{\eta}\sigma_{z}\vert{\Xi_{c}^{\prime+}{\downarrow}}\rangle_{^{2}{\lambda}}$  &$\{-\frac{\sqrt{2}\cos\zeta}{3},\frac{2\sin\zeta}{3}\}$  &$\langle{\Xi_{c}^{\prime+}{\downarrow}}\vert\hat{I}_{\eta}\sigma_{z}\vert{\Xi_{c}^{\prime+}{\downarrow}}\rangle_{^{4}{\lambda}}$  &$\{\frac{\cos\zeta}{3},-\frac{\sqrt{2}\sin\zeta}{3}\}$\\
$\langle{\Xi_{c}^{+}{\downarrow}}\vert\hat{I}_{\eta^{\prime}}\sigma_{z}\vert{\Xi_{c}^{+}{\downarrow}}\rangle$  &$\{0,0\}$  &$\langle{\Xi_{c}^{+}{\downarrow}}\vert\hat{I}_{\eta^{\prime}}\sigma_{z}\vert{\Xi_{c}^{+}{\downarrow}}\rangle_{^{2}{\lambda}}$  &$\{0,0\}$
&$\langle{\Xi_{c}^{+}{\downarrow}}\vert\hat{I}_{\eta^{\prime}}\sigma_{z}\vert{\Xi_{c}^{\prime+}{\downarrow}}\rangle$  &$\{\frac{\sin\zeta}{\sqrt{6}},-\frac{\cos\zeta}{\sqrt{3}}\}$  &$\langle{\Xi_{c}^{+}{\downarrow}}\vert\hat{I}_{\eta^{\prime}}\sigma_{z}\vert{\Xi_{c}^{\prime+}{\downarrow}}\rangle_{^{2}{\lambda}}$  &$\{\frac{\sin\zeta}{\sqrt{6}},-\frac{\cos\zeta}{\sqrt{3}}\}$  &$\langle{\Xi_{c}^{+}{\downarrow}}\vert\hat{I}_{\eta^{\prime}}\sigma_{z}\vert{\Xi_{c}^{\prime+}{\downarrow}}\rangle_{^{4}{\lambda}}$  &$\{\frac{\sin\zeta}{\sqrt{3}},-\sqrt{\frac{2\cos\zeta}{3}}\}$\\
$\langle{\Xi_{c}^{\prime+}{\downarrow}}\vert\hat{I}_{\eta^{\prime}}\sigma_{z}\vert{\Xi_{c}^{+}{\downarrow}}\rangle$  &$\{\frac{\sin\zeta}{\sqrt{6}},-\frac{\cos\zeta}{\sqrt{3}}\}$  &$\langle{\Xi_{c}^{\prime+}{\downarrow}}\vert\hat{I}_{\eta^{\prime}}\sigma_{z}\vert{\Xi_{c}^{+}{\downarrow}}\rangle_{^{2}{\lambda}}$  &$\{\frac{\sin\zeta}{\sqrt{6}},-\frac{\cos\zeta}{\sqrt{3}}\}$
&$\langle{\Xi_{c}^{\prime+}{\downarrow}}\vert\hat{I}_{\eta^{\prime}}\sigma_{z}\vert{\Xi_{c}^{\prime+}{\downarrow}}\rangle$  &$\{-\frac{\sqrt{2}\sin\zeta}{3},-\frac{2\cos\zeta}{3}\}$  &$\langle{\Xi_{c}^{\prime+}{\downarrow}}\vert\hat{I}_{\eta^{\prime}}\sigma_{z}\vert{\Xi_{c}^{\prime+}{\downarrow}}\rangle_{^{2}{\lambda}}$  &$\{-\frac{\sqrt{2}\sin\zeta}{3},-\frac{2\cos\zeta}{3}\}$  &$\langle{\Xi_{c}^{\prime+}{\downarrow}}\vert\hat{I}_{\eta^{\prime}}\sigma_{z}\vert{\Xi_{c}^{\prime+}{\downarrow}}\rangle_{^{4}{\lambda}}$  &$\{\frac{\sin\zeta}{3},\frac{\sqrt{2}\cos\zeta}{3}\}$\\
$\langle{\Lambda_{c}^{+}{\downarrow}}\vert\hat{I}_{\eta}\sigma_{z}\vert{\Lambda_{c}^{+}{\downarrow}}\rangle$  &$0$  &$\langle{\Lambda_{c}^{+}{\downarrow}}\vert\hat{I}_{\eta}\sigma_{z}\vert{\Lambda_{c}^{+}{\downarrow}}\rangle_{^{2}{\lambda}}$  &$0$
&$\langle{\Lambda_{c}^{+}{\downarrow}}\vert\hat{I}_{\eta}\sigma_{z}\vert{\Sigma_{c}^{+}{\downarrow}}\rangle$  &$0$  &$\langle{\Lambda_{c}^{+}{\downarrow}}\vert\hat{I}_{\eta}\sigma_{z}\vert{\Sigma_{c}^{+}{\downarrow}}\rangle_{^{2}{\lambda}}$  &$0$  &$\langle{\Lambda_{c}^{+}{\downarrow}}\vert\hat{I}_{\eta}\sigma_{z}\vert{\Sigma_{c}^{+}{\downarrow}}\rangle_{^{4}{\lambda}}$  &$0$\\
$\langle{\Sigma_{c}^{+}{\downarrow}}\vert\hat{I}_{\eta}\sigma_{z}\vert{\Lambda_{c}^{+}{\downarrow}}\rangle$  &$0$  &$\langle{\Sigma_{c}^{+}{\downarrow}}\vert\hat{I}_{\eta}\sigma_{z}\vert{\Lambda_{c}^{+}{\downarrow}}\rangle_{^{2}{\lambda}}$  &$0$
&$\langle{\Sigma_{c}^{+}{\downarrow}}\vert\hat{I}_{\eta}\sigma_{z}\vert{\Sigma_{c}^{+}{\downarrow}}\rangle$  &$-\frac{2\sqrt{2}\cos\zeta}{3}$  &$\langle{\Sigma_{c}^{+}{\downarrow}}\vert\hat{I}_{\eta}\sigma_{z}\vert{\Sigma_{c}^{+}{\downarrow}}\rangle_{^{2}{\lambda}}$  &$-\frac{2\sqrt{2}\cos\zeta}{3}$  &$\langle{\Sigma_{c}^{+}{\downarrow}}\vert\hat{I}_{\eta}\sigma_{z}\vert{\Sigma_{c}^{+}{\downarrow}}\rangle_{^{4}{\lambda}}$  &$\frac{2\cos\zeta}{3}$\\
$\langle{\Lambda_{c}^{+}{\downarrow}}\vert\hat{I}_{\eta^{\prime}}\sigma_{z}\vert{\Lambda_{c}^{+}{\downarrow}}\rangle$  &$0$  &$\langle{\Lambda_{c}^{+}{\downarrow}}\vert\hat{I}_{\eta^{\prime}}\sigma_{z}\vert{\Lambda_{c}^{+}{\downarrow}}\rangle_{^{2}{\lambda}}$  &$0$
&$\langle{\Lambda_{c}^{+}{\downarrow}}\vert\hat{I}_{\eta^{\prime}}\sigma_{z}\vert{\Sigma_{c}^{+}{\downarrow}}\rangle$  &$0$  &$\langle{\Lambda_{c}^{+}{\downarrow}}\vert\hat{I}_{\eta^{\prime}}\sigma_{z}\vert{\Sigma_{c}^{+}{\downarrow}}\rangle_{^{2}{\lambda}}$  &$0$  &$\langle{\Lambda_{c}^{+}{\downarrow}}\vert\hat{I}_{\eta^{\prime}}\sigma_{z}\vert{\Sigma_{c}^{+}{\downarrow}}\rangle_{^{4}{\lambda}}$  &$0$\\
$\langle{\Sigma_{c}^{+}{\downarrow}}\vert\hat{I}_{\eta^{\prime}}\sigma_{z}\vert{\Lambda_{c}^{+}{\downarrow}}\rangle$  &$0$  &$\langle{\Sigma_{c}^{+}{\downarrow}}\vert\hat{I}_{\eta^{\prime}}\sigma_{z}\vert{\Lambda_{c}^{+}{\downarrow}}\rangle_{^{2}{\lambda}}$  &$0$
&$\langle{\Sigma_{c}^{+}{\downarrow}}\vert\hat{I}_{\eta^{\prime}}\sigma_{z}\vert{\Sigma_{c}^{+}{\downarrow}}\rangle$  &$-\frac{2\sqrt{2}\sin\zeta}{3}$  &$\langle{\Sigma_{c}^{+}{\downarrow}}\vert\hat{I}_{\eta^{\prime}}\sigma_{z}\vert{\Sigma_{c}^{+}{\downarrow}}\rangle_{^{2}{\lambda}}$  &$-\frac{2\sqrt{2}\sin\zeta}{3}$  &$\langle{\Sigma_{c}^{+}{\downarrow}}\vert\hat{I}_{\eta^{\prime}}\sigma_{z}\vert{\Sigma_{c}^{+}{\downarrow}}\rangle_{^{4}{\lambda}}$  &$\frac{2\sin\zeta}{3}$\\
\bottomrule[0.50pt]
\bottomrule[1.50pt]
\end{tabular}
\end{sidewaystable}

\end{document}